\RequirePackage{fix-cm}
\RequirePackage{fixltx2e} 
\documentclass[aps,pre,twocolumn,superscriptaddress,10pt,nofootinbib]{revtex4-1}

\usepackage{subfigure}
\usepackage{amsfonts, amssymb, amsmath}
\usepackage{bm}
\usepackage{graphics}
\usepackage{graphicx}
\usepackage{wasysym}
\usepackage{natbib}
\usepackage{mathtools}
\usepackage[utf8x]{inputenc}
\usepackage[T1]{fontenc}
\usepackage{esint}
\usepackage{bbold}
\usepackage{comment}
\usepackage{efbox}
\usepackage{paralist}
\usepackage{soul}
\usepackage[usenames,dvipsnames]{xcolor}
\usepackage{color}
\definecolor{darkblue}{rgb}{0,0,0.6}
\definecolor{darkred}{rgb}{0.6,0,0}
\usepackage{tikz}
\usetikzlibrary{matrix,patterns}
\usetikzlibrary{decorations.pathreplacing}
\usetikzlibrary{shapes.misc}
\tikzset{cross/.style={cross out, draw=black, minimum size=2*(#1-\pgflinewidth), inner sep=0pt, outer sep=0pt},
cross/.default={1pt}}

\usepackage{hyperref}
\hypersetup{
bookmarksopen=true,
pdftitle= Dynamic renormalization group theory for open Floquet systems, 
pdfauthor= Steven Mathey et al., 
pdftoolbar=false, 
pdfstartview={FitH},		
pdfmenubar=true,			
pdfhighlight=/O,			
colorlinks=true,			
urlcolor=darkblue,
citecolor=darkblue,		
linkcolor=MidnightBlue,	
}

\renewcommand{\vec}[1]{{\boldsymbol{#1}}}
\newcommand{\tx}{{t,\vec{x}}}
\newcommand{\txp}{{t',\vec{x}'}}
\newcommand{\op}{{\omega,\vec{p}}}
\newcommand{\I}{\text{i}}
\newcommand{\ie}{i.e.~}
\newcommand{\Ie}{I.e.~}
\newcommand{\eg}{e.g.~}

\newcommand{\Fig}[2]{Fig.~\ref{#1}\textcolor{MidnightBlue}{#2}}
\newcommand{\fig}[2]{\ref{#1}\textcolor{MidnightBlue}{#2}}
\newcommand{\Figs}[2]{Figs.~\ref{#1}\textcolor{MidnightBlue}{#2}}
\newcommand{\Sect}[1]{Sect.~\ref{#1}}
\newcommand{\Sects}[1]{Sects.~\ref{#1}}
\newcommand{\sect}[1]{\ref{#1}}
\newcommand{\Eq}[1]{Eq.~(\ref{#1})}
\newcommand{\App}[1]{App.~\ref{#1}}

\newcommand{\eq}[1]{(\ref{#1})}
\newcommand{\Eqs}[1]{Eqs.~(\ref{#1})}

\usepackage{abbrevs}
\usepackage{etoolbox}
\newabbrev\RG{Renormalization Group (RG)}[RG]
\newabbrev\UV{Ultraviolet (UV)}[UV]
\newabbrev\WF{Wilson--Fisher (WF)}[WF]
\newabbrev\FBZs{Floquet--Brillouin Zones (FBZ)\vphantom{\FBZ}}[FBZs]
\newabbrev\FBZ{Floquet--Brillouin Zone (FBZ)\vphantom{\FBZs}}[FBZ]
\newabbrev\KZ{Kibble--Zurek (KZ)}[KZ]
\newabbrev\IRD{Infinitely Rapidly Driven (IRD)}[IRD]
\newabbrev\FDRs{Fluctuation-Dissipation Relations (FDR)\vphantom{\FDR}}[FDRs]
\newabbrev\FDR{Fluctuation-Dissipation Relation (FDR)\vphantom{\FDRs}}[FDR]
\robustify{\RG}
\robustify{\UV}
\robustify{\WF}
\robustify{\FBZ}
\robustify{\FBZs}
\robustify{\KZ}
\robustify{\IRD}
\robustify{\FDR}
\robustify{\FDRs}

\makeatletter 
\renewcommand\maybe@space@{%
  \maybe@ictrue 
  \expandafter   \@tfor
    \expandafter \reserved@a
    \expandafter :%
    \expandafter =%
                 \nospacelist
                 \do \t@st@ic
  \ifmaybe@ic 
    \space
  \fi
}
\makeatother

\begin{document}

\title{Dynamic renormalization group theory for open Floquet systems}

\author{Steven Mathey}
\email[]{smathey@thp.uni-koeln.de}
\affiliation{Institut f\"ur Theoretische Physik, Universit\"at zu K\"oln, 50937 Cologne, Germany}

\author{Sebastian Diehl}
\affiliation{Institut f\"ur Theoretische Physik, Universit\"at zu K\"oln, 50937 Cologne, Germany}

\date{\today}

\begin{abstract}
We develop a comprehensive \RG approach to criticality in open Floquet systems, where dissipation enables the system to reach a well-defined Floquet steady state of finite entropy, and all observables are synchronized with the drive. We provide a detailed description of how to combine Keldysh and Floquet formalisms to account for the critical fluctuations in the weakly and rapidly driven regime. A key insight is that a reduction to the time-averaged, static sector, is not possible close to the critical point. This guides the design of a perturbative \textit{dynamic \RG} approach, which treats the time-dependent, dynamic sector associated to higher harmonics of the drive, on an equal footing with the time-averaged sector. Within this framework, we develop a weak drive expansion scheme, which enables to systematically truncate the \RG flow equations in powers of the inverse drive frequency $\Omega^{-1}$. This allows us to show how a periodic drive inhibits scale invariance and critical fluctuations of second order phase transitions in rapidly driven open systems: Although criticality emerges in the limit $\Omega^{-1}=0$, any finite drive frequency produces a scale that remains finite all through the phase transition. This is a universal mechanism that relies on the competition of the critical fluctuations within the static and dynamic sectors of the problem.
\end{abstract}

\maketitle

\ResetAbbrevs{All}
\section{Introduction}

Critical dynamics is well known to emerge at second order phase transitions of equilibrium many-body systems \cite{tauber2014critical}. Out of equilibrium, the situation is even richer, giving rise to novel universal effects without equilibrium counterparts. Forms of genuine non-equilibrium criticality are realized in diverse physical systems, and range from turbulence in quantum \cite{Pruefer2018,Erne2018,Schmied2018,Madeira2019} and classical \cite{Frisch2004a} systems, over interface dynamics \cite{Kardar1986a} like the spreading of fire fronts \cite{Miettinen2005}, to reaction-diffusion dynamics governing chemical reactions \cite{Nicolis2007}. Non-equilibrium driving conditions can even lead to full blown self-organized criticality \cite{Bak1987a}, where the system exhibits critical dynamics without a need for fine-tuning. 

A particular class of non-equilibrium systems that moved into the focus of intense research recently are many-body Floquet ensembles \cite{Eckardt2017,Moessner2017}. These are thermodynamically large systems of interacting particles subjected to a periodic drive. They have recently gained a lot of interest, because of their fundamentally new properties. For example, periodic drives can be used to generate artificial gauge fields for ultracold atoms \cite{Struck2013,Goldman2014,Jotzu2014,Schweizer2019,Wintersperger2020}, time crystals in atomic \cite{Choi2017}, ionic \cite{Zhang2017} or dissipative \cite{Yao2020,Gambetta2019,Lazarides2019,RieraCampeny2019,Seibold2020} systems, novel topological states without equilibrium counterparts \cite{Lindner2011,Karzig2015,po2016,Nathan2017,Yates2018,McIver2018,Wintersperger2020,Bandyopadhyay2020,Decker2019}, driven analogs of many-body localization \cite{DAlessio2013,Abanin2014,Lazarides2015,Ponte2015} or Floquet spinor Bose gases \cite{Fujimoto2019,Clark2019,Li2019}.

In the present work, we focus on the nature of criticality in open Floquet systems. To this end, we investigate the interplay of critical dynamics with a drive implemented through periodically time-dependent couplings within a Keldysh-Floquet \RG approach. As main physics result, already exposed in our previous letter \cite{mathey2018a}, we uncover a new universal mechanism specific to Floquet systems: the drive generates a scale that remains finite even when the system approaches a symmetry breaking phase transition. This means that criticality is effectively suppressed by a rapid (although not infinitely rapid) periodic drive. Here, we present the broader field-theoretical framework behind these results, which may also be applied to other many-body Floquet problems, and include several new results and refinements (see \Sect{subsec:key}).

In particular, we are interested in Floquet steady states. \Ie the dynamics at asymptotically long times, where the memory of initial conditions is lost, and observables are synchronized with the drive.\footnote{For an observable to be ``synchronized with the drive'', means that it has a periodic time dependence with the same period as the drive. The observable may however have harmonics (Fourier modes with frequencies given by integer multiples of $\Omega$) that are not explicitly included in the drive.} In the absence of dissipation, a generic periodically driven system typically heats up to infinite temperature, since the system has no means to evacuate the energy that it gains. Nontrivial Floquet steady states can however be reached when such heating dynamics is hindered. This scenario can emerge, \eg in the presence of integrability \cite{Shirai2015,Khemani2016,Shirai2018}, disorder \cite{Abanin2014,Nathan2017}, many-body scars \cite{Mukherjee2019,haldar2019,mizuta2020} or at large drive frequencies when a ``prethermal'' state emerges on intermediate time scales \cite{DAlessio2014,Chandran2016,Canovi2016,Bukov2015b,Genske2015,Mori2016,Bukov2015c,Weidinger2017,Clark2017,Kandelaki2017,Howell2018,Sun2018,Boulier2018,Shibata2018,Mallayya2019,citro2020}. Another option, which is the one we investigate here, is to append a dissipation mechanism to the system. Baths occur quite naturally in quantum systems \cite{Breuer2000,rini2007,Ketzmerick2010,Kaiser2014,Iadecola2015,Schnell2018,nuske2020} such as phonons in solid state superfluids \cite{Dehghani2014,Babadi2017,Murakami2017}, for example. Such open Floquet systems are realized as, \eg quantum dots and optical cavities \cite{Xu2015,Chitra2015,Lemonde2016,Stehlik2016,Gong2018}, Brownian motors \cite{Hanggi2009,Salger2013,Denisov2014}, spin chains \cite{Lazarides2017,Lerose2018,Kennes2018b,sahoo2019} or cold atoms in optical lattices \cite{Li2016,Iwahori2017,Tomita2017,Chong2018,Qin2018,brown2020}.

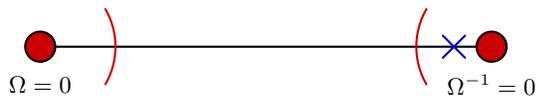
\begin{figure}[t]
 \begin{center}
\begin{tikzpicture}[scale=1]
\draw[thick,-] (-3,0) -- (3,0);
\draw[thick,fill=black!20!red] (-3,0) circle (0.2);
\draw [black!20!red,thick,line cap=round,domain=-30:30] plot ({-3+cos(\x)}, {sin(\x)});
\draw[thick,fill=black!20!red] (3,0) circle (0.2);
\draw [black!20!red,thick,line cap=round,domain=150:210] plot ({3+cos(\x)}, {sin(\x)});
\node[] at (-3,-0.5) {$\Omega=0$};
\node[] at (3,-0.5) {$\Omega^{-1}=0$};
\draw (2.5,0) node[cross=5pt,thick,line cap=round,rotate=0,black!20!blue]{};
\end{tikzpicture}
\end{center}
\caption{Embedding of the present work in the spectrum of driven systems. Shown is the driving frequency range, with undriven (equilibrium) and infinitely rapidly driven (IRD, non-equilibrium) limits. Both extremes exhibit criticality. In their vicinity, criticality is masked: A slow drive cuts off asymptotic scaling via the Kibble--Zurek mechanism, a fast drive via the mechanism elaborated on in this work.}
\label{fig_two_dots}
\end{figure}

To embed our work in a broader perspective on driven systems, consider \Fig{fig_two_dots}{}, where the vertical line signals driving frequency. Critical physics can only emerge in the two extreme limits of a vanishing ($\Omega = 0$) and infinite ($\Omega^{-1}=0$) drive frequencies. The system is at thermal equilibrium when $\Omega = 0$ (left-hand side), and exhibits equilibrium criticality at a temperature $T= T_c$. When $\Omega$ is small but finite, the drive is adiabatic far away from the critical point, but introduces a new scale leading to the breakdown of criticality once the latter is approached: this is the essence of the \KZ mechanism. In the opposite $\Omega^{-1} = 0$ limit, \IRD criticality takes place \cite{Sieberer2013a,Sieberer2013b,Tauber2013a}. This limit is distinct from the opposite one physically by the absence of detailed balance, and operationally by a different universal speed at which the system looses coherence at the critical point. Here we explore the vicinity of this opposite extreme case, close to the \IRD critical point. We find that, phenomenologically similar to the slowly driven regime, a finite scale emerges, suppressing scaling behavior. Despite this similarity at first sight, the mechanism is vastly different, including on the level of observable consequences. In particular, while the \KZ mechanism only probes the set of equilibrium critical exponents, the opposite limit hosts new and independent universal exponents, as we will demonstrate. 

\subsection{Key results}
\label{subsec:key}

In this work, we develop a comprehensive \RG approach to capture near critical, open Floquet systems. In particular, the peculiarities of such systems lead us to the construction of a \textit{dynamic} version of the \RG applied to Floquet systems \cite{mathey2018a}. By this, we refer to the following: Common \textit{static} renormalization methods account for the rapid periodic drive through a renormalization of the static, time-independent description, such as the Floquet Hamiltonian (see, \eg \cite{Grifoni1998,Bukov2015,Higashikawa2018,Citro2015,Fujimoto2019}). These powerful methods are able to capture fundamental phenomena, like the stabilization of the Kapitza pendulum. A main technical insight of this work is, however, that while this scheme is appropriate far away from the critical point, it necessarily breaks down once pushing an open Floquet system close criticality: A reduction to an effective static, time-averaged description is not possible, and we crucially need to include the renormalization of the dynamic components and their interplay with the static parts, accounting for the coupling between the different \FBZs. Such a dynamic \RG scheme is developed here, which affords a  controlled expansion for large drive frequencies exceeding the other scales in the problem. In this framework, and combining it with the $\epsilon$-expansion in \RG, we confirm and refine our previous results \cite{mathey2018a}, most notably the fact that a finite drive frequency cuts off the divergence of the correlation length as the system crosses the phase transition.

We implement this scheme for a $d$-dimensional gas of interacting bosons in contact with a bath that induces loss and gain of particles. The system is equipped with the usual bosonic $U(1)$ symmetry that breaks spontaneously as the system is tuned through its critical point. We include the drive by allowing the couplings of this theory to periodically depend on time, and let the system undergo the phase transition with this drive switched on.

\subsubsection{Basic physical picture}

A hint of this main physics result is found by inspecting the form of the mass or gap term, which measures the distance from the critical point, and now is time dependent. A more thorough consideration shows that the relevant mass term is the imaginary part of this quantity, describing damping, see \Sect{sec_poles} and \Fig{fig_poles}{a}. In the notation below, it reads
\begin{eqnarray}
\mu^I(t) = \mu^I_0 + \sum_{n\neq0} \mu^I_n \, \text{e}^{-\I \Omega t} \, .
\label{eq_mut}
\end{eqnarray}
The time dependence of this coupling signals synchronization in the Floquet steady state. There are two limiting cases where the driving scale drops out: the undriven equilibrium limit $\Omega =0$, but also the \IRD non-equilibrium regime $\Omega^{-1}=0$, where the higher Fourier modes average out, and the rotating wave approximation becomes exact. Formally, both cases are then characterized by a continuous time-translation invariance. Imposing this symmetry in the undriven equilibrium and the driven \IRD limit then rules out mass terms other than $\mu^I_0$. In contrast, giving up this symmetry allows for more mass terms, \ie \RG relevant couplings with the potential to modify and hamper the critical physics. From an \RG viewpoint, this leads to \KZ physics at slow drive, and our result at rapid drive. We come back to the fundamental difference between these cases at the end of this section.

Our result at rapid drive lends itself to the following simple picture: Even when the system is critical, \ie $\mu^I_0 \rightarrow 0$, the synchronized mass still oscillates, as $\mu^I(t)$ only vanishes on period average. The mass is then being rapidly dragged across the phase transition periodically. We find that this induces a finite scale even when $\mu^I_0 = 0$. We can think of this as a blurring of the phase boundary as a result of the non-vanishing $\mu^I_{n\neq0}$ (see \Fig{fig_phase_diagram}{a}, light-red area). The importance of the mass oscillation also emphasizes the dynamic nature of the effect, not captured by a renormalization of the static sector alone. 

\subsubsection{Weak drive expansion scheme}

We assume that the drive frequency $\Omega$, is larger than all the other scales involved in the dynamics (but not infinite). We capture the limit of a rapid drive through an asymptotic expansion in powers of $\Omega^{-1}$ (see \Sect{sec_asymptotic_Omega_expansion}). This expansion reproduces the rotating wave approximation as the order zero contribution (or at $\Omega^{-1}=0$), and provides access to the effects of a large but finite drive frequency. The non-equilibrium critical physics in the \IRD limit $\Omega^{-1}=0$ \cite{Sieberer2013a,Sieberer2013b,Tauber2013a} is then recovered. Our result emerges from the inclusion of the $\Omega^{-1}$-corrections. In the present work, we consolidate the previous picture \cite{mathey2018a} by pushing this expansion up to $\mathcal{O}(\Omega^{-2})$.

A direct expansion of the loop integrals that emerge in perturbation theory involves inverting Green functions that are not time-translation invariant. Even if the Floquet formalism turns this task into inverting non-diagonal infinite-dimensional matrices, this remains a formidable task that we circumvent in \Sect{sec_poles}. Instead we find that a systematic weak drive asymptotic expansion in powers of $\Omega^{-1}$ can be obtained via a preliminary expansion in powers of the drive amplitude. Indeed, we show that the loop integrals actually depend on the drive amplitude $E$, through the ratio $E/\Omega$. In the weak drive limit $E/\Omega=0$, the system is undriven and the loop integrals are easily computed. A systematic expansion in powers of $\Omega^{-1}$ then emerges by first expanding in powers of $E/\Omega$, computing the expanded loop integrals, and finally re-expanding the obtained result in powers of $\Omega^{-1}$. This procedure, provides an algorithm that can be applied to reach any order in an $\Omega^{-1}$-expansion, and this order is directly tied to the order of the underlying weak drive $E/\Omega$-expansion. See \Sect{sec_approximating_G} and \App{app_second_order_Greens}, where we go up to $\mathcal{O}((E/\Omega)^2)$ and \Sect{sec_rg_flow_equations_simplified}, where include terms up to $\mathcal{O}(\Omega^{-2})$ in the \RG flow equations.

\subsubsection{Dynamic \texorpdfstring{\RG}{RG}}

\begin{figure}[t]
\begin{center}
\begin{tikzpicture}[scale=.99,transform shape]
 \node[] at (0,0) {\includegraphics[height=0.9\columnwidth]{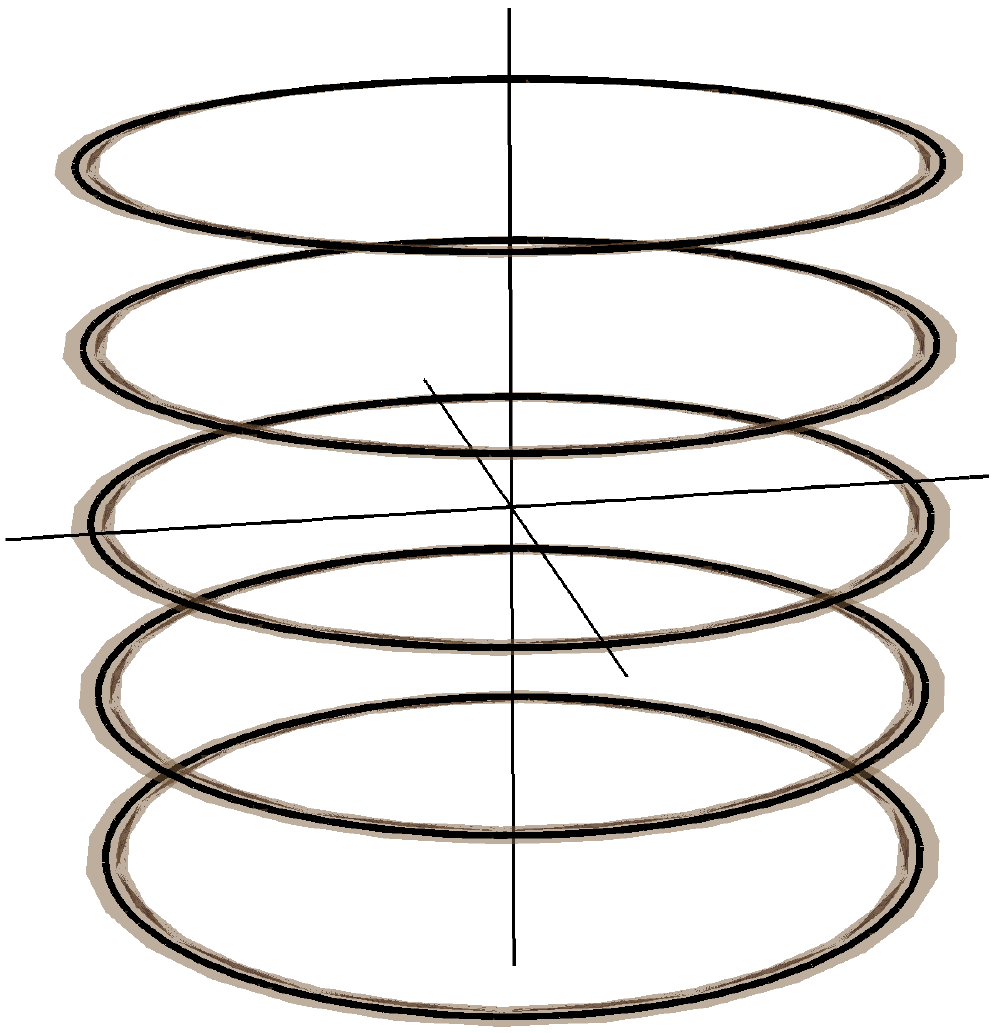}};
 \node[] at (3.8,0.5) {$p_1$};
 \node[] at (1.2,-1.2) {$p_2$};
 \node[] at (-0.1,3.9) {$\omega$};
  \draw[<->,thick] (-3.7,-0.1) -- (-3.7,1.2);
 \node[] at (-4,0.6) {$\Omega$};
 \node[] at (2.2,-0.2) {$k-dk$};
 \path[draw] (1.8,0) -- (2.6,0) -- (3.0,0.2);
 \node[] at (3.8,-0.7) {$k+dk$};
 \path[draw] (3.3,-0.5) -- (4.1,-0.5) -- (3.4,0.2);
\end{tikzpicture}
\end{center}
\caption{Momentum shells in the dynamic RG. Similarly to traditional RG, we integrate out fluctuations on $k$-dependent momentum shells (represented here as gray hoops and for $d=2$). In the absence of drive, there is a single momentum shell and the frequency is fixed to be $\omega \sim k^2$ (middle shell). The drive induces a coupling between different Floquet Brillouin Zones (FBZ) (\ie frequencies that are separated by integer multiples of $\Omega$), and forces us to include an \textit{a priori} infinite tower of momentum shells. The choice of the zeroth FBZ ($p_1-p_2$-plane) corresponds to the choice of a rotating frame.}
\label{fig_momentum_shells}
\end{figure}
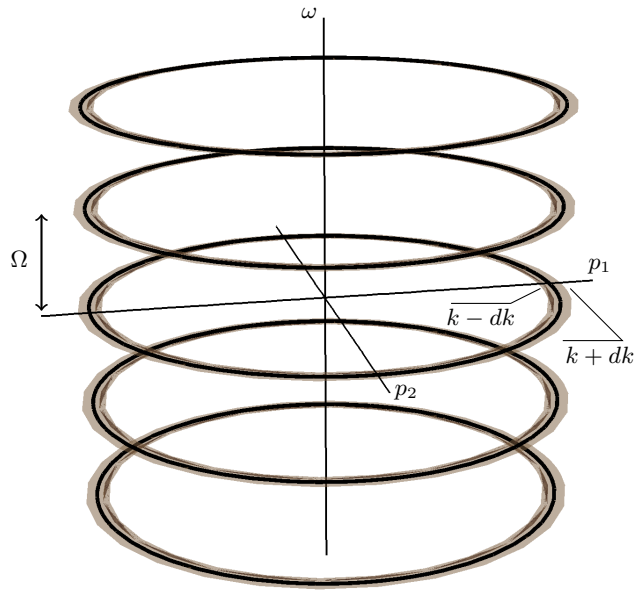

In order to access the critical physics, we generalize the usual momentum shell $1$-loop \RG \cite{tauber2014critical} to periodically driven systems. We develop a dynamic version of the \RG, where all the Fourier modes of the periodic couplings are renormalized together. This means that not only the time-averaged static description is renormalized by the drive, but also the time-dependent parts ($n\neq0$ Fourier modes). Our \RG approach is constructed by introducing a running cut off scale $k$, and integrating out fluctuations on $k$-dependent momentum shells (see \Figs{fig_momentum_shells}{} and \fig{fig_diagramms}{}). The absence of energy conservation inherent to the periodic drive induces a direct coupling between the different \FBZs. This leads to a tower of coupled momentum shells (as opposed to a single shell in the absence of drive) in momentum and frequency space, with the rungs separated by integer multiples of $\Omega$, \Fig{fig_momentum_shells}{}. This implies that, for any choice of rotating frame, there are rapid fluctuations that still contribute even when all but the large spatial scales are integrated out. This plays an especially important role at criticality, where the large-scale (and usually slow) modes take over. Our result is rooted in the fact that the system continuously absorbs and emits energy at frequencies that are integer multiples of $\Omega$. Modes within different \FBZs easily interact with each other. The periodic drive enables modes with frequencies being multiples of $\Omega$ to enter on an equal footing (in what concerns large scale fluctuations) with the slow fluctuations, and dramatically affect the critical physics even though the drive period is very short.

In \Sect{sec_rg_flow_equations_general} we compute the \RG flow of the periodically time-dependent couplings. We use a real-time representation of the flow equations [\Eqs{eq_real_time_flow_equations_mu} and \eq{eq_real_time_flow_equations_g}], where the entire time dependence of the couplings is renormalized. The periodicity of the drive is then included by making use of the Floquet formalism for the Green functions. This provides \RG flow equations for all the Fourier modes of the periodically time-dependent couplings which are expressed in terms of loop frequency integrals [\Sect{sec_rg_flow_equations_wigner} and \Eq{eq_dimensionful_flow_coupling}].

Although we work at large drive frequencies, we go beyond the common approaches based on the Floquet Hamiltonian and/or variants of the Magnus expansion (see, \eg \cite{Grifoni1998,Bukov2015,Higashikawa2018,Citro2015,Fujimoto2019}) in two distinct ways: First, we include loop integrals (as opposed to tree-level processes) to account for fluctuations, and second, we include the renormalization of the dynamic sector, \ie higher Fourier modes $n\neq0$, without integrating them out. Indeed, focusing on the Floquet Hamiltonian alone (without including an analysis of the kick operators \cite{Grifoni1998,Bukov2015,Higashikawa2018} and fluctuations on top of this Hamiltonian) only provides mean-field stroboscopic information on the system. Without further treatment, this is a tree-level approximation, that exclusively includes static renormalization effects. For example, the famous analysis of the Kapitza pendulum provides an effective Floquet potential with a new local minimum when the drive is strong enough. The static potential is renormalized as a result of the drive. Including fluctuations into this analysis would enable one to ask which minimum is actually stable against fluctuations. The kick operators in turn provide a dynamic renormalization effects through the full periodic time dependence of the pendulum as it rests in each of these minima. Our approach includes both of these features combined. Indeed, loop perturbation theory is specifically designed to include such fluctuations and, although we do not resolve the time-dependent order parameter, we do not integrate out any of the periodic degrees of freedom.

\subsubsection{Absence of criticality}\label{subsub:abs}

Our main physics result is obtained by analyzing the \RG flow close to the critical point in the presence of the periodic couplings. In the absence of drive, and also in the \IRD limit, the critical physics is governed by the well-known \WF \RG fixed point. This fixed point characterizes all the critical exponents together with the divergence of the correlation length. In the presence of the drive, additional couplings (proportional to $\Omega^{-1}$) become available and have to be included in the \RG flow. Our result is a consequence of the fact that some of these couplings are relevant and thus destabilize the \WF fixed point (see \Fig{fig_rg}{}). This means that it is impossible to observe a diverging correlation length experimentally in a rapidly driven system, because this can only happen when the \RG flow reaches its fixed point. In practice, we find that the correlation length grows (as it would without the drive) as the system approaches the phase transition, but then saturates at a value that depends on the ratio of the drive amplitudes and frequency (see \Fig{fig_phase_diagram}{a}). As $\Omega^{-1} \rightarrow 0$, this emergent cut-off scale is removed, and we recover \IRD criticality. On a more technical note, this shows that the rotating wave approximation -- setting $\Omega^{-1} =0$ and thus discarding the dynamic effects of a rapid drive from the outset -- necessarily breaks down in the vicinity of a critical point. 

Although we consider purely relaxational dynamics with a monochromatic drive \Eq{eq_mug_I} for simplicity, we show that this is a universal mechanism. Indeed, we expect that (except in fine tuned cases) criticality is regularized as soon as a periodic drive is switched on.

As we discuss above, the drive allows for new relevant couplings. Actually, a new critical exponent is associated to each of these couplings [see \Eq{eq_eigenvalues}]. These exponents are independent from the \IRD critical exponents, and are an original property of the periodically driven system. In the present work, we find that increasing the order of the truncation of the asymptotic $\Omega^{-1}$-expansion produces additional couplings, and therefore gives access to the additional original exponents.

\subsubsection{Interpretation and relation to the slowly driven limit}

Our result can be interpreted in the framework of fluctuation induced first order transitions \cite{Coleman1973,Halperin1974,Fisher1974,Nelson1974}, see \Sect{sec_discussion}. This means that the interplay of critical and dynamic fluctuations change the phase transition from a second to a first order one, without an explicit symmetry breaking as usually present in first order phase transitions with a critical endpoint of higher symmetry. Technically, what our scenario shares with other instances of fluctuation induced first order transitions is the presence of multiple near gapless modes, which interact strongly. While these may be Goldstone or gauge modes in the traditional instances of this scenario, here they are realized by the poles of the different \FBZs, all reaching criticality jointly as dictated by the Floquet formalism, \Fig{fig_poles}{a}. Indeed, we also find the \RG phenomenology common to such scenarios: the drive enables the \RG flow to take the system to an unstable range of parameters (negative interaction parameter). This signals that higher order couplings (that are not included here) are responsible for the system's stability and opens the door for $\phi^6$ phenomenology, where a first order transition is expected.

This discussion shows that the mechanism established here for a rapid drive is very different from the \KZ scenario for a slow drive (cf.~\Fig{fig_two_dots}{}) -- even operationally: In the \KZ scenario, the drive provides information on the underlying equilibrium critical point via the set of equilibrium critical exponents. Instead, here we obtain new, independent exponents, see \Sects{subsub:abs} and \sect{sec_scaling_behaviour}. This is rationalized from the fact that in the former case, we deal with an infrared modification of the critical physics (a slow driving scale is introduced, so slow that the periodic functions in \Eq{eq_mut} can be expanded in powers of $\Omega$ and the periodicity is never probed on the accessible time scales), while in the latter case the modification is in the ultraviolet (a fast driving scale is introduced, and the periodicity is crucial) [see also the additional mass scales in \Eq{eq_mut}]. It is a very basic insight of \RG theory \cite{tauber2014critical} that it is such ultraviolet scales that are able to modify and add critical exponents to the observable phenomenology.

\subsubsection{Plan of the paper}

Our paper is organized as follows. Our model, physical setup and formalism are defined in \Sect{sec_setup}. Next, we briefly review the Floquet formalism in \Sect{sec_floquet_formalism}. In particular, our weak drive asymptotic expansion is described in \Sect{sec_asymptotic_Omega_expansion}. Our main result is derived in \Sect{sec_floquet_rg}. We first discuss the dynamic \RG, before we derive perturbative \RG flow equations in \Sect{sec_rg_flow_equations_general}. In particular, the most general form of the \RG flow equations is given in \Eq{eq_dimensionful_flow_coupling}. These equations are then simplified and asymptotically expanded in powers of $\Omega^{-1}$ in the following sections. This ultimately leads to \Eqs{eq_dimensionful_flow_equations_Omega2} and \eq{eq_dimensionful_flow_drive_parameters_final}, which describe the flow of the Fourier modes of the couplings to $\mathcal{O}(\Omega^{-2})$. The critical physics is analyzed in \Sect{sec_scaling_behaviour}, and we finally derive and analyze our main result in \Sect{sec_critical_exponents}. \Sect{sec_discussion} is devoted the interpretation of our result in terms of a fluctuation-induced first order transition.

\section{Set-up}
\label{sec_setup}

In this section, we define our physical model and set-up. Moreover, we introduce the Keldysh field theory that will constitute our theoretical framework. For definiteness and in order to make contact with the current discussion of Floquet systems, we first consider a fully quantum mechanical ``parent model''. We note, however, that for the discussion of criticality in driven open systems, a semi-classical description is appropriate, because decoherence takes place close to the critical point \cite{Sieberer2013a,Sieberer2013b,Tauber2013a}. Our main result is eventually derived from the dynamics of \Eq{eq_action}, which captures the mesoscopic physics close to the critical point.

We consider driven open quantum many-body Floquet systems described by a Lindblad master equation
\begin{align*}
 \partial_t \hat{\rho} = - \I \left[ \hspace{-1pt} \hat{H}(t),\hat{\rho}\right] \hspace{-2pt} + \hspace{-3pt} \int_{\vec{x}} \hspace{-2pt} \sum_{i=plt} \kappa_i(t) \left( \hspace{-2pt} \hat{L}_i \hat{\rho} \hat{L}_i^\dagger - \frac{1}{2} \left\{\hat{L}_i^\dagger \hat{L}_i,\hat{\rho}\right\} \hspace{-2pt} \right) \hspace{-2pt} ,
\end{align*}
with $\hat{\rho}$ the many-body density matrix. Microscopically, our system is characterized by a Hamiltonian
\begin{align}
 \hat{H} = \int_{\vec{x}} \hat{\psi}^\dagger \left(\frac{-\nabla^2}{2m}\right) \hat{\psi} + \frac{g}{2} [\hat{\psi}^\dagger]^2 \hat{\psi}^2 \, ,
 \label{eq_hamiltonian}
\end{align}
($\hat{\psi}$ and $\hat{\psi}^\dagger$ are bosonic field operators) coupled to an external bath through jump operators
\begin{align}
 \hat{L}_p = \hat{\psi}^\dagger \, , && \hat{L}_l = \hat{\psi} \, , && \hat{L}_t = \hat{\psi}^2 \, ,
 \label{eq_jump_operators}
\end{align}
that model single-body pumping, loss and two-body losses respectively (with $\kappa_i(t)$ the corresponding drive/dissipation rates). See, \eg \cite{Sieberer2013b,Sieberer2015b} for additional information on this model without explicit time dependence. The periodic time dependence (with period $T=2\pi/\Omega$) can occur through an explicit time dependence of the Hamiltonian $H(t+2\pi/\Omega) = H(t)$ and/or through periodic excitations of the external bath $\kappa_i(t+2\pi/\Omega)=\kappa_i(t)$. The system is invariant under the internal $U(1)$ symmetry, \ie under the phase rotation $\hat{\psi} \rightarrow \text{e}^{\I \alpha} \hat{\psi}$. When the $U(1)$ symmetry is spontaneously broken, $\langle \hat{\psi} \rangle$ acts as an order parameter.

A Floquet steady state is realized at asymptotically long times, when the information on the initial conditions is lost and the dynamics is synchronized with the drive. The steady state is determined by the time dependence of the couplings alone. In particular, this implies that the system is not invariant under continuous time-translations. Then, observables, which depend on relative-times as is the case in any steady state, depend also on the absolute (or mean) time periodically and with the same period as the external drive (see \Sect{sec_green_functions}). The drive frequency $\Omega$, appears as a parameter that can be tuned (see \Fig{fig_two_dots}{}). In the trivial case $\Omega = 0$, the system is undriven and the dissipation ensures that the system is at thermal equilibrium. In the opposite \IRD limit $\Omega^{-1}=0$ (rotating wave approximation), the periodicity of the dynamics also disappears, and the system becomes generically invariant under time-translations. It violates however the conditions of detailed balance, characteristic of thermal equilibrium \cite{Sieberer2015}. In this work, we focus on the dynamics at large $\Omega$ and elucidate the effect of a rapid drive on the stationary \IRD critical physics (see \Fig{fig_phase_diagram}{}, where $x=0$ corresponds to the \IRD system).

We use the Schwinger--Keldysh formalism, which describes the Floquet steady state in terms of an action functional instead of the equivalent Lindblad description \cite{Rammer2007a,kita2010,Sieberer2013b,Sieberer2015b}. With \Eq{eq_jump_operators}, the dissipative dynamics comes in the form of one-body pumping and losses as well as two-body losses. Then the Keldysh action can be written as\footnote{Here and in the following, we use the short-hand notation ${\int_{t,\boldsymbol{x}} = \int_{-\infty}^{\infty}\text{d}t \int_{-\infty}^{\infty}\text{d}^dx}$.}
\begin{align}
 S =  & \int_{t,\boldsymbol{x}} \Phi^\dagger \left(\begin{array}{cc} 0 & G_A^{-1} \\ G_R^{-1} & P_K \end{array} \right) \Phi  + 2 \, \text{Im}(g) |\tilde{\phi}|^2 |\phi|^2  \nonumber \\
 &+ \int_{t,\boldsymbol{x}} g \, \left(\tilde{\phi}^* \phi \left|\phi\right|^2 + {\phi}^* \tilde{\phi} \tilde{\left|\phi\right|}^2\right)  + \text{c.c.}  \, .
 \label{eq_action_quantum}
\end{align}
${\Phi = (\phi,\tilde{\phi})}$ contains the 'classical' field $\phi$, as well as the 'response' or 'quantum' field $\tilde{\phi}$ that is inherent to the dynamical functional formalism. The field operator expectation value is the order parameter in the field theory $\langle \hat{\psi} \rangle = \langle \phi \rangle$. The retarded, advanced and Keldysh inverse propagators are
\begin{align}
 & G_R^{-1}(t,t') = \left[\I \partial_t + K p^2 + \mu\right] \, \delta(t-t')  \, , \nonumber \\
 & G_A^{-1}(t,t') = \left[\I \partial_t + K^* p^2 + \mu^*\right] \, \delta(t-t')  \, , \nonumber \\
 & P_K(t,t') = \I \gamma \, \delta(t-t')  \, .
 \label{eq_def_gamma}
\end{align}
$\vec{p}$ is the $d$ dimensional momentum and $p = \sqrt{\vec{p}^2}$ is its norm. The combination of coherent and dissipative dynamics is encoded in the couplings $K,\mu,g$, which are complex valued. Their real parts account for the coherent dynamics inherited from the underlying Hamiltonian, and the imaginary parts of $\mu$ and $g$ emerge from $\kappa_i(t)$ and have an interpretation as incoherent one- and two-body pumping and losses of particles resulting from the coupling to the bath \cite{Carusotto2013a,Sieberer2013b,Sieberer2015b}.

In the absence of an explicit time dependence, the system is stationary and its $U(1)$ symmetry can be spontaneously broken. It then undergoes a second order phase transition, close to which large-scale fluctuations dominate the physics. This regime allows for an effective mesoscopic description of the system, where the fluctuations on short temporal and spatial scales are integrated out. There, $\phi$ is interpreted as a fluctuating order parameter field, and the order parameter is obtained as its expectation value $\langle \phi\rangle$. Physically, one finds that decoherence takes place close to the critical point \cite{Sieberer2013a,Sieberer2013b,Tauber2013a}. Then the corresponding mesoscopic model,
\begin{align}
 S =  \int_{t,\boldsymbol{x}} \Phi^\dagger \left(\begin{array}{cc} 0 & G_A^{-1} \\ G_R^{-1} & P_K \end{array} \right) \Phi  +  \left( g \, \tilde{\phi}^* \phi \left|\phi\right|^2 + \text{c.c.} \right) \, ,
 \label{eq_action}
\end{align}
describes the semi-classical non-conservative dynamics of the order parameter,\footnote{Formally, this description is obtained from \Eq{eq_action_quantum} by neglecting the second and fourth terms on the right-hand side and re-interpreting the couplings as being mesoscopic effective quantities. Its emergence can be traced down to the difference in canonical dimension of the fields $\phi$ and $\tilde{\phi}$, which implies that the neglected terms are irrelevant in sufficiently high space dimension and die out at criticality.} which is a scalar complex field. The small-scale fluctuations together with most of the microscopic details are encapsulated in effective couplings $K,\mu,g$, which are complex valued, and retain their interpretation as real and imaginary parts, accounting for the coherent and dissipative dynamics, respectively. This mesoscopic description inherits the periodic nature of the microscopic drive in its most general form. Indeed, the non-linearity of the model make the coarse graining of the small scales a complex procedure that affects all the effective couplings without distinction. Then, the effective couplings are all time-dependent in the Floquet steady state.
Here we choose\footnote{Here and (unless stated otherwise) in the following, $\sum_n$ is a short-hand notation for $\sum_{n=-\infty}^\infty$.}
\begin{align}
\mu = \sum_n \text{e}^{-\I n \Omega t} \, \mu_n \, , && g = \sum_n \text{e}^{-\I n \Omega t} \, g_n \, ,
\label{eq_def_mn_gn}
\end{align}
but keep $K$ and $\gamma$ constant for simplicity. Additionally to the bosonic model of \Eq{eq_hamiltonian}, this mesoscopic model can also be applied for example, to superfluids in solid state systems. Indeed, the order parameter has the same $U(1)$ symmetry and may not be conserved as a result of the coupling to a bath of phonons.

A way to recover semi-classical thermal equilibrium dynamics in \Eq{eq_action} is to turn off the time-dependent drive and choose purely imaginary couplings. Indeed, in this case we obtain the relaxation dynamics of the $2$-component order parameter (model A of \cite{Hohenberg1977a}, see \App{app_wilson_fisher}). This vanishing of the real parts of the couplings is a consequence of the full decoherence that emerges on large scales, where the dynamics is purely dissipative. Even when all the couplings are real in the microscopic theory describing Hamiltonian reversible evolution, imaginary parts are generated by the coarse graining (and ultimately take over near the critical point). Focusing on the mesoscopic description, we can use these imaginary parts to determine the phase structure of the system's stationary state close to the critical point. $\text{Im}(\mu)$ and $\text{Im}(g)$ represent the one- and two-particle loss rates respectively. When $\text{Im}(\mu)$ is large and positive, the system can not sustain a condensate and is in the symmetric (or disordered) phase. As $\text{Im}(\mu)$ is lowered and becomes negative it becomes a single-particle injection rate, and the driven-dissipative steady state is only stabilized by the two-particle losses $\text{Im}(g)$. The $U(1)$ symmetry spontaneously breaks at a zero crossing of the (properly renormalized) $\text{Im}(\mu)$, and a condensate is established.

We emphasize that the Floquet steady state that we describe is far from equilibrium. In particular, this means that many high-frequency modes are macroscopically occupied in the symmetry broken phase. Indeed, using the Floquet formalism, we find that the kinetic part of the action can be written as
\begin{align}
 S^{(2)} = \sum_{nm} \int_{-\Omega/2}^{\Omega/2} \text{d}\omega \, \Phi_n^\dagger G^{-1}_{nm} \Phi_m \, ,
 \label{eq_floquet_modes}
\end{align}
where $(\phi_n(\omega),\tilde{\phi}_n(\omega)) = \Phi_n(\omega) = \Phi(\omega+n\Omega)$ is only defined for $|\omega|<\Omega/2$, and $G^{-1}_{nm}$ is obtained by combining \Eqs{eq_greens} and \eq{eq_gnm}. In the presence of the periodic drive, the $U(1)$ symmetry breaks spontaneously (by continuity to the \IRD limit, see \Sect{sec_discussion}) and generates a Floquet condensate, where the order parameter is an oscillating function of time \cite{Eckardt2005,Hai2008,Creffield2008,Xie2009,Arimondo2012,Vorberg2013,Choudhury2014,Gertjerenken2014,Bukov2015b,Heinisch2016}. This means that all the fields $\Phi_n$ are macroscopically occupied. These characterize the different Fourier modes of the order parameter in the steady state,
\begin{align}
 \langle \phi_n(\omega) \rangle = \delta(\omega) f_n \, , && \langle \phi(t) \rangle = \sum_n  \text{e}^{\I n \Omega t} f_n \, .
 \label{eq_oscillating_op}
\end{align}
We see that the order parameter synchronizes with the drive, \ie it has components oscillating at all of its harmonics. This in turn enables the strong effect of the drive that we find at large values of $\Omega$. Energy can be exchanged between any pair of Floquet bands since they are all macroscopically occupied (see \Fig{fig_poles}{b}).

\section{Floquet formalism and Green functions}
\label{sec_floquet_formalism}

In this section, we show how the Green functions can be computed (\Sect{sec_green_functions}), and approximated (\Sects{sec_poles} and \sect{sec_approximating_G}) within the Floquet formalism. Moreover, we discuss our asymptotic $\Omega^{-1}$-expansion in \Sect{sec_asymptotic_Omega_expansion}.

\subsection{Floquet formalism}
\label{sec_green_functions}

Here, we discuss the Floquet formalism and the computation of the single-particle Green functions. These are the essential elements of perturbation theory (see \Fig{fig_diagramms}{}). The two-point correlation functions are defined as\footnote{Here and in the following, we suppress the spatial dependence when possible.}
\begin{align}
 G(t,t') = -\I \left(\begin{array}{cc} \langle \phi(t) \phi^*(t')\rangle & \langle \phi(t) \tilde{\phi}^*(t')\rangle \\
                      \langle \tilde{\phi}(t) \phi^*(t')\rangle & \langle \tilde{\phi}(t) \tilde{\phi}^*(t')\rangle
                     \end{array} \right) \, ,
\label{eq_defg}
\end{align}
with correlation functions involving only $\tilde{\phi}$ vanishing. In the absence of interactions, the elements of $G$ are computed by inverting the kinetic part of the action \Eq{eq_action},
\begin{align}
G(t,t') = \left(\begin{array}{cc} - \left[G_R P_K G_A\right](t,t') & G_R(t,t') \\ G_A(t,t') & 0 \end{array} \right) \, .
\label{eq_greens}
\end{align}
The time dependence of $\mu$ complicates the computation of $G(t,t')$ since we can not simply convert to Fourier space and take the algebraic inverse. For this reason we resort to the Floquet formalism \cite{Floquet1883a}, where the above inversion can be performed by first re-casting it as a matrix inversion of the Floquet representation of the Green functions [see \Eqs{eq_gnm} and \eq{eq_inverting_greens}, and \Fig{fig_floquet_formalism}{}], before the result is re-converted to its real-time representation. See \cite{Grifoni1998,Bukov2015} and references therein for an introduction and \cite{Arrachea2005,wu2008,Stefanucci2008,Tsuji2008,Genske2015,Walldorf2018} for the combination of the Floquet and Keldysh formalisms.

In the Floquet steady state, the Green functions take the following form,
\begin{align}
 G(t,t') = \sum_n \int_\omega \text{e}^{-\I\left[\omega (t-t') + n \Omega (t+t')/2\right]} \, G_n(\omega) \, ,
 \label{eq_Gn_to_Gt}
\end{align}
They are periodic in the mean time $t_a = (t+t')/2$ and can be represented in terms of the Wigner Green functions\footnote{Here and in the following, we use the short-hand notation ${\int_\omega = 1/(2\pi) \int_{-\infty}^{\infty}\text{d}\omega}$ and ${\fint_{t} = \Omega/(2\pi) \int_0^{2\pi/\Omega}\text{d}t}$.} \cite{Tsuji2008},
\begin{align}
 G_n(\omega) = \int_\tau \fint_{t_a} \text{e}^{\I\left[\omega \tau + n \Omega t_a \right]} \, G(t_a+\tau/2,t_a-\tau/2) \, ,
 \label{eq_wigner_greens}
\end{align}
which encode the periodicity in $t_a$ with a discrete index and the standard relative-time dependence with a continuous frequency. The Wigner Green functions are directly related to the real-time Green functions through \Eq{eq_Gn_to_Gt}. The periodic drive produces an infinite set of Green functions that encode the time-periodicity of the single-particle sector of the system.

We now introduce the Floquet Green functions. These are defined by introducing the \FBZ $-\Omega/2<\omega<\Omega/2$, and folding the frequency dependence of $G_n(\omega)$ onto itself. This produces a second index that compensates for a restricted frequency dependence. Precisely, the Floquet Green functions are \cite{Tsuji2008},
\begin{align}
 G_{nm}(\omega) = G_{n-m}\left(\omega+\frac{n+m}{2}\Omega\right) \, , && \left|\omega\right| \leq \frac{\Omega}{2} \, ,
 \label{eq_gnm}
\end{align}
which are two-index Green functions constructed from the single-index Wigner Green functions. These definitions are directly applicable to the inverse propagators \eq{eq_def_gamma}. The Floquet Green functions provide a means to compute the Green functions from the inverse propagators. Indeed, $G$ is computed from the latter through \Eq{eq_greens}, which contains functional inverses. The Floquet representation \eq{eq_gnm}, has the great advantage that it turns functional inverses into matrix inverses. In other words, the following statements are equivalent,
\begin{align}
 & \int_z G(t,z) G^{-1}(z,t') = \delta(t-t') \, , \nonumber \\
 & \sum_{s} G_{ns}(\omega) G^{-1}_{sm}(\omega) = \delta_{nm} \, .
 \label{eq_inverting_greens}
\end{align}
\Eq{eq_gnm} provides an alternative representation for $G_n(\omega)$ that can be inverted in a straightforward way.

In summary (see \Fig{fig_floquet_formalism}{}), the Green functions are computed from their inverse in the following way: First, the Wigner inverse propagators, $G^{-1}_{R/A;n}(\omega)$ are computed from their real-time representations with \Eq{eq_wigner_greens}. Second, $G^{-1}_{R/A;n}(\omega)$ are converted to the Floquet inverse propagator [\Eq{eq_gnm}]. The Floquet Green functions are then obtained by taking the matrix inverse of $G^{-1}_{R/A;nm}(\omega)$. The next step is to convert the Floquet Green functions back to their Wigner representation. \Eq{eq_gnm} provides a one-to-one mapping that directly provides $G_{nm}(\omega)$ from $G_n(\omega)$. The inverse mapping is not as straightforward. For $n$ even, we can use
\begin{align}
 G_n(\omega) = G_{m+\frac{n}{2},m-\frac{n}{2}}(\omega-m \Omega) \, .
 \label{eq_Gnm_to_Gn}
\end{align}
$m$ is the number of times $\Omega$ has to be subtracted from $\omega$ such that $\left|\omega-m\Omega\right|\leq \Omega/2$. When $n$ is odd, $\omega$ is shifted by an additional half-integer multiple of $\Omega$ and \Eq{eq_Gnm_to_Gn} is applied with $m\rightarrow m \pm 1/2$. The direction of the shift is determined by the sign of $\omega -m \Omega$. Finally, $G(t,t')$ is obtained by inserting \Eq{eq_Gnm_to_Gn} back into \Eq{eq_Gn_to_Gt}.

\begin{figure}[t]
 \begin{center}
 \begin{tikzpicture}[scale=1]
\node at (0,0.5) {$G^{-1}(t,t')$};
\node at (0,-0.5) {$G(t,t')$};
\node at (2.5,0.5) {$G^{-1}_n(\omega)$};
\node at (2.5,-0.5) {$G_n(\omega)$};
\node at (6,0.5) {$G^{-1}_{nm}(\omega)$};
\node at (6,-0.5) {$G_{nm}(\omega)=(G^{-1})^{-1}_{nm}(\omega)$};
\path[draw,->,line width = 1.5 pt] (1,0.5) -- (1.5,0.5);
\path[draw,->,line width = 1.5 pt] (1.5,-0.5) -- (1,-0.5);
\path[draw,->,line width = 1.5 pt] (3.5,0.5) -- (5,0.5);
\path[draw,->,line width = 1.5 pt] (4,-0.5) -- (3.5,-0.5);
\path[draw,->,line width = 0.5 pt,gray] (6,0.2) -- (6,-0.2);
\end{tikzpicture}
\end{center}
\caption{Path to compute the single-particle Green functions within the Floquet formalism. Starting from the inverse Green functions [top left, \Eq{eq_def_gamma}] we define the Wigner $G_n(\omega)$ [top middle, \Eq{eq_wigner_greens}], and then Floquet [top right, \Eq{eq_gnm}] inverse Green functions. The Floquet Green functions are then obtained as the matrix inverse of $G_{nm}^{-1}(\omega)$ [right-hand-side, thin gray arrow, \Eq{eq_inverting_greens}]. Then the real-time Green functions are obtained by first converting the Floquet Green functions [bottom right, \Eq{eq_gnm}] to their Wigner form [bottom middle, \Eq{eq_Gnm_to_Gn}] and finally converting the latter to the real-time representation [bottom left, \Eq{eq_defg}]. We implement this program in \Sect{sec_approximating_G}, where both conversions (real-time $\leftrightarrow$ Wigner and Wigner $\leftrightarrow$ Floquet, black arrows) are performed exactly, while the matrix inversions (thin gray arrow) are approximated.}
\label{fig_floquet_formalism}
\end{figure}

In this work, we will mainly focus on the Wigner Green functions. We treat the Floquet representation as a tool to compute $G_n(\omega)$. The latter is indeed easier to interpret since $G_n(\tau) = \int_\omega G_n(\omega) \text{e}^{-\I \omega \tau}$ provides the discrete Fourier modes of the Green function with respect to the mean time $t_a$. $G_n(\omega)$ is also closer to the Green functions that are often used in perturbation theory, in that their frequency dependence in unbounded. In particular, we can use the residue theorem and benefit from our understanding of the pole structure of $G_n(\omega)$ in this representation.

\subsection{Pole structure}
\label{sec_poles}

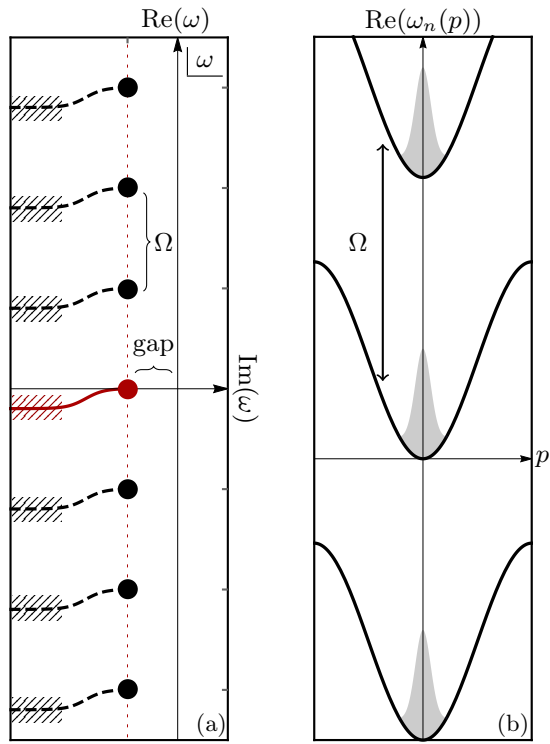
\begin{figure}[t]
\begin{center}
\begin{tikzpicture}[scale=.99,transform shape]
 \node[] at (-2,0) {\begin{tikzpicture}[scale=1.1,transform shape]
\node[] at (0,0) {\includegraphics[width=\columnwidth,angle=-90]{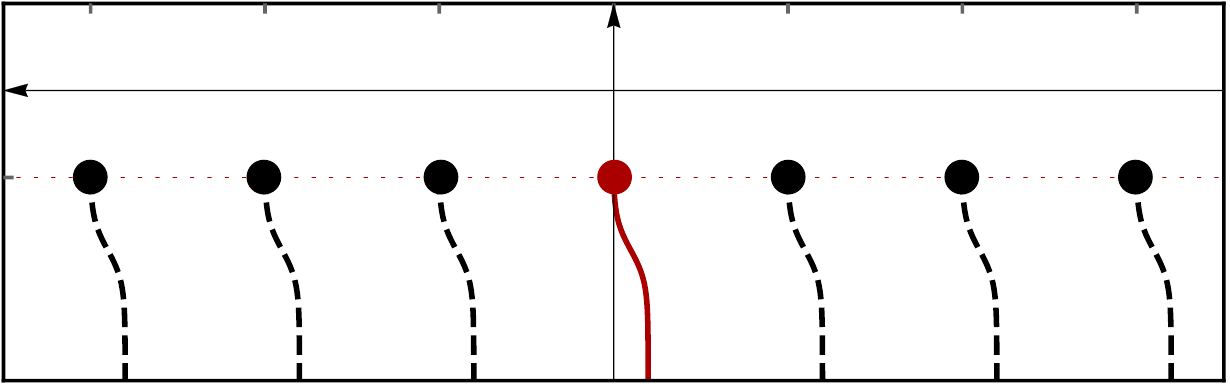}};
\draw [decorate,decoration={brace,mirror}] (0.3,1.2) -- (0.3,2.4) node [black,midway,xshift=0.25cm] {$\Omega$};
\draw [decorate,decoration={brace}] (0.2,0.2) -- (0.65,0.2) node [black,midway,yshift=0.3cm] {gap};
\fill[pattern=north east lines,pattern color=red!60!Black] (-1.3,0.1-0.47) rectangle (-0.7,0.4-0.47);
\fill[pattern=north east lines,pattern color=black] (-1.3,1.3-0.47) rectangle (-0.7,1.6-0.47);
\fill[pattern=north east lines,pattern color=black] (-1.3,2.525-0.47) rectangle (-0.7,2.825-0.47);
\fill[pattern=north east lines,pattern color=black] (-1.3,3.75-0.47) rectangle (-0.7,4.05-0.47);
\fill[pattern=north east lines,pattern color=black] (-1.3,-1.15-0.47) rectangle (-0.7,-0.85-0.47);
\fill[pattern=north east lines,pattern color=black] (-1.3,-2.35-0.47) rectangle (-0.7,-2.05-0.47);
\fill[pattern=north east lines,pattern color=black] (-1.3,-3.6-0.47) rectangle (-0.7,-3.3-0.47);
\node[] at (1.05,4.0) {\efbox[rightline=false,topline=false]{$\omega$}};
\node[] at (0.7,4.5) {$\text{Re}(\omega)$};
\node[rotate=-90] at (1.55,0) {$\text{Im}(\omega)$};
\end{tikzpicture}};
 \node[] at (2,0) {\begin{tikzpicture}[scale=1.099,transform shape]
 \node[] at (0,0) {\includegraphics[height=\columnwidth]{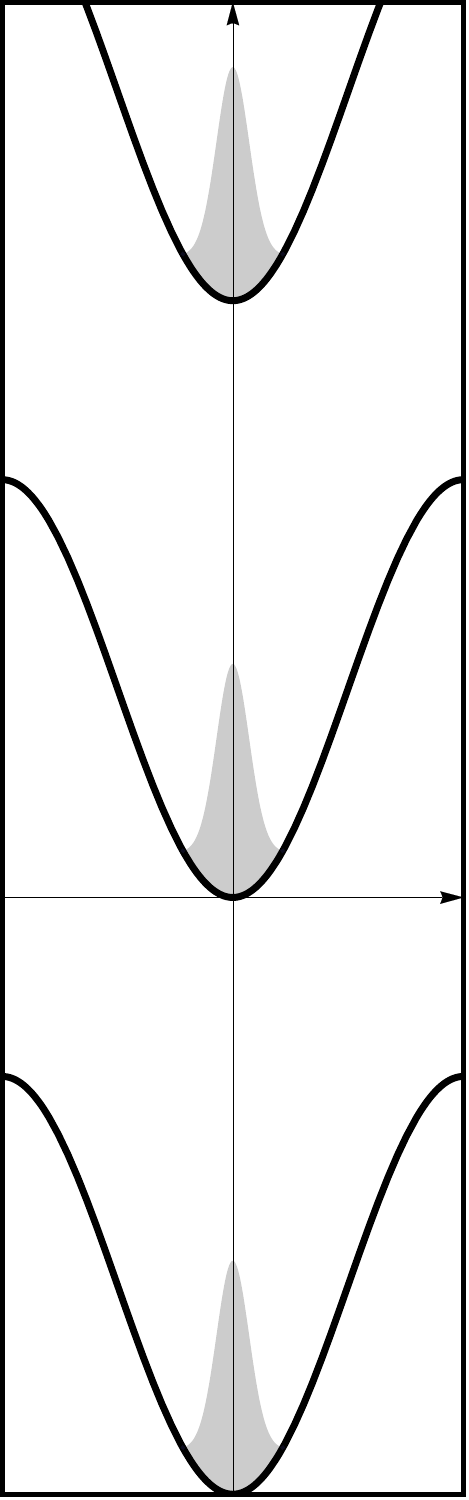}};
 \node[] at (1.45,-0.85) {$p$};
 \node[] at (0,4.5) {$\text{Re}(\omega_n(p))$};
 \draw[<->,thick] (-0.5,0.1) -- (-0.5,3);
 \node[] at (-0.8,1.8) {$\Omega$};
\end{tikzpicture}};
\node[] at (-0.95,-4.7) {(a)};
\node[] at (3.1,-4.7) {(b)};
\end{tikzpicture}
\end{center}
\caption{Poles of the retarded Green function. (a) Location of the poles in the complex frequency plane. The absence of energy conservation gives rise to lines of poles spaced by $\Omega$. The imaginary part of the poles is the damping rate of the corresponding mode. In a Floquet system, all the modes have the same damping rate and reach criticality simultaneously. (b) We represent the Floquet bands through the (non-interacting) dispersion relations of the different Floquet modes $\Phi_n$ as a function of momentum. In the Floquet steady state, all the fields $\Phi_n$ interact and contribute to the periodically time-dependent condensate.}
\label{fig_poles}
\end{figure}

In this section, we compute $G_n^R(\omega)$ exactly for a monochromatic drive. This provides valuable insights into the analytic structure of the Green functions (see \Fig{fig_poles}{}). Specifically, we find that $G_n^R(\omega)$ has infinitely many poles with real parts separated by integer multiples of $\Omega$ and identical imaginary parts. This is a consequence of Floquet's theorem for linear periodic differential operators. This will be relevant in the following sections, where we perform perturbation theory. The main result of this section is \Eq{eq_sum_with_half_poles}, where the sum is constrained so that only terms where $m$ has the same parity as $n$ contribute. In this section, we choose a monochromatic drive,
\begin{align}
\mu = \mu_0 + \mu_{1} \text{e}^{-\I \Omega t} + \mu_{-1} \text{e}^{\I \Omega t} \, ,
\label{eq_monochromatic_poles}
\end{align}
which enables the exact evaluation of $G_n^R(\omega)$.

The semi-classical nature of our mesoscopic system enables a representation of \Eq{eq_action} as a Langevin equation \cite{tauber2014critical},
\begin{align}
& \I \partial_t \phi = \left[ K \nabla^2  - \mu  - g |\phi|^2 \right] \phi + \xi \, . 
\label{eq_ddgpe}
\end{align}
$\xi$ is a Gaussian white noise, which has correlation ${\langle \xi(t,\boldsymbol{x}) \xi^*(t',\boldsymbol{x}') \rangle = 2 \gamma \delta(t-t') \delta(\boldsymbol{x}-\boldsymbol{x}')}$, with $\gamma>0$, and vanishes on average. This makes it possible to compute the single-particle retarded Green function in real time by solving \Eq{eq_ddgpe} with $g=0$,
\begin{align}
 \phi(t) = \phi(t_0) \text{e}^{\I \int_{t_0}^t M(t') \text{d}t'} \hspace{-2pt} - \I \int_{t_0}^t \hspace{-4pt} \text{e}^{\I \int_{t'}^t M(t'') \text{d}t''}\xi(t') \text{d}t' .
 \label{eq_real_time_solution}
\end{align}
$M(t) = Kp^2+\mu(t)$ is the real-time representation of the right-hand side of the equation of motion, with its Fourier modes given by
\begin{align}
M_0 = Kp^2+\mu_0 \, , &&  M_{n\neq0} \equiv E_n =\mu_{n} \, .
\label{eq_Mt_fourier}
\end{align}
Only the $n=0$ Fourier mode depends on the momentum $p$, because $K$ does not depend on time. We introduce the notation $E_n$, for the drive amplitude, to clearly separate the dynamic sectors $n\neq 0$ from the static one, $n=0$. The Fourier modes of $E$ coincide with those of $M(t)$ for $n\neq 0$. In real-time, it is defined as $E(t) = M(t)-M_0$ (with a vanishing average $E_0 = 0$). This definition extends beyond the monochromatic drive that we use in this section (see \Sect{sec_approximating_G}). $G_R(t,t')$ is obtained from
\begin{align}
 G_R(t,t') = \left. \frac{\delta \langle \phi(t) \rangle_f}{\delta f(t')} \right|_{f=0} \, .
 \label{eq_linear_response}
\end{align}
$\langle \dots \rangle_f$ is the average over a shifted noise ${\xi'=\xi+f}$, which is still Gaussian and has the same variance as $\xi$, but does not average to zero. Instead its average is ${\langle \xi'(t) \rangle_f = f(t)}$. We see that the retarded Green function can be interpreted as a linear response to a weak noise with non-vanishing average. Then, taking the average of \Eq{eq_real_time_solution} and differentiating with respect to $f$ provides
\begin{align}
 G_R(t,t') = -\I \theta(t-t') \,  \text{e}^{\I \int_{t'}^t M(t'') \text{d}t''} \, .
\end{align}
We have sent $t_0 \rightarrow - \infty$ at the end of the calculation\footnote{The terms containing $t_0$ drop out because because ${\text{Im}(M_0) >}0$.} because the initial conditions drop out in the Floquet steady state. 

Inserting the periodic time dependence of $M$ and using the Jacobi--Anger expansion provides
\begin{align*}
 & G_R(t_a+\tau/2,t_a-\tau/2) = \nonumber \\
 & -\I \theta(\tau) \text{e}^{\I M_0 \tau} \sum_m J_m\bigg[\frac{2\left( E_{-1} \text{e}^{\I \Omega t_a} + E_{1} \text{e}^{-\I\Omega t_a}\right)}{\Omega} \bigg] \text{e}^{\I m \Omega \frac{\tau}{2}} ,
\end{align*}
with $J_m(x)$, the $m$-th Bessel function of the first kind. Finally we partially convert $G_R(t,t')$ to its Wigner form and obtain
\begin{align}
 G_R(\omega,t_a) & = \int_\tau \text{e}^{\I \omega \tau} G_R(t_a+\tau/2,t_a-\tau/2) \nonumber \\
 &= \sum_m \frac{J_m\bigg[\frac{2\left( E_{-1} \text{e}^{\I \Omega t_a} + E_{1} \text{e}^{-\I\Omega t_a}\right)}{\Omega} \bigg]}{\omega+M_0+\frac{m \Omega}{2}}  \, .
 \label{eq_green_omega_t}
\end{align}
This seems to indicate that the poles are spaced by half-integer multiples of $\Omega$. This is however not the case because only half of the terms of the above sum contribute to the Wigner Green function. Indeed, an expansion of the Bessel function together with a Fourier transform of each term provides,
 \begin{align}
 & G_{R;n}(\omega) = \sum_{s=0}^\infty \frac{(-1)^s \Omega^{-2s}}{s!} \Bigg[ \frac{H_{n,0,s}}{s!(\omega+M_0)}  \nonumber \\
 & \hspace{-5pt} + \hspace{-4pt} \sum_{m=1}^{\infty} \frac{\Omega^{-m}}{(s+m)!} \hspace{-2pt} \left( \hspace{-2pt} \frac{ H_{n,m,s}}{\omega+M_0+\frac{m \Omega}{2}} + \frac{(-1)^m  H_{n,m,s}}{\omega+M_0-\frac{m \Omega}{2}} \hspace{-2pt} \right) \hspace{-3pt} \Bigg] .
 \label{eq_sum_with_half_poles}
\end{align}
The Fourier transform produces a constraint that can only be satisfied if $n+m$ is even. This implies that $n$ and $m$ must have the same parity so that half of the terms in the above sum vanish. The constraint is enforced through
\begin{align*}
 H_{n,m,s} = \sum_{r=0}^{2s+m} \left(\begin{array}{c} 2s+m \\ r \end{array}\right) E_{-1}^{2s+m-r} E_{1}^{r} \delta_{2s-2r +m +n} \, .
\end{align*}

\Eq{eq_sum_with_half_poles} provides an exact expression for the single-particle retarded Green function that elucidates the analytic structure of the Wigner Green functions (see \Fig{fig_poles}{a}). This is clearly visible in \Eq{eq_green_omega_t}: $G_R(\omega,t_a)$ is an infinite sum of poles that all have the same imaginary parts and are spaced by integer multiples of $\Omega$. This means that all the poles produce divergences simultaneously as the system becomes critical. Moreover, the residues of the poles are smooth functions of $E_{\pm1}/\Omega$ and are suppressed by increasing powers of $\Omega^{-1}$ as $|n|$ grows.

In our system, $\omega$ can be shifted to any value by an appropriate choice of rotating frame. This amounts to shifting the real part of $\mu$. Although this implies that there is not absolute meaning to large and small frequencies, large and small values of $n$ can be defined relatively to a choice of rotating frame. For example, setting $\text{Re}(\mu)=0$, provides a definition of the position of the central pole in \Eq{eq_green_omega_t}, relatively to which all the other poles are placed.

Here, we are able to compute $G_{R;n}(\omega)$ because we chose a simple monochromatic drive \Eq{eq_monochromatic_poles}. Our dynamic \RG approach (see \Sect{sec_floquet_rg}) will however require us to handle general drive protocols. In that case, the analytical calculation of this section become much more involved and is too complex to be of any use. For this reason we resort to an asymptotic $\Omega^{-1}$-expansion, which is detailed in the following section.

\subsection{Asymptotic \texorpdfstring{$\Omega^{-1}$}{1/Omega}-expansion}
\label{sec_asymptotic_Omega_expansion}

We now explain how an asymptotic $\Omega^{-1}$-expansion can be obtained for perturbation theory. Indeed, we will ultimately be interested in the case of a fast drive. We will asymptotically expand our problem in powers of $\Omega^{-1}$ and truncate the expansion to second order. This expansion provides corrections on top of the well-known rotating wave approximation, which is valid when $\Omega^{-1}=0$. It must be carried out at the level of the real-time Green functions or in loop integrals such as \Eq{eq_dimensionful_flow_coupling}, where the frequency variable of the Green functions (or a product thereof) is integrated out. A straightforward expansion of the Wigner (or Floquet) Green functions is not applicable because there is no way to ensure that $\omega \ll \Omega$ in our problem. Indeed, the loop contributions come with indefinite integrals over the domain $\omega \in (-\infty, \infty)$, so that $\omega/\Omega$ can not be neglected.

The asymptotic $\Omega^{-1}$-expansion is a double expansion (in powers of the drive amplitude $E$ and $\Omega^{-1}$). We see from \Eq{eq_sum_with_half_poles} that the Green functions are composed of an infinite series of poles. The amplitude of the drive and (more importantly for us) its frequency $\Omega$, appear in the residues of the different poles with different powers. This produces an ordering principle and enables us to expand the flow equations in powers of $\Omega^{-1}$. For a weak drive, the Green functions can be expanded in powers of $E$ before the loop integrals are performed. Indeed, the coefficients of such an expansion are all functions of $\omega$ that produce convergent loop integrals. This is because this expansion (unlike a direct $\Omega^{-1}$-expansion) preserves the pole structure of $G_{R,n}(\omega)$ (\Fig{fig_poles}{a}). It is clear from \Eq{eq_sum_with_half_poles} that any term of order $\mathcal{O}(E^n)$ will come with a corresponding factor of $\Omega^{-n}$. The expansion in powers of $E$ is actually a weak drive expansion in powers of $E/\Omega$, which can be truncated when $E \ll \Omega$. Then, the terms of $\mathcal{O}(E^n)$ can only produce terms of order $\mathcal{O}(\Omega^{-n})$ or higher in the loop integrals. We can therefore construct an expansion truncated to order $\mathcal{O}(\Omega^{-n})$ by first carrying out the weak drive expansion to order $\mathcal{O}(E^n)$, then performing the loop integrals and finally re-expanding the result to $\mathcal{O}(\Omega^{-n})$. This produces an expansion where all the terms up to $\mathcal{O}(\Omega^{-n})$ are systematically included.

In summary, the expansion is carried out in three steps:\begin{inparaenum}[(i)] \item The Green functions are computed up to a given order in powers of $E$ [see \Eqs{eq_GRn} and \eq{eq_GKn}]. \item The obtained expressions are inserted into the loop integrals \Eq{eq_dimensionful_flow_coupling}, which are evaluated analytically. \item The integrals are re-expanded in powers of $\Omega^{-1}$ up to the same order as for the $E$-expansion. The preliminary $E$-expansion provides a great simplification that renders the loop integrals analytically tractable. This in turns makes it possible to expand the end result in powers of $\Omega^{-1}$.\end{inparaenum}

We perform this procedure in the following. We expand the Green functions in powers of $E$ below and insert them in the loop integrals in \Sect{sec_rg_flow_equations_simplified}. See in particular \Eq{eq_dimensionful_flow_equations_Omega2}, where we obtain the \RG flow equations, and where the different contributions from each expansion are identified. Additional details can also be found in \App{app_expansion_to_Omega2}. The end result is an asymptotic expansion that is useful when $\Omega$ is larger than all the Fourier coefficients of $M(t)$ [defined below in \Eq{eq_Mt_fourier}]. Indeed, the first is a weak drive expansion that can be truncated when $\Omega$ is large enough, \ie when $E \ll \Omega$. It contains however no assumption on the magnitude of the $n=0$ Fourier mode of $M(t)$. The second expansion can be truncated when $M_0 \ll \Omega$. This will be automatically the case in our work, because we focus on the critical physics where $M_0$ is tuned to zero. The full asymptotic $\Omega^{-1}$ expansion has three important advantages:\begin{inparaenum}[(i)] \item It greatly simplifies the final \RG flow equations. Compare \Eq{eq_dimensionful_flow_equations} to \eq{eq_flow_mun_Eexpansion}, where the second step ($\Omega^{-1}$-expansion) is omitted. \item It provides a single expansion parameter $\Omega^{-1}$, which controls the effects of the drive. \item It contains the rotating wave approximation as the limiting case $\Omega^{-1}=0$.\end{inparaenum}

In this work, we are interested in the critical regime, where $M_0 \ll \Omega$. It is however possible to look at the opposite regime within the weak drive expansion, \ie $E \ll \Omega \ll M_0$. An expansion in powers of $M_0^{-1}$ can be obtained in the same way as our asymptotic $\Omega^{-1}$-expansion. First, the weak drive expansion is performed. Then the loop integrals are computed, and, finally the result is expanded in powers of $M_0^{-1}$. In this regime, we find that, after the loop integrals are computed [see, \eg \Eqs{eq_flow_g_E0} and \eq{eq_flow_mun_Eexpansion}], the terms of order $E^n$ come with a power of $M_0^{-n-\alpha}$, where $\alpha$ is the power of $M^{-1}$ that appears in the $\mathcal{O}(E^0)$ term. Then, taking the limit $M_0 \rightarrow \infty$ effectively turns off the drive since the elements of the weak drive expansion are suppressed. This means that far away from the phase transition, where there is a large mass, the drive only has a weak, entirely perturbative effect.

\subsection{Weak drive expansion of the Green functions}
\label{sec_approximating_G}

In this section, we show how the Green functions are expanded in powers of the drive amplitude. The obtained expressions will be used later on to derive approximate \RG flow equations for fast drives. Although the calculation of \Sect{sec_poles} provides a physical picture, \Eq{eq_sum_with_half_poles} can not be used directly because it was obtained for a monochromatic drive and is not easily generalized.

Here, we write the Wigner Green functions \Eq{eq_wigner_greens}, as a perturbative series in $E_n$. We start with $G_{R;n}(\omega)$. The Floquet inverse propagator is written as a sum of two matrices
\begin{align}
G^{-1}_R(\omega) = S  + E  \, .
\label{eq_floquet_inverse_propagator}
\end{align}
$S_{nm} = (\omega + m \Omega + M_0) \delta_{nm}$ is defined as the diagonal part of $G^{-1}_R(\omega)$, and can be inverted straightforwardly, while the effect of the drive is included in $E_{nm} = E_{n-m}$. The $n=0$ part of $G^{-1}_{R;n}$ is contained in $M_0 = K p^2+\mu_0$. We write the inverse of $G^{-1}_R(\omega)$ as
\begin{align}
 & G_{R}(\omega) = \frac{1}{S} \sum_{N=0}^\infty (-1)^N \left[E \frac{1}{S}\right]^N \, .
 \label{eq_series}
\end{align}
We now truncate this expression to a finite order in $E$ and convert the outcome to the Wigner representation order by order. To first order in $E$ we have,
\begin{align}
 & G_{R;n}(\omega) = \frac{\delta_{n0}}{\omega+M_0} - \frac{E_n}{\left(\omega+M_0\right)^2-\left(\frac{n\Omega}{2}\right)^2} \, .
 \label{eq_GRn} 
\end{align}
The advanced Green function is obtained through the relation ${G_{A;n}(\omega)=G_{R;-n}(\omega)^*}$. The Wigner Keldysh Green function is obtained by truncating the expansion of $G_R$ and $G_A$ [in \Eq{eq_series}] independently, sandwiching $P_K$ between the results [see \Eq{eq_greens}] and re-expanding (in powers of $E$) at the end. Then \Eq{eq_Gnm_to_Gn} provides (to first order)
\begin{align}
 & G_{K;n}(\omega) = - \frac{\I \gamma \delta_{n0}}{\left|\omega+M_0\right|^2}  \nonumber \\
 & + \I \gamma  \frac{E_n (\omega+M_0^* + \frac{n}{2}\Omega)+E_{-n}^* (\omega+M_0 - \frac{n}{2}\Omega)}{\left|\omega+M_0+\frac{n}{2}\Omega\right|^2 \, \left|\omega+M_0^*-\frac{n}{2}\Omega\right|^2} \, .
 \label{eq_GKn}
\end{align}
We stop at the first order here, but the second order expansion, which we use in \Sect{sec_rg_flow_equations_simplified}, is given in \App{app_second_order_Greens}.

\section{Dynamic \texorpdfstring{\RG}{RG}}
\label{sec_floquet_rg}

In this section, we detail our \RG approach to Floquet dynamics. Our main result, which is explained in \Sect{sec_scaling_behaviour}, is that the periodic drive actually destroys criticality. \Sects{sec_rg_flow_equations_general} and \sect{sec_rg_flow_equations_simplified} are dedicated to deriving the appropriate \RG flow equations with a very general expression being obtained in \Sect{sec_rg_flow_equations_general} [see \Eq{eq_dimensionful_flow_coupling}]. The asymptotic $\Omega^{-1}$-expansion is carried out in \Sect{sec_rg_flow_equations_simplified}.

We start however by comparing our dynamic \RG to the traditional \RG procedure and commenting on their conceptual differences and technical similarities. Both approaches are based on the idea of a mesoscopic effective description with an \UV cut-off in momentum space. \Eq{eq_action} provides this description, which is only valid on spatial scales larger than the inverse of the \UV momentum cut-off $\Lambda$. For example, $1/\Lambda$ is interpreted as a discrete lattice spacing or a small interaction range below which the physics is more complicated.

\begin{figure}[t]
\begin{center}
\begin{tikzpicture}[]
 \node at (-2.5,0) { \begin{tikzpicture}[scale=0.61]
\draw[color=black, line width = 1.5pt] (-10, 0.02) circle (1);
\draw[color=black, fill, line width = 1.5pt] (-10, -1) circle (0.2);
\path[draw,color=black, line width = 1.5pt,dashed,line cap=round] (-11.5,-1.2) -- (-10,-1.02);
\path[draw,color=black, line width = 1.5pt,line cap=round] (-10,-1.02) -- (-8.5,-1.2);
\node at (-12.5,0) {$k\partial_k \mu =$};
\end{tikzpicture}};
 \node at (2,0) {\begin{tikzpicture}[scale=0.61]
\draw[color=black, line width = 1.5pt,line cap=round] (4,0) arc(0:270:1);
\draw[color=black, line width = 1.5pt,dashed,line cap=round] (4,0) arc(0:-90:1);
\path[draw, color=black, line width = 1.5pt,line cap=round] (1,1) -- (2,0);
\path[draw, color=black, line width = 1.5pt,dashed,line cap=round] (2,0) -- (1,-1);
\path[draw, color=black, line width = 1.5pt,line cap=round] (5,1) -- (4,0) -- (5,-1);
\draw[color=black, fill,line width = 1.5pt] (1.97, 0) circle (0.2);
\draw[color=black, fill,line width = 1.5pt] (4.03, 0) circle (0.2);
\node at (0,0) {$k\partial_k g =$}; 
\end{tikzpicture}};
\end{tikzpicture}
\end{center}
\caption{Diagrammatic representation of the RG flow equations. We obtain the RG flow of the Fourier modes of the couplings $\mu_n$ and $g_n$ \Eq{eq_dimensionful_flow_coupling}, with $1$-loop perturbation theory. The external solid and dashed lines represent $\phi$ and $\tilde{\phi}$ indexes respectively. In the loops, the solid and the dashed-solid lines represents the Keldysh and retarded/advanced Green functions respectively. The black dots represent the coupling.}
\label{fig_diagramms}
\end{figure}
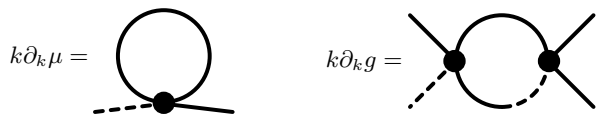

An \RG transformation is implemented by changing $\Lambda$ and reabsorbing this change in effective couplings (\ie integrating out high-momentum shells). Lowering $\Lambda$ to an effective running cut-off scale $k$, amounts to coarse graining the small-scale degrees of freedom. The \RG flow equations enable the computation of an effective action at any value of the running cut-off, starting from \Eq{eq_action} at $k=\Lambda$. This change is obtained via a differential equation, where the derivative of the couplings with respect to the running cut-off is given by loop integrals (see \Fig{fig_diagramms}{}). The cut-offs are momentum scales and limit the momenta available in the loop integrals. As we discuss below, there is no need to implement a cut-off in the frequency integrals, which are unbounded.

The difference between traditional and dynamic \RG lies in the handling of the renormalization of the $n\neq0$ Fourier modes. In static \RG, only the effect of the drive on the renormalization of the Fourier modes $\mu_0$ and $g_0$ is accounted for, while in dynamic \RG all the Fourier modes $\mu_n$ and $g_n$, are treated on an equal footing and renormalized together. Fundamentally, the difference resides in the handling of the frequency integrations. In both cases, these are performed by closing the integral path in the complex plane and using the residue theorem. For this reason the poles of the Green functions play an important role. As we discuss in \Sect{sec_poles}, these are located along a line in the complex plane with identical imaginary parts and real parts separated by integer multiples of the drive frequency, ${\omega_n = -K k^2 - \mu_0 + n\Omega}$, see \Fig{fig_poles}{a}. We see that these poles dictate the frequencies that play an important role in the loop integrals. In the \RG approach to undriven systems, there is only one pole (with $n=0$). Then, the high frequencies do not play an important role since only frequencies close to $\omega_0 = -K k^2 - \mu_0$ contribute to the loop integrals. In the dynamic \RG however, there are infinitely many poles in the first place, and arbitrarily high frequencies contribute to the loop integrals. The Floquet formalism forces us to consider all the poles on equal footing and there is no cut-off on the frequency axis. The theory is coarse grained on a spatial scale given by $k$, but fast scales ($\sim \Omega$) remain in the game. Although this is technically very similar to undriven \RG (find the poles and use residue theorem), it is physically very different. Frequencies are only defined modulo $\Omega$ in Floquet systems because their energy is not conserved. For this reason arbitrarily high frequencies (separated by integer multiples of $\Omega$) contribute.

In particular, this plays an important role in the critical physics. As we see from \Fig{fig_poles}{a}, the loop integrals are convergent as long as $\text{Im}(\mu_0) < 0$. Fluctuations play an increasingly important role as $|\text{Im}(\mu_0)|$ decreases, and the critical point is reached as $\text{Im}(\mu_0) \rightarrow 0$. When the system is undriven this phenomenon takes place on large space and time scales. Only the small momenta and frequencies contribute significantly. In the presence of a  periodic drive however, the Floquet formalism forces the large frequencies to play a role at criticality. More precisely, all the poles contribute to the loop integrals with the same degree of divergence, but different residues. This explains how the effect of a very fast scale such as $\Omega$ is not averaged out at large scales, and can still affect the critical physics.

The \IRD system is recovered if the limit $\Omega^{-1} \rightarrow 0$ is taken before the system is sent to its critical point. Indeed, a rapid periodic drive has a small quantitative effect as long as the system is gapped, because all the poles introduced by the drive are subject to the same gap, and are suppressed as $\Omega^{-1}$.

In the current work, we take the periodic drive into account within our dynamic \RG approach. In addition, we use a well-known loop expansion similar to \cite{DeSarkar2014}. This provides access to the critical physics of a generic interacting system in $4-\epsilon$ spatial dimensions. This is a systematic expansion in powers of $\epsilon = 4-d$ that is equivalent to the loop expansion at criticality. In particular, the \RG fixed point and its properties smoothly depend on $\epsilon$ [and are simple when $\epsilon = 0$, see \Eq{eq_wf_fixed_point}]. For this reason we extend the applicability our our results down to three spatial dimensions ($\epsilon = 1$).

The renormalization of periodically driven systems was already undertaken in different situations. In particular the \RG was used as a resummation-tool for $0d$ systems such as quantum dots \cite{Eissing2015,Eissing2016,Kennes2018}, where non-linearities were treated nonpertubatively, and single molecules \cite{Rozhkov2005}, where the \RG was used to treat secular terms. These approaches did not focus on critical physics. Alternatively, disordered $1d$ periodically driven critical systems were studied in \cite{Monthus2017,Monthus_2018,Berdanier2018b,Berdanier2018}. There the \RG was implemented either exactly \cite{Monthus2017,Monthus_2018} or through a Schrieffer–Wolff \cite{Luttinger1955,Schrieffer1966} transformation \cite{Berdanier2018b,Berdanier2018}. Both cases involve a static approach to \RG where the Floquet Hamiltonian is renormalized and used to infer the critical properties of the system. Finally the periodically driven bosonic $\phi^4$ model was renormalized in \cite{DeSarkar2014} within a momentum shell $1$-loop approach. This is an early dynamic \RG approach where a very general time-dependent interaction was included, and its effect on the renormalization of the couplings is computed in a simplified way. This produces a set of \RG flow equations that lead to a break down of the underlying approximation, which is interpreted as the signature of a crossover to a state dominated by the drive.

In this work, we extend the analysis performed in \cite{mathey2018a} by going to the next order in the asymptotic expansion in powers of $\Omega^{-1}$. This has enabled us not only to check that our result persists when the additional $\mathcal{O}(\Omega^{-2})$ contributions are included but also to further develop the formalism put forward in \cite{mathey2018a}. Moreover we investigate the following key questions: \begin{inparaenum}[(i)]
                                                           \item Is there a Floquet \RG fixed point which drives a phase transition from the \IRD phase to another far from equilibrium phase (see, \eg \cite{Dykman1990,Citro2015,Chitra2015,Sen2016,Russomanno2016,Berdanier2018b,Berdanier2018})? Such a fixed point would implicate the drive in an essential way and would be fundamentally impossible to observe at (or close to) thermal equilibrium. Including $\mathcal{O}(\Omega^{-2})$ terms provide an opportunity for competition between these and the $\mathcal{O}(\Omega^{-1})$ terms, which could yield a new fixed point. We do not find any new fixed point within the $1$-loop approximation. We interpret this as a strong hint that such a new fixed point does not exist in the regime of large driving frequencies.
                                                           \item The new critical exponent that emerges at $\mathcal{O}(\Omega^{-1})$ takes a particularly simple form $\nu_d = 1/(4-d)$. Is this the leading term in the $\Omega^{-1}$ asymptotic expansion with further corrections at smaller values of $\Omega$? We find a different picture, where the terms of $\mathcal{O}(\Omega^{-2})$ produce additional couplings that each come with their own new exponent [see \Eq{eq_eigenvalues}], but no exponent depends on the non-universal scale $\Omega$.
                                                          \end{inparaenum}

\subsection{General \texorpdfstring{\RG}{RG} flow equations}
\label{sec_rg_flow_equations_general}

In this section, we perform perturbation theory to $1$-loop order in the presence of a periodic drive. The main result is \Eq{eq_dimensionful_flow_coupling}, where loop integrals are given in terms of the Wigner Green functions. The only approximation leading to \Eq{eq_dimensionful_flow_coupling} is the loop expansion (in turn controlled by the $\epsilon$-expansion). The effects of the drive are not approximated yet.

Our theory is defined through its microscopic action \Eq{eq_action}, which can be used to compute correlation functions such as,
\begin{align}
 \langle A \rangle = \int \underset{\omega,p<\Lambda}{\Pi} \text{d}\phi(\op) \text{d}\tilde{\phi}(\op) \, \, \text{e}^{\I S }  A  \, .
 \label{eq_funtional_integral}
\end{align}
$A$ represent a generic observable, which is typically a product of fields $\phi$ and $\tilde{\phi}$. For example $G_R(t,t')$ is obtained with $A = -\I \phi(t) \tilde{\phi}^*(t')$ [see \Eq{eq_defg}]. The integral measure and the momentum integrals of \Eq{eq_action} are cut off at a large momentum scale $\Lambda$, which we denote as the \UV cut-off.

We start by defining the flowing effective action $\Gamma_k$ which encodes the coarse graining inherent to the \RG. $\Gamma_k$ functions just like the microscopic action $S$, although with the momenta $p>k$ being integrated out. $k$ is the running cut-off that decreases along the \RG flow, thus including fluctuations on increasingly larger spatial scales. In order to compute $\Gamma_k$, we use an exact representation of the \RG flow as a starting point. We emphasize however that the \RG flow equations \Eq{eq_dimensionful_flow_coupling} can be equivalently obtained in a traditional momentum shell loop expansion \cite{tauber2014critical}, with the Floquet structure taken into account, of course. See for example \App{app_wilson_fisher}, where we recover the \WF fixed point to leading order in the $\epsilon=4-d$ expansion from \Eq{eq_dimensionful_flow_coupling}. In this exact representation, the renormalization of $S$ (\ie the change of $\Gamma_k$ with $k$) is given by Wetterich's flow equation \cite{Wetterich:1992yh}
\begin{align}
 k \partial_k \Gamma_k = \frac{\I}{2} \text{Tr}\left[ \frac{k \partial_k R_k}{\Gamma_k^{(2)}+R_k} \right] \, .
 \label{eq_wetterich}
\end{align}
$\Gamma_k$ is identified with $S$ at large values of $k$, $\Gamma_\Lambda = S$, and $\Gamma_k^{(2)}$ is the second field derivative of $\Gamma_k$ which has two sets of (field, space and time) indexes and depends on the fields $\phi$ and $\tilde{\phi}$. The trace as well as the inversion and multiplication by $k\partial_k R_k$ in \Eq{eq_wetterich} are functional. They contain a sum over the discrete indexes and integrals over the continuous ones (see \App{app_flow_equations}). $R_k$ is a cut-off operator that we will specify below. It is diagonal in Fourier and Floquet spaces, and is a function of momentum that is large for $p\ll k$ and small otherwise. The role of $R_k$ is to suppress fluctuations on momentum scales smaller than $k$. \Eq{eq_wetterich} is a non-linear functional differential equation for $\Gamma_k$. It does not contain any approximation and (together with the initial condition $\Gamma_\Lambda = S$), it provides an alternative, equivalent functional differential representation complementing the functional integral representation of \Eq{eq_funtional_integral}.

In order to solve \Eq{eq_wetterich} we must pick an approximation scheme and a cut-off operator. We choose a sharp cut-off operator. \Ie $R_k$ is zero for $p>k$ and infinity otherwise. Moreover, we resort to traditional perturbation theory to $1$-loop order. The corresponding \RG flow equation is obtained from \Eq{eq_wetterich} by neglecting the derivative of $\Gamma_k^{{\scalebox{0.55}{(2)}}}$ with respect to the running cut-off on the right-hand side \cite{Papenbrock1995,Litim2002,Codello2014} and including a single momentum shell in the trace,
\begin{align}
 k \partial_k \Gamma_k = - \frac{\I}{2} \text{Tr}\left\{\text{ln}\left(\Gamma_k^{(2)}[\tilde{\phi},\phi]\right)\right\}_k \, .
 \label{eq_oneloop}
\end{align}
The $\text{Tr}\{ \dots \}_k$ contains a loop integration over all frequencies and discrete field indexes, but the momenta have a fixed modulus given by $p=k$, see \Eq{eq_trace}. This restriction is a result of the sharp cut-off operator. \Eq{eq_oneloop} can be interpreted as integrating out momentum shells one after the other as $k$ is lowered from $\Lambda$ to zero.

\RG flow equations are obtained for the different couplings by expanding \Eq{eq_oneloop} in powers of $\phi$ and $\tilde{\phi}$ and identifying the expansion coefficients on both sides. We focus on the terms of order $2$ and $4$. The flowing inverse retarded Green function is defined as\footnote{Here and in the following, we often omit the $k$ index in order to shorten the notation.}
\begin{align}
 \left. \frac{\delta^2 \Gamma_k}{\delta \tilde{\phi}^*(t) \delta \phi(t')} \right|_{\Phi=0} = G^{-1}_{R}(t,t') \, ,
 \label{eq_proj_gammaR}
\end{align}
and the flowing two-body coupling is
\begin{align}
 \hspace{-4pt} \left. \frac{\delta^4 \Gamma_k[\phi,\tilde{\phi}]}{\delta \tilde{\phi}^*(t_1) \delta \phi^*(t_2) \delta \phi(t_3) \delta \phi(t_4)}\right|_{\Phi=0} \hspace{-4pt}  = \Gamma^{(4)}(t_1,t_2,t_3,t_4) .
 \label{eq_proj_gamma4}
\end{align}
The \RG flow of $G^{-1}_{R}(t,t')$ and $\Gamma^{(4)}(t_1,t_2,t_3,t_4)$ are obtained by taking two and four derivatives of the right-hand side of \Eq{eq_oneloop} respectively, and evaluating the result at zero field,
\begin{align}
 k\partial_k G^{-1}_{R}(t,t') = - \frac{\I}{2} \left. \text{Tr}\left\{ G  \frac{\delta^2 \Gamma^{(2)}}{\delta \tilde{\phi}^*(t) \delta \phi(t')} \right\}_k   \right|_{\Phi=0} \, ,
 \label{eq_flow_GR}
\end{align}
and similarly for $k\partial_k \Gamma^{(4)}(t_1,t_2,t_3,t_4)$. The right-hand side of the above equation is the trace of the product of the matrix of Green functions $G = 1/\Gamma^{(2)}[\Phi=0]$, [see \Eq{eq_action}, where $\Gamma^{(2)}$ is the matrix in the first term of \Eq{eq_greens}] with the second field derivative of $\Gamma^{(2)}$, and is evaluated at zero field $\Phi = 0$. There is no term with three derivatives of $\Gamma_k$, because these vanish when $\Phi=0$. The flow of $G^{-1}_{R}(t,t')$ and $\Gamma^{(4)}(t_1,t_2,t_3,t_4)$ are represented diagrammatically in \Fig{fig_diagramms}{}, with the lines representing different elements of the $G$ matrix and the vertexes representing the interaction.

The \RG flow of $\mu$ and $g$ is extracted from these equations by integrating them over their relative space and time variables (or equivalently setting the external momenta and frequencies to zero). The flowing dissipative mass $\mu_k$, is defined as
\begin{align}
 \mu_k(t) = \hspace{-2pt} \int_\tau \int_{\vec{r}} \hspace{-2pt} \Gamma_k^{(2)}\hspace{-3pt}\left(t+\frac{\tau}{2},\vec{x}+\frac{\vec{r}}{2},t-\frac{\tau}{2},\vec{x}-\frac{\vec{r}}{2}\right) ,
 \label{eq_def_muk}
\end{align}
which coincides with $\mu(t)$ at $k=\Lambda$. The \RG flow of $\mu_k$ is then obtained by taking the derivative of \Eq{eq_def_muk} with respect to the running cut-off $k$, and inserting \Eq{eq_flow_GR} on the right-hand side. The \RG flow of the interaction parameter $g_k(t)$ is obtained in a similar way (see \App{app_flow_equations} for additional details).

We see that the flow of $G^{-1}_{R}(t,t')$ [right-hand side of \Eq{eq_flow_GR}] is a (non-linear) function of the couplings $\mu$ and $g$. Indeed, the right-hand side of \Eq{eq_flow_GR} is obtained from $G$, which is the inverse of $\Gamma^{(2)}$, $G^{-1} = \Gamma^{(2)}$. For $\mu$ and $g$ given, $\Gamma^{(2)}$ is readily obtained, because it takes the same form as the second field-derivative of the action $S$, \Eq{eq_action} (at $1$-loop), and is thus linear in $\mu$. Then computing the flow of $G^{-1}_{R}(t,t')$ is a matter of inverting $\Gamma^{(2)}$ to obtain $G$. This is where the Floquet formalism [and \Eq{eq_inverting_greens}] will become very useful.  For given time-dependent couplings $\mu(t)$ and $g(t)$, the loop integrals are computed in three steps: \begin{inparaenum}[(i)]
                                                                                                         \item $\Gamma^{(2)}$ is computed from $\mu(t)$ [see \Eq{eq_def_muk}].
                                                                                                         \item $G$ is computed from $\Gamma^{(2)}$ with the Floquet formalism [and \Eq{eq_def_gamma}].
                                                                                                         \item Everything is assembled according to the right-hand side of \Eq{eq_flow_GR} (or \Fig{fig_diagramms}{}) and the frequency integrals are performed. This is where $g(t)$ enters.
                                                                                                        \end{inparaenum}

\subsubsection{Real-time representation}

At $1$-loop, the flow of $G^{-1}_{R}(t,t')$ is quite simple: Only the dissipative mass is renormalized. Then, we can insert \Eq{eq_def_gamma} on the left-hand side of \Eq{eq_flow_GR}, integrate over $\tau$ and obtain,
\begin{align}
 k\partial_k \mu(t) & = -2 \I S_d k^d g(t) G_K(t,t) \, .
 \label{eq_real_time_flow_equations_mu}
\end{align}
${S_d = 2 \pi^{d/2}/[(d/2-1)! (2\pi)^{d}]}$ is the area factor that emerges from the momentum integration. Similarly, we obtain the flow of $g$ as (see \App{app_flow_equations})
\begin{align}
 & k\partial_k g(t) \hspace{-2pt} = \hspace{-2pt} 4 \I S_d k^d \hspace{-4pt} \int_\tau \hspace{-4pt} g(\tau) g(2t \hspace{-2pt} - \hspace{-2pt} \tau) G_R(\tau,2t \hspace{-2pt} - \hspace{-2pt} \tau) G_K(\tau,2t \hspace{-2pt} - \hspace{-2pt} \tau)  \nonumber \\
&  + \hspace{-1pt} 8 \I S_d k^d \hspace{-4pt} \int_\tau \hspace{-4pt} g(2t-\tau) g(\tau)^* G_A(\tau,2t-\tau) G_K(2t-\tau,\tau) \nonumber \\
 & + \hspace{-1pt}  8 \I S_d k^d \hspace{-4pt} \int_\tau \hspace{-4pt} g(2t\hspace{-1pt}-\hspace{-1pt}\tau) g(\tau) G_R(2t\hspace{-1pt}-\hspace{-1pt}\tau,\tau) G_K(\tau,2t\hspace{-1pt}-\hspace{-1pt}\tau)  .
 \label{eq_real_time_flow_equations_g}
\end{align}
These equations are written in terms of the retarded, advanced and Keldysh Green functions. The matrix multiplication and traces in the Keldysh sector have been performed. The only remnant of the trace of \Eq{eq_oneloop} is the integration over $\tau$ in the second equation. Here and in the following, the \RG flow equations involve Green functions evaluated at $p=k$, which (after including the $k$-dependence) are given by \Eq{eq_greens}. The Green functions that appear here are analogous to the Green functions discussed in \Sect{sec_green_functions}, with the only difference that $p$ is replaced by $k$ and $\mu$ by $\mu_k$, as a consequence of the \RG procedure.

The two above equations provide a real-time representation of the \RG flow equations. They describe the coarse graining of the entire time dependence of $\mu(t)$ and $g(t)$. The periodicity of the drive is not included here, and no assumption or approximation is made with respect to the drive frequency. This representation of the \RG flow equations is very general and can be used, for example, also as a basis for an adiabatic expansion (see \App{app_slow_drive} and \cite{mathey2019a}). It is however not the most efficient representation for our purpose, because the periodicity of the drive is not built in. Moreover, it is difficult to use without an additional approximation, because the relation between the coupling $\mu$ and the Green functions is not yet worked out.

\subsubsection{Wigner representation}
\label{sec_rg_flow_equations_wigner}

We incorporate the periodicity of the drive in the real-time \RG flow equations \Eqs{eq_real_time_flow_equations_mu} and \eq{eq_real_time_flow_equations_g} by converting the Green functions to their Wigner representation and including the time-periodicity of $g$. Inserting \Eqs{eq_def_mn_gn} and \eq{eq_Gn_to_Gt} into \Eqs{eq_real_time_flow_equations_mu} and \eq{eq_real_time_flow_equations_g} provides the following set of coupled \RG flow equations for $\mu_n$ and $g_n$,
\begin{align}
 k&\partial_k {g}_n = 2 \I S_d k^d \sum_{\substack{m_1,m_2\\m_3,m_4}}  {g}_{m_4} \delta_{n,m_{1234}} \int_\omega \nonumber \\
&  \times \Bigg\{ G_{R;m_1}(\omega,k) G_{K;m_2}\left[-\omega - \frac{m_3-m_4}{2} \Omega ,k\right] {g}_{m_3} \nonumber \\
& \quad + 2 G_{A;m_1}(\omega,k) \, G_{K;m_2}\left[\omega - \frac{m_4-m_3}{2} \Omega ,k\right] ({g}_{-m_3})^* \nonumber \\
 &  \quad + 2 G_{R;m_1}(\omega,k) G_{K;m_2}\left[\omega - \frac{m_3-m_4}{2} \Omega ,k\right] {g}_{m_3}  \Bigg\}\, , \nonumber \\
 k&  \partial_k {\mu}_n  = - 2 \I S_d k^d \sum_m \int_\omega {g}_m G_{K;n-m}(\omega,k) \, .
 \label{eq_dimensionful_flow_coupling}
\end{align} 
${m_{1234} = \sum_{i=1}^4 m_i}$, and $\delta_{ij}$ is the Kronecker delta. As before the right-hand sides of these equations are non-linear functions of $\mu_n$ and $g_n$. The complexity of these functions stems from the relation between the couplings and the Green functions $\Gamma^{(2)} = G^{-1}$, where $\Gamma^{(2)}$ depends on $\mu_n$ linearly. We will reduce this complexity with our asymptotic $\Omega^{-1}$-expansion in \Sect{sec_rg_flow_equations_simplified}. \Eq{eq_dimensionful_flow_coupling} provides a set of differential equations for the effective couplings as the running cut-off $k$ is lowered. They are obtained to first order in perturbation theory and with momentum shell \RG \cite{tauber2014critical}.

Our main result [based on \Eq{eq_dimensionful_flow_equations_Omega2}] will be derived from this representation. We reiterate that the Wigner Green functions have the advantage of exploiting the Floquet formalism (and thus incorporating the periodic drive automatically) while retaining an infinite frequency range to integrate over. This makes it possible to treat (and interpret) the loop integrals in a way that is standard for undriven problems.

\subsection{Simplified \texorpdfstring{\RG}{RG} flow equations}
\label{sec_rg_flow_equations_simplified}

We now discuss how the \RG flow equations \Eq{eq_dimensionful_flow_coupling}, can be simplified in the presence of a fast drive. We start by obtaining general equations to $\mathcal{O}(\Omega^{-1})$, \Eq{eq_dimensionful_flow_equations}. Next we further simplify the \RG flow equations by choosing purely imaginary couplings. This simplification allows us to include the terms up to $\mathcal{O}(\Omega^{-2})$ [leading to \Eq{eq_dimensionful_flow_equations_Omega2}], and eventually confirms the picture that emerges at $\mathcal{O}(\Omega^{-1})$. Finally we choose a monochromatic drive and obtain \Eq{eq_dimensionful_flow_drive_parameters_final}.

\subsubsection{Asymptotic \texorpdfstring{$\mathcal{O}(\Omega^{-1})$}{O(1/Omega)}-expansion}

As we have discussed in \Sect{sec_poles}, the drive produces infinitely many poles in the Green functions that must all be taken into account, since they all produce divergences of the same order. This can be done systematically within the asymptotic $\Omega^{-1}$-expansion, where we identify all the terms at a given order (and lower) in $\Omega^{-1}$ and neglect the rest. See \Sect{sec_asymptotic_Omega_expansion} for details on how this expansion is performed. For general couplings, and to order $\Omega^{-1}$, the asymptotic expansion of the \RG flow equations is obtained by inserting \Eqs{eq_GKn} and \eq{eq_GRn} into \Eq{eq_dimensionful_flow_coupling}. This leads to
\begin{align}
& k \partial_k \mu_n = -\frac{S_d k^d \gamma}{\text{Im}(M_0)} \left(g_n + \frac{\I X_n}{\Omega} \right) \, , \nonumber \\
& k \partial_k g_{2r} = \frac{S_d k^d \gamma}{\text{Im}(M_0)} \nonumber \\
& \quad \times \left\{\left[\frac{g_r}{2 M_0} +  \frac{g_r-(g_{-r})^*}{\I  \text{Im}(M_0)} \right] \left(g_r+\frac{\I X_r}{\Omega}\right) + \frac{Y_{2r}}{\Omega}\right\} \, , \nonumber \\
&  k \partial_k g_{2r+1} = \frac{S_d k^d \gamma}{\text{Im}(M_0)} \frac{Y_{2r+1}}{\Omega} \, .
\label{eq_dimensionful_flow_equations}
\end{align}
Here and in the following we use $M_0 = K k^2 + \mu_0$. $Y_n$ and $X_n$ are the pre-factors of the $\mathcal{O}(\Omega^{-1})$ corrections
\begin{align}
 & X_n = \I \sum_{m\neq0} \frac{g_{n-m} \left( \mu_{m}-\mu_{-m}^*\right)}{m} \, , \nonumber \\
 & Y_n =  4 \sum_{m\neq -n/2} \frac{g_{n+m} (g_m)^* }{2m+n} \, .
 \label{eq_defXY}
\end{align}
We see from \Eq{eq_dimensionful_flow_equations} that even and odd Fourier modes are not renormalized in the same way. For any given microscopic drive, Fourier modes with arbitrarily high values of $n$ are generated by the \RG. We see however that the \RG preferentially doubles the drive frequency by generating mode $n=2r$ from mode $n=r$ already at $\mathcal{O}(\Omega^{0})$. See \App{app_expansion_to_Omega2} for additional details.

\subsubsection{Imaginary couplings and \texorpdfstring{$\mathcal{O}(\Omega^{-2})$}{O(1/(Omega*Omega))} corrections}

It was observed in \cite{Sieberer2013a,Sieberer2013b,Tauber2013a} for time-independent couplings (\IRD limit) that the critical physics is governed by a fixed point with purely imaginary couplings. The real parts of the couplings flowing to zero is a signal of the full decoherence that emerges on large scales, where the dynamics is purely dissipative. We expect this phenomenology to be applicable to the present case, and therefore neglect the real parts to begin with. This provides a great simplification and has enabled us to include corrections up to order $\Omega^{-2}$. The imaginary couplings provide real parameters that can be used to represent the system's phase diagram. In particular this means that we can think of the $n=0$ Fourier mode $\text{Im}(\mu_0)$, as a dissipative mass that triggers the phase transition when it is lowered below a threshold (see \Fig{fig_phase_diagram}{a}).

Purely imaginary couplings are given by $K = \I K^I$, $\mu(t) = \I \mu^I(t)$ and $g(t) = \I g^I(t)$, with $K^I$, $\mu^I(t)$ and $g^I(t)$ real in the time domain. Then the Fourier modes of the couplings are complex numbers that satisfy
\begin{align}
  \mu_{-n}^I = (\mu_n^I)^* \, , && g_{-n}^I = (g_n^I)^* \, .
  \label{eq_imaginary_couplings}
\end{align}
As can be checked from \Eq{eq_dimensionful_flow_equations}, in the case of purely imaginary couplings, no real parts are generated by the \RG. This is related to the following symmetry of the mesoscopic action
\begin{align}
 \hat{\phi}(t) = \phi(t)^* \, , && \hat{\tilde{\phi}}(t) = - \tilde{\phi}(t)^* \, ,
\end{align}
which is realized when the couplings are all imaginary. Physically, this is interpreted by the fact that no reversible dynamics can emerge out of purely dissipative dynamics (except in topologically non-trivial systems at the boundary \cite{Tonielli2020}). 

The corresponding flow equations are:
\begin{align}
 & k \partial_k \mu^I_n \hspace{-2pt} =  \hspace{-2pt} \frac{S_d k^d \gamma}{M_0^I} \hspace{-2pt} \left[ \frac{4 M_0^I R_n}{\Omega^2} \hspace{-1pt} - \hspace{-1pt} g^I_n \left(1 + \frac{4 M_2}{\Omega^2} \right) \hspace{-1pt} - \hspace{-1pt} \frac{X_n}{\Omega} \hspace{-1pt} + \hspace{-1pt} \frac{2S_n}{\Omega^2} \right] , \nonumber \\
 & k\partial_k g^I_{2r}  = \frac{5 S_d k^d \gamma }{2[M_0^I]^2} \Bigg\{ g_{r}^I \left[ g_{r}^I \left( 1 + \frac{23 M_2}{10 \Omega^2}\right) + \frac{X_{r}}{\Omega} \right. \nonumber \\
 & \left.  - \frac{3 S_{r}}{10 \Omega^2}\right] + \frac{X_{r}^2}{2\Omega^2}  \Bigg\} + \frac{40 S_d k^d \gamma}{\Omega^2} \left[G_{2r} - \frac{Q_{2r}}{2M_0^I} - \frac{g_{r}^I R_{r}}{2M_0^I} \right] \, ,\nonumber \\
 & k\partial_k g^I_{2r+1}  = \frac{40 S_d k^d \gamma}{\Omega^2} \left[G_{2r+1} - \frac{Q_{2r+1}}{2M_0^I} \right] \, .
 \label{eq_dimensionful_flow_equations_Omega2}
\end{align}
The drive is encapsulated into the drive parameters that depend on $\mu_n$ and $g_n$,
\begin{align}
&  X_n = -2 \I \sum_{m\neq0} \frac{\mu_m^I g_{n-m}^I}{m} \, , \quad R_n = \sum_{m\neq0} \frac{\mu_m^I g_{n-m}^I}{m^2} \, , \nonumber \\
&   M_2 = \sum_{m\neq0} \frac{\mu_m^I \mu_{-m}^I}{m^2} \, , \quad \hspace{5pt} G_n = \sum_{m\neq n/2} \frac{g_m^I g_{n-m}^I}{(-2m+n)^2} \, , \nonumber \\
& S_n = \sum_{\substack{m_1\neq0,m\neq0\\m_1+ m\neq0}} \frac{\mu_m^I \mu_{m_1}^I g_{-m-m_1+n}}{m m_1} \, , \nonumber \\
& Q_n = \sum_{\substack{2m\neq n,2m_1\neq n\\m+m_1\neq n}} \frac{\mu_{n-m-m_1}^I g_m^I g_{m_1}^I}{(2m-n)(2m_1-n)} \, .
\label{eq_drive_coefficients}
\end{align}
Here $X_n$ is the same as the one defined in \Eq{eq_defXY}, but with the purely imaginary couplings inserted. There is no term with $Y_n$ in \Eq{eq_dimensionful_flow_equations_Omega2} because $Y_n=0$ when the couplings are purely imaginary. See \App{app_expansion_to_Omega2} for further details on the derivation of \Eq{eq_dimensionful_flow_equations_Omega2}.

The above equations result from a double expansion: First, the Green functions are expanded in powers of $\mu_{n\neq 0}$ up to order $2$, then the resulting loop integrals are expanded in powers of $\Omega^{-1}$. See \Sect{sec_asymptotic_Omega_expansion} for the details. This structure is visible in the above equation: The $\Omega^{-1}$-expansion of the terms of $\mathcal{O}(E^0)$ starts at $\mathcal{O}(\Omega^{0})$. This produces the terms with $g_n$ and $G_n$ above. The expansion of the $\mathcal{O}(E)$ terms starts at $\mathcal{O}(\Omega^{-1})$ and produces the terms with $R_n$, $X_n$ and $Q_n$. Finally, the expansion of the $\mathcal{O}(E^2)$ terms starts at $\mathcal{O}(\Omega^{-2})$ and produces the remaining terms.

\Eqs{eq_dimensionful_flow_equations} and \eq{eq_dimensionful_flow_equations_Omega2} are obtained with two independent approximations. \Eq{eq_dimensionful_flow_coupling} is controlled for a weak coupling. More precisely, it is systematic to order one in the $\epsilon = 4-d$ expansion. As in a standard $\phi^4$ analysis, our results depend smoothly on $\epsilon$ and the critical physics is exactly captured for $d=4$. Then we can (at least qualitatively) extend our results down to $d=3$. \Eqs{eq_dimensionful_flow_equations} and \eq{eq_dimensionful_flow_equations_Omega2} are the result of a further asymptotic expansion in powers of $\Omega^{-1}$ and are systematic to order $1$ and $2$ respectively. Our results are therefore systematic to $\mathcal{O}(\epsilon) \times \mathcal{O}(\Omega^{-2})$.

\subsubsection{Monochromatic drive}

We now choose a specific model, where the drive is monochromatic. The microscopic couplings (at $k=\Lambda$) are chosen to be [see \Eq{eq_imaginary_couplings}]
\begin{align}
& \mu =  \I \left(\mu_0^I + \mu_1^I \, \text{e}^{\I \Omega t} + (\mu_1^I)^* \, \text{e}^{-\I \Omega t} \right) \, , \nonumber \\
& g =  \I \left(g_0^I + g_1^I \, \text{e}^{\I \Omega t} + (g_{1}^I)^* \, \text{e}^{-\I \Omega t} \right) \, ,
\label{eq_mug_I}
\end{align}
with $\mu_{1}^I$ and $g_{1}^I$ complex valued. Then the \RG flow equations can be greatly simplified. In particular we can focus on the flow of $\mu_0^I$ and $g_0^I$ and write simplified flow equations for the drive parameters $M_2$, $X_0$, $S_0$, $R_0$, $G_0$ and $Q_0$. These parameters are all real for purely imaginary couplings and quantify the importance of the drive since they vanish together with the drive amplitude. $X_0$, which was already identified in \cite{mathey2018a}, is the only drive parameter that remains at $\mathcal{O}(\Omega^{-1})$. The simplified equations are obtained by inserting \Eq{eq_dimensionful_flow_equations_Omega2} into the derivatives of the drive parameters with respect to the running cut-off. A detailed analysis of the obtained equations provides
\begin{align}
 & k \partial_k M_2 = -\frac{2 S_d k^d \gamma}{M_0^I} R_0  && k \partial_k U = k \partial_k X_0 = k \partial_k G_0  = 0 \nonumber \\
 & k \partial_k S_0 = - \frac{5 S_d k^d \gamma}{2[M_0^I]^2} U && k \partial_k R_0 = -\frac{4 S_d k^d \gamma}{M_0^I} G_0  \, ,
 \label{eq_dimensionful_flow_drive_parameters_final}
\end{align}
with $Q_0 = 0$ and $U$ given in \Eq{eq_defU}. The details on their derivation are given in \App{app_flow_drive_coefficients}. Together with \Eq{eq_dimensionful_flow_equations_Omega2}, these equations are reduced to the flow equations used in \cite{mathey2018a}, when they are truncated to $\mathcal{O}(\Omega^{-1})$.

For a monochromatic drive, the drive coefficients can be related to the complex phases and amplitudes of $\mu_1^I$ and $g_1^I$
\begin{align}
 & \mu_1^I = |\mu_1^I|\text{e}^{\I\theta_{\mu}} \, , && g_1^I=|g_1^I|\text{e}^{\I\theta_{g}} \, .
 \label{eq_mg_amplitude_phase}
\end{align}
The amplitudes $|\mu_1^I|$ and $|g_1^I|$, provide the amplitude of the oscillations and the complex phases $\theta_\mu$ and $\theta_g$, the phase of the time dependence of the couplings. Indeed, inserting the above equation into \Eq{eq_mug_I} provides
\begin{align}
& \mu^I(t) = \mu_0^I + 2 |\mu_1^I| \cos( \Omega t+\theta_{\mu}) \, , \nonumber \\
& g^I(t) = g_0^I + 2 |g_1^I| \cos(\Omega t+\theta_g) \, .
\end{align}
In turn, inserting \Eq{eq_mug_I} in \Eq{eq_drive_coefficients} provides simpler expressions for the drive coefficients
\begin{align}
 & X_0 = 4 \text{Im}(\mu_1^I [g_1^I]^*) \, , && R_0 = 2 \text{Re}(\mu_1^I [g_1^I]^*) \, , \nonumber \\
 & M_2 = 2 \left|\mu_1^I\right|^2 \, , && G_2 = \frac{1}{2} \left|g_1^I\right|^2 \, , \nonumber \\
 & S_0 = 0 \, , && U = - 2 \text{Re}([\mu_1^I [g_1^I]^*]^2) \, . 
 \label{eq_drive_coefficients_monochromatic}
\end{align}
With \Eq{eq_mg_amplitude_phase}, we see that only the difference between the two phases $\theta_\mu-\theta_g$ appear in the \RG flow equations. For example, we have $X_0 = 4 |\mu_1^I| |g_1^I| \sin(\theta_\mu-\theta_g)$. This expresses the fact that the Floquet steady state is unchanged if the time dependences of all the couplings are shifted together.

We emphasize that the above equation provide the microscopic (monochromatic) drive parameters at the beginning of the \RG flow. As scales are integrated out, higher order Fourier modes are generated and the sums in \Eq{eq_drive_coefficients} must be accounted for (see \App{app_expansion_to_Omega2}).

\subsection{Critical physics}
\label{sec_scaling_behaviour}

In the \IRD limit, our system undergoes a second order phase transition as the dissipative mass $\text{Im}(\mu_0)$ is lowered below a critical value. We will see in this section, that the inclusion of even a weak drive has a dramatic effect on this transition: In the presence of a periodic drive a new scale enters, and it becomes impossible for the correlation length to diverge.

\subsubsection{Re-scaling}

Second order phase transitions and critical physics are characterized by scaling solutions to the \RG flow equations. The physics is fully scale invariant (and the correlation length is infinite) if the flowing couplings are proportional to powers of the running cut-off scale: $\mu_n \sim k^{D_n^\mu}$ and $g_n \sim k^{D_n^g}$. The exponents $D_n^\mu$ and $D_n^g$ are the scaling dimensions of the couplings. Furthermore, the critical physics (with a large yet finite correlation length) can be extracted from the reaction of the flow to small perturbations of these scaling solutions. These take a particularly simple form when they are written in terms of the rescaled couplings: $\hat{\mu}_n = \mu_n k^{-D_n^\mu}$ and $\hat{g}_n = g_n k^{-D_n^g}$. The above scaling solutions turn into a fixed point (for $\hat{\mu}_n$ and $\hat{g}_n$), where the flow of the rescaled couplings stops. Different fixed points (\ie universality classes) can have different scaling dimensions and therefore correspond to different rescaling choices.

The scaling dimensions are computed from the \RG flow equations. It is convenient to write the scaling dimensions as a sum of their canonical dimensions, and an anomalous correction. The canonical dimensions are fixed and given by the canonical scaling at the Gaussian fixed point, where the spatial and temporal coordinates are rescaled as ${\hat{\boldsymbol{q}}= \boldsymbol{q}/k}$ and ${\hat{\omega}= \omega /(K^I k^2)}$. In the \IRD case (and at equilibrium) the relevant fixed point is the interacting Wilson--Fisher (WF) fixed point (see \App{app_wilson_fisher}). There the anomalous dimensions vanish at $1$-loop order in perturbation theory. In this work, we are interested in the effect of the drive on the \IRD criticality and therefore extend this \WF rescaling to all the couplings
\begin{align}
 \hat{\mu}_n = k^{-2} \frac{\mu_n^I}{K^I} \, , && \hat{g}_n = k^{d-4} \frac{\gamma g_n^I}{4 [K^I]^2} \, .
 \label{eq_rescaling}
\end{align}
Then we also rescale ${\hat{\Omega} = \Omega/(K^I k^2)}$, and the flow equations are given by
\begin{align}
 & k \partial_k \hat{\mu}_0 = -2 \hat{\mu}_0 - 4 S_d \left[\frac{\hat{g}_0 \left(1 + 4 m \right) + x - 2s}{1+\hat{\mu}_0} -4 r \right] \, ,\nonumber \\
 & k\partial_k \hat{g}_{0}  = -\epsilon \hat{g}_0 + 10 S_d \left[ 16 g - \frac{8\hat{g}_0 r}{1+\hat{\mu}_0} \right. \nonumber \\
 & \qquad + \left. \frac{\hat{g}_{0} \left[ \hat{g}_{0} \left( 10 + {23 m}\right) + 10 x  - {3 s} \right] + {5x^2}}{10 (1+\hat{\mu}_0)^2}  \right] \, , \nonumber \\
   & k\partial_k u = -2\epsilon u  \, ,\quad \qquad \hspace{6pt} \, k\partial_k s = -\epsilon s - \frac{20 S_d u}{(1+\hat{\mu}_0)^2} \, , \nonumber \\
 & k\partial_k x = -\epsilon x  \, , \qquad \qquad \, k\partial_k r = (d-2) r - \frac{16 S_d}{1+\hat{\mu}_0} g \, ,\nonumber \\
&  k\partial_k g = (2d-4) g \, , \qquad k\partial_k m = -\frac{8 S_d}{1+\hat{\mu}_0} r \, ,
 \label{eq_flow_equations_Omega2_zero}
\end{align}
with $\epsilon = 4-d$. The rescaled drive parameters are obtained by inserting \Eq{eq_rescaling} [and ${\hat{\Omega} = \Omega/(K^I k^2)}$] in \Eq{eq_drive_coefficients}. They are given explicitly in \Eq{eq_rescaled_drive_parameters}. This choice of rescaling produces flow equations, where neither the running cut-off scale, nor $\gamma$, nor $K^I$ appear explicitly.

We emphasize that this choice is tailored to capture the rapidly driven system close to the \IRD criticality, \ie close to the corresponding \WF fixed point. It would not be a good choice to look for a new hypothetical non-equilibrium Floquet fixed point, which could describe a second order phase transition at a finite drive frequency. We see no indication of such a fixed point in our flow equations. Yet, a hypothetical perturbatively (in $\Omega^{-1}$) inaccessible fixed point cannot be fully excluded based on our $1$-loop analysis. In this case however, we do not expect it to play a role in the present regime of asymptotically large $\Omega$. Indeed, such a fixed point would depend on the drive frequency as a fixed parameter, because $\Omega$ enters in the discrete time-translation invariance of the system, and can thus not be renormalized. Then the fixed point couplings (such as $\mu^*$ and $g^*$) would take extreme values when $\Omega^{-1}$ is very small. This means that the \RG flow would need to bring the system across a wide range of couplings before this fixed point is felt. Such a Floquet criticality could only play a role at asymptotically large scales when $\Omega$ is asymptotically large.

\subsubsection{Critical exponents}
\label{sec_critical_exponents}

We find that, in the \IRD limit, the \RG flow equations \Eq{eq_flow_equations_Omega2_zero}, coincide with the known equilibrium \RG flow of $O(2)$ models (see \App{app_wilson_fisher}). In particular \Eq{eq_flow_equations_Omega2_zero} contains a single fixed point, where all the drive parameters vanish and
\begin{align}
\hat{\mu}_0=\hat{\mu}^* \cong - \frac{\epsilon}{5} \, , && \hat{g}_0=\hat{g}^* \cong \frac{4 \pi^2 \epsilon}{5} \, ,
\label{eq_wf_fixed_point}
\end{align}
to order $\mathcal{O}(\epsilon)$. This is the \WF fixed point.\footnote{These values coincide with the equilibrium fixed point, but there is a universal fine structure in the approach to it.} The finite drive frequency plays an important role by providing additional couplings that destabilize the \WF fixed point and eventually prevents the \RG flow from ever reaching it. We start by describing this mechanism qualitatively and explain how the drive can produce a finite correlation length even when the equilibrium couplings are tuned to criticality. Quantitative predictions and a precise consistency with the result of \cite{mathey2018a} will come next.

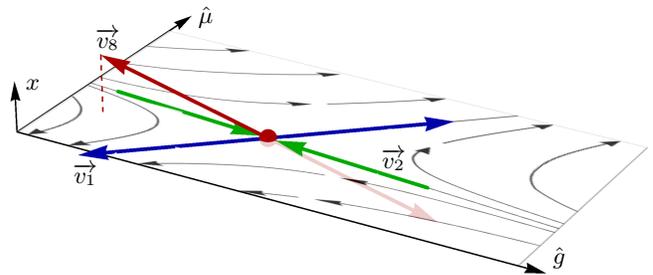
\begin{figure}[t]
\begin{tikzpicture}[scale=1]
\node[] at (0,0) {\includegraphics[width=\columnwidth]{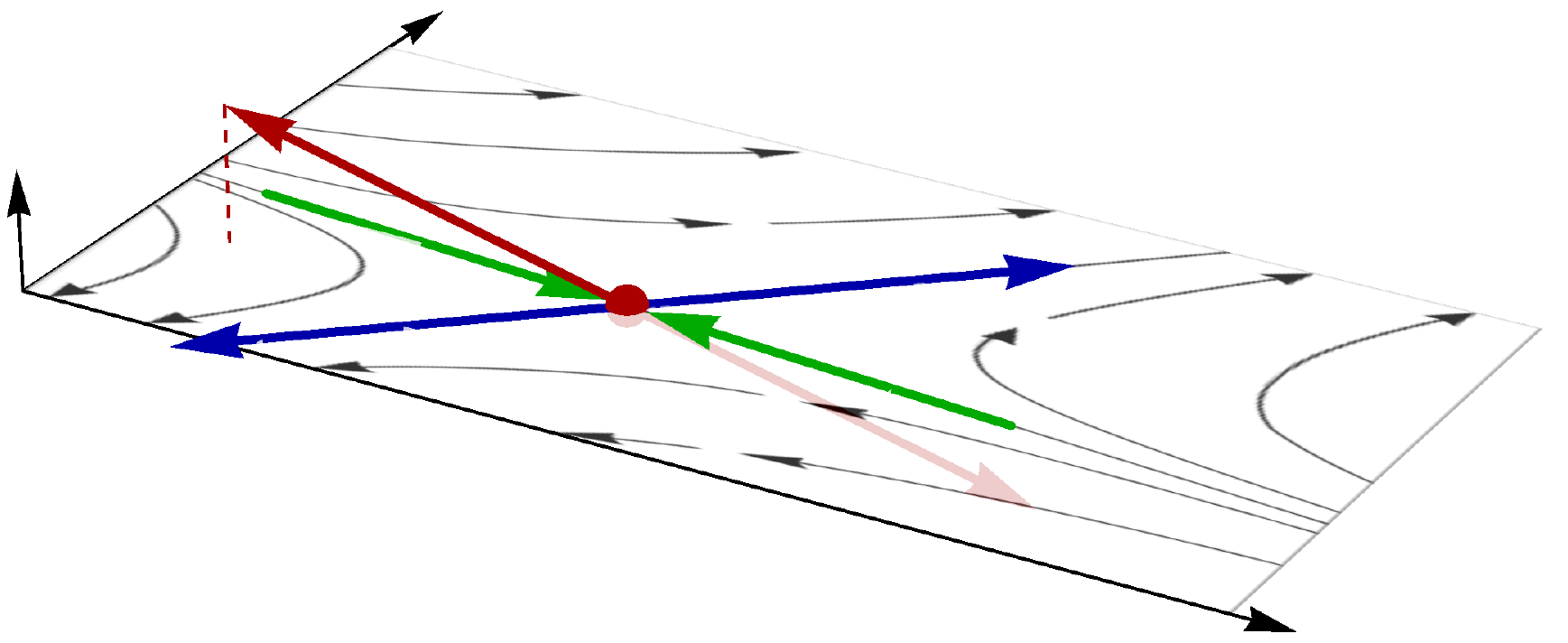}};
\node[] at (-1.7,1.7) {$\hat{\mu}$};
\node[] at (-4,0.8) {$x$};
\node[] at (3.0,-1.5) {$\hat{g}$};
\node[] at (-3.3,-0.4) {$\protect\overrightarrow{v_1}$};
\node[] at (-3.0,1.4) {$\protect\overrightarrow{v_8}$};
\node[] at (0.8,-0.2) {$\protect\overrightarrow{v_2}$};
\end{tikzpicture}
\caption[]{The \RG flow is represented, with the \WF fixed point (red dot) in the middle. The colored arrows represent the eigenvectors of the stability matrix. $\protect\overrightarrow{v_1}$ (blue, pointing in the lower-left direction) and $\protect\overrightarrow{v_8}$ (red, left) are the relevant directions, and $\protect\overrightarrow{v_2}$ (green, lower-right) is irrelevant\textsuperscript{\ref{ftn_footnote_eigensystem}}. The phase boundary is the plane generated by $\protect\overrightarrow{v_2}$ and $\protect\overrightarrow{v_8}$. A non-vanishing value of $x$ shifts the location of the transition, because $\protect\overrightarrow{v_8}$ is not perpendicular to $\protect\overrightarrow{v_1}$ and $\protect\overrightarrow{v_2}$.}
\label{fig_rg}
\end{figure}

When the system is tuned to the critical region of the phase diagram, the \RG flow takes the system very close to its fixed point, where it stays for a wide range of values of the running cut-off scale: $k\in[k_0,k_1]$, with $k_0 \ll k_1$ by definition of the critical region (see \Fig{fig_rg}{}). Both the first (with $\Lambda>k>k_1$) and last (with $k_0>k$) parts of the flow are not universal. Large and small values of $k$ characterize the small- and large-scale physics respectively. The physics is critical when there is a wide range between these two extremes. In particular, tuning the system to its critical point is equivalent to sending $k_0$ to zero and extending the scale-invariant regime to arbitrarily large scales. We will identify $k_0$ with the system's correlation length $\xi \sim k_0^{-1}$, because it provides a scale beyond which non-universal physics kicks in. Then $\xi$ diverges as the system is tuned to the critical point and the scaling of $\xi$ with the distance to criticality is directly related to the rate at which the \RG flows away from its fixed point.

Quantitative predictions are obtained by looking at the flow close to its fixed point. To this end we define the following vector
\begin{align}
 \overrightarrow{G} = \left(\hat{\mu}_0-\mu^*,\hat{g}_0-g^*,m,r,g,s,u,x\right) \, ,
 \label{eq_defG_couplings}
\end{align}
which denotes the displacement from the \WF fixed point in the space of couplings. The critical physics is fully characterized by the \RG flow close to this fixed point. For this reason we focus on systems where the coordinates of $\overrightarrow{G}$ are small, and linearize the \RG flow
\begin{align}
 k\partial_k \overrightarrow{G} = A \overrightarrow{G} \, .
 \label{eq_linearised_flow}
\end{align}
$A$ is the stability matrix of the flow equations evaluated at the \WF fixed point, which can be extracted from \Eq{eq_flow_equations_Omega2_zero}. Its eigenvalues provide the escape rate of the \RG flow from the fixed point. We find that it is an upper-triangular matrix (see \App{app_scaling_behaviour}), so that its eigenvalues can be read off its diagonal. They are [to $\mathcal{O}(\epsilon)$]
\begin{align}
\overrightarrow{\lambda} = \hspace{-2pt} \left( \hspace{-2pt} -2 + \frac{2\epsilon}{5}\, , \epsilon\, ,0 \, , 2-\epsilon \, , 4-2\epsilon \, , -\epsilon \, , -2\epsilon\, , -\epsilon \right) .
 \label{eq_eigenvalues}
\end{align}

The stability matrix must have an upper-triangular structure at the fixed point. Indeed, when all the drive coefficients vanish at the beginning of the \RG flow (as is the case in the \IRD limit), then they are zero for the entire flow. Close to the fixed point this is reflected in the upper-triangular structure of $A$, which imposes that, if the drive coefficients vanish on the right-hand side, then they do not change as $k$ is lowered. Here we will also benefit from the fact that $A$ is fully upper-triangular. This will be used later to easily show that criticality can only emerge when $s=u=x=0$.

In particular, \IRD criticality together with the corresponding divergence of the correlation length $\xi \sim |\Delta|^{-\nu}$ is fully contained in \Eq{eq_eigenvalues}. The mass gap, $\Delta$ represents the distance to the critical point in the \IRD limit. This is a result of the upper-triangular block structure of $A$. Indeed, in the \IRD limit it is possible to leave the drive coefficients out of the problem and restrict $A$ to its upper-left block, which is a $2\times2$ matrix. Then the critical exponents are the first two entries in \Eq{eq_eigenvalues}  ${\lambda_1 = -2 + \frac{2\epsilon}{5}}$ and $\lambda_2 = \epsilon$ (see \App{app_wilson_fisher}). In that case, the fixed point is made unstable by the negative (so-called relevant) eigenvalue $\lambda_1$. In the critical region the couplings eventually flow away from the fixed point as $k$ is lowered (see \Fig{fig_rg}{}) unless the system is tuned to hit it exactly. This identifies the phase boundary as a separatrix between the two attractive (under \RG) regions of the parameter space. We use $\Delta$ to denote the distance from the phase boundary in the \IRD limit\footnote{At equilibrium, $\Delta = (T-T_c)/T_c$ would be the reduced temperature.} and find that the flow behaves as $|\overrightarrow{G}| \sim |\Delta| \, k^{\lambda_1}$ for $k_1\gg k \geq k_0$. The running cut-off scale where the flow leaves the vicinity of the fixed point is $k_0$ and is extracted by setting $|\overrightarrow{G}(k_0)| \sim 1$. Identifying the correlation length with the inverse of this scale $\xi \sim 1/k_0$, leads to the well-known scaling $\xi \sim |\Delta|^{-\nu}$ with $\nu = -1/\lambda_1 \cong 1/2 + \epsilon/10$ \cite{tauber2014critical}.

We now return to the rapidly driven system, where the above argument has to be generalized. In the presence of a drive all the eigenvalues of $A$ become available, and there are $4$ negative eigenvalues. The fixed point is therefore greatly destabilized, and $4$ independent couplings have to be tuned to a specific value for $\xi$ to diverge. Our main result is based on the fact that (as we show below) this tuning amounts to either setting the drive amplitude to zero or, equivalently, going to the \IRD limit.

It can be seen from the stability matrix \Eq{eq_stability_matrix}, that the fixed point can only be reached when $s=u=x=0$. Indeed, the solution of the flow equations is
\begin{align}
 \overrightarrow{G} = \sum_{i=1}^{8} c_i(k) \overrightarrow{v_i} \, , && c_i(k) = c_i \, \left(\frac{k}{\Lambda}\right)^{\lambda_i} \, ,
 \label{eq_flow_general}
\end{align}
close to the fixed point. This equation is obtained by using the eigenvectors of the stability matrix $\overrightarrow{v_i}$, to represent the flow.\footnote{The eigensystem is defined as $A  \overrightarrow{v_i} = \lambda_i  \overrightarrow{v_i}$. The eigenvalues and eigenvectors are labeled from $i=1$ to $i=8$ in the specific order of \Eq{eq_eigenvalues}, because this choice makes the stability matrix upper-triangular [see \Eq{eq_stability_matrix}]. For this reason $x$ is associated with $\lambda_8$ and $\overrightarrow{v}_8$.\label{ftn_footnote_eigensystem}} $c_i$ are the projections of $\overrightarrow{G}(\Lambda)$ onto $\overrightarrow{v_i}$. They depend on the microscopic couplings and can be used to label the position of the system in the phase diagram. The upper triangular form of $A$ makes it possible to straightforwardly compute $c_i$ as functions of the underlying couplings used in \Eq{eq_defG_couplings}. In particular, we find that $c_8$ is proportional to $x$, that $c_7$ is a linear combination of $x$ and $u$, and that $c_6$ is a linear combination of $x$, $u$ and $s$. If $c_{678}=0$, then also $x=u=s=0$. $c_{678}$ (and $c_1$) are the coefficients that are associated to negative eigenvalues, and are therefore relevant couplings. This implies that the fixed point can only be reached (and therefore the correlation length only diverge) if $s$, $u$ and $x$ are set to zero. In that case, we recover \IRD criticality as discussed above.

The above \IRD argument leading to an estimation of the correlation length can be directly generalized to the driven case. Indeed, the only effect of the drive is to provide additional flow directions along which the \WF fixed point is unstable (see \Fig{fig_rg}{}). Then, the scale $k_0 \sim \xi^{-1}$ is defined as before as the value of the running cut-off scale where the flow leaves the vicinity of the fixed point
\begin{align}
 k_0 = \Lambda \, \text{Max}\left[|c_1|^{\nu},|c_6|^{1/\lambda_6},|c_7|^{1/\lambda_7},|c_8|^{1/\lambda_8}\right] \, .
\end{align}
The drive introduces multiple scaling regimes. We illustrate this by going back to $\mathcal{O}(\Omega^{-1})$, where all the drive coefficients but $x$ drop out. Then only $c_1$ and $c_8$ (with $\lambda_{1,8}$) remain in the above equation, and $c_8 \sim x$. We find that for $|c_1|$ large enough, we have $|c_1|^{\nu}>|c_8|^{1/\lambda_8}$ and the correlation length scales as $\xi \sim |c_1|^{-\nu}$. Then the $\IRD$ scaling, $\xi \sim |\Delta|^{-\nu}$ is recovered when $\Delta$ is large enough (see \App{app_scaling_behaviour}). When $|c_1|$ decreases however, $|c_8|^{1/\lambda_8}$ takes over and the correlation length scales as $\xi \sim |x|^{-1/\epsilon}$. The only way to have $\xi \rightarrow \infty$ as $c_1 \rightarrow 0$ is to set $x=0$.

\begin{figure}[t]
 \begin{center}
  \begin{tikzpicture}[scale=.9,transform shape]
   \node[] at (-2.3,0) {\includegraphics[width=0.48\columnwidth]{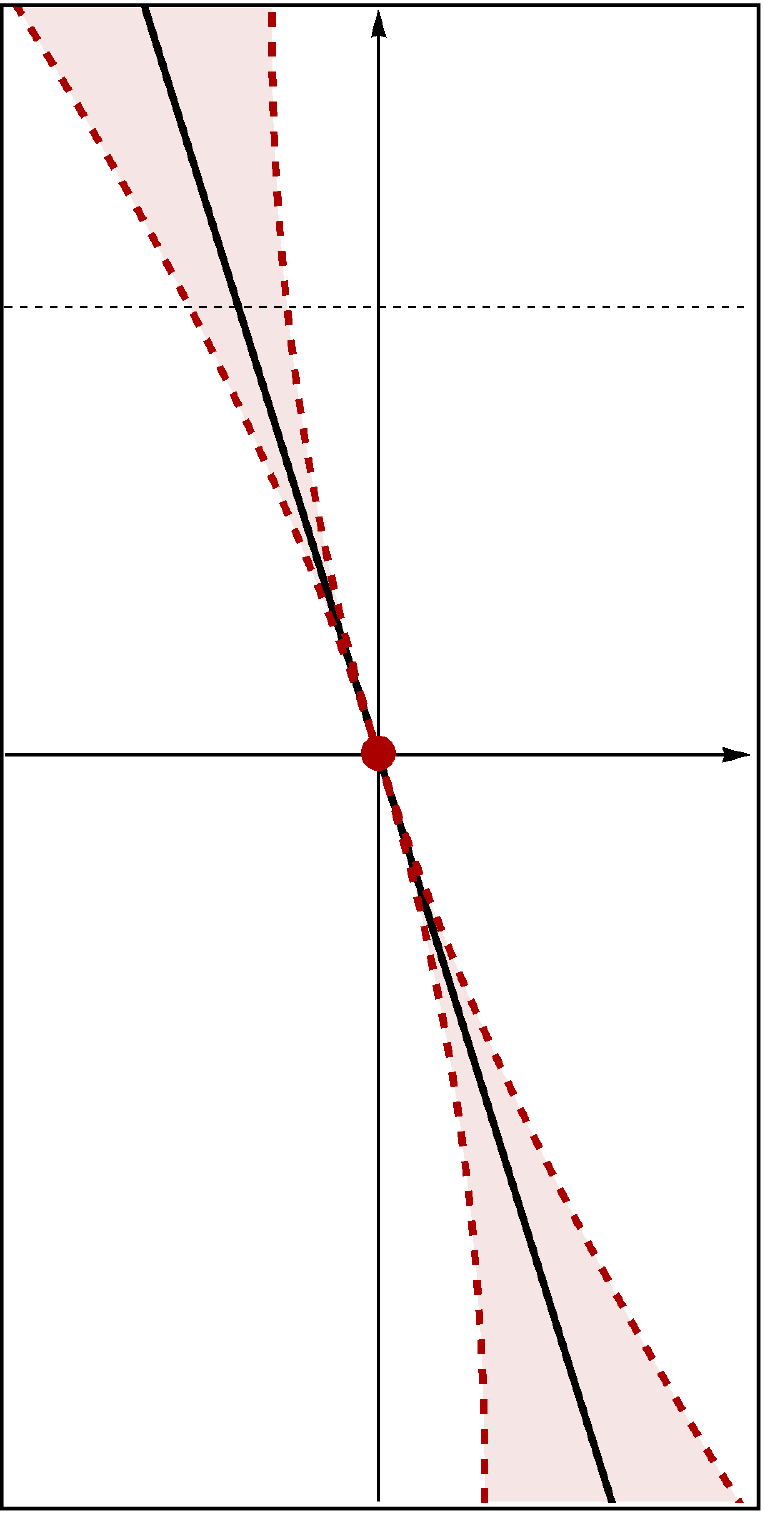}};
      \draw[white,opacity=0,fill=blue,fill opacity=0.25,shading=axis, left color=blue, right color=white] (-4.35,0) -- +(-90:0.6) arc(-90:90:0.6) -- cycle;
   \draw[white,opacity=0,fill=blue,fill opacity=0.2,shading=axis, left color=white, right color=blue] (-0.25,0) -- +(90:0.6) arc(90:270:0.6) -- cycle;
   \node[text width=1] at (-2.3+0.3,1.5) {Symmetric phase};
   \node[text width=1] at (-2.3-1.5,-1.5) {Ordered phase};
   \node[] at (-2.3+1.8,-0.2) {$\Delta$};
   \node[] at (-2.3+0.7,3.8) {$x \sim \Omega^{-1}$};
   \draw[<->,thick] (-2.3+0.6,-3.5) -- (-2.3+1,-3.5);
   \node[] at (-2.3+0.85,-3.8) {$c_1$};
   \node[] at (2.3,0) {\includegraphics[width=0.48\columnwidth]{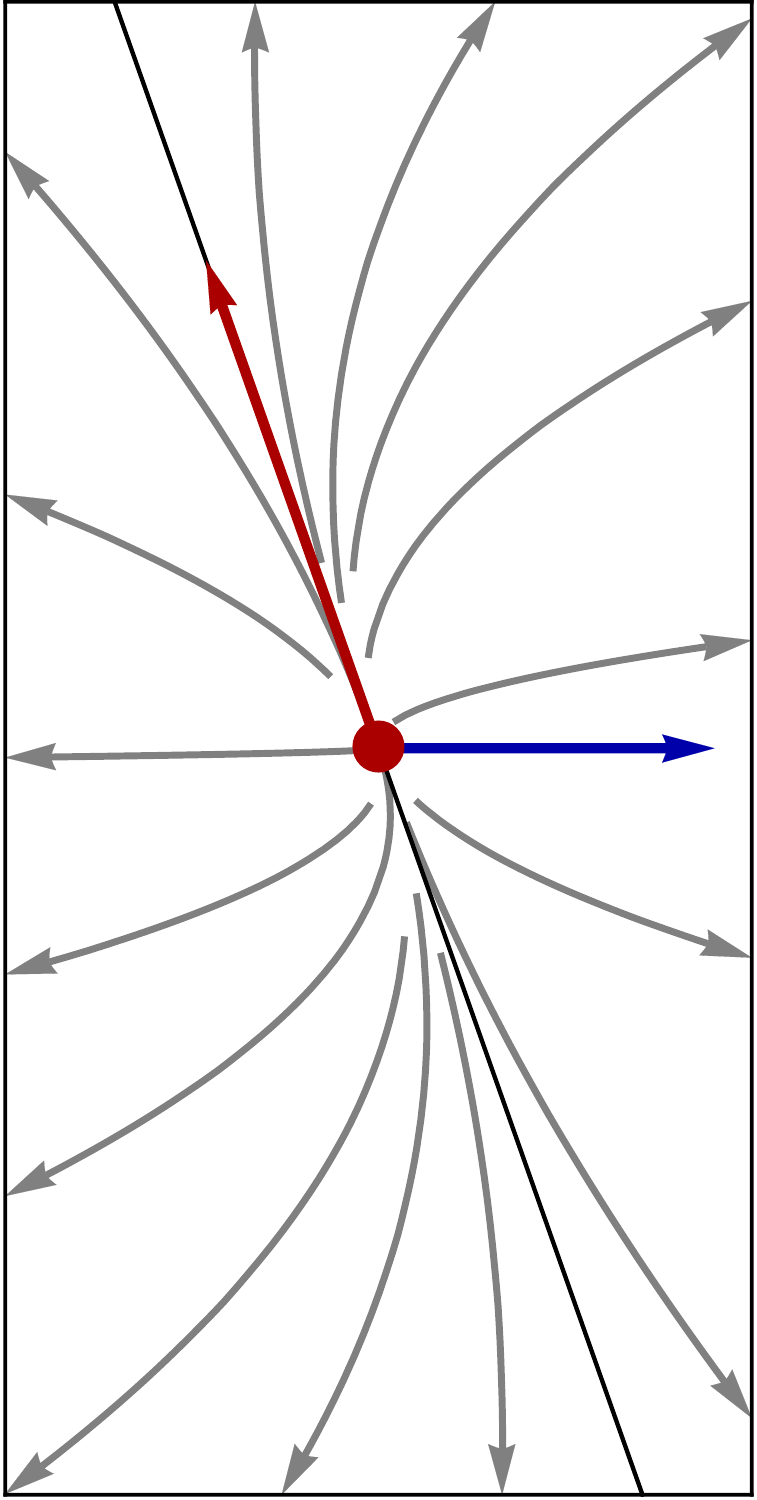}};
   \node[] at (3.6,-0.25) {$\protect\overrightarrow{v_1}$};
  \node[] at (1.2,2.5) {$\protect\overrightarrow{v_8}$};
   \node[] at (-2.3-1.85,-3.9) {\footnotesize (a)};
   \draw[white,fill=white] (2.3-2.0,-4.05) rectangle (2.3-1.7,-3.75);
   \node[] at (2.3-1.85,-3.9) {\footnotesize (b)};
  \end{tikzpicture}
 \end{center}
  \caption{(a) Schematic phase diagram of the open Floquet Bose system in $4-\epsilon$ dimensions. $\Delta$ is the distance from the phase transition in the \IRD system, $x\sim \Omega^{-1}$ is the leading rescaled drive coefficient [cf.~\Eq{eq_drive_coefficients_monochromatic}]. The symmetry breaking phase transition occurs at the solid black line. It is second order only at $\Omega^{-1} =0$ (red dot). Otherwise, fluctuations associated to the periodic drive transform the transition to a weakly first order one. The dashed red lines represent a crossover region between the known $\Omega^{-1}=0$ scaling regime and one where scaling is frozen out (light-red). The black dotted line represents a typical experimental path through the phase diagram, \ie in the presence of a rapid but finite drive frequency $\Omega$. The blue shaded areas far away from the phase boundary represent the domain of applicability of perturbation theory. (b) Schematic projection of the RG flow onto the space spanned by the mass gap $\Delta$ and the drive coefficient $x$. The projection of the corresponding eigenvectors $\protect\overrightarrow{v_1}$ and $\protect\overrightarrow{v_8}$ are represented in blue and red respectively\textsuperscript{\ref{ftn_footnote_eigensystem}}. The fixed point is represented in red in the middle and the RG flow is illustrated by the gray arrows. The phase boundary is shown as a thin black line.}
  \label{fig_phase_diagram}
\end{figure}
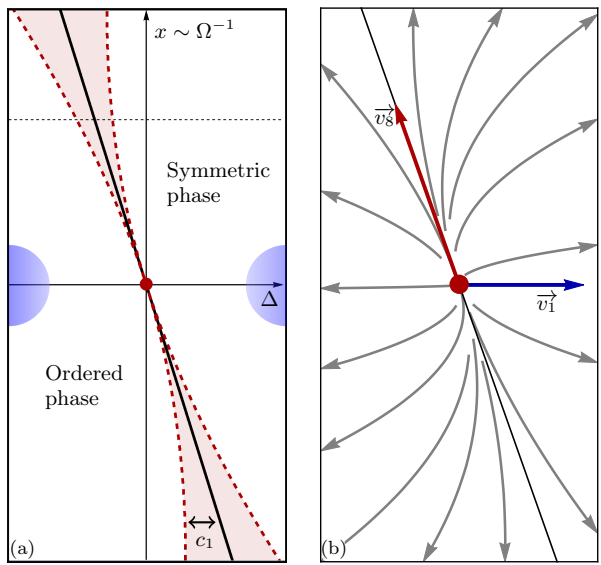

Moreover, we can estimate the point at which the system crosses over from one scaling to another by equating the correlation lengths in the two regimes. We find that this happens when $|c_1|^\nu \sim x^{1/\epsilon}$ (see \Fig{fig_phase_diagram}{a}). We can picture this as the critical phase boundary being blurred as a result of the system being rapidly and periodically dragged across the phase transition. Indeed, when the system is tuned to the phase boundary, the dissipative mass vanishes on period average, but retains oscillating components because of the drive [see \Eq{eq_mut}]. These in turn allow for the system to exhibit a finite correlation length even when $\Delta = 0$.

We can now identify the role of the additional drive parameters that enter at $\mathcal{O}(\Omega^{-2})$. These produce additional scaling regimes. The picture remains however the same: Far from the \IRD critical point ($|c_1|$ large enough), the scaling is the equilibrium one $\xi \sim |\Delta|^{-\nu}$. As we approach the critical point $|c_1|$ decreases and the scaling crosses over to one of the $\xi \sim |c_{6,7,8}|^{1/\lambda_{6,7,8}}$. The correlation length never diverges. We emphasize that the $3$ relevant exponents $\nu_{di}=-1/\lambda_{i}$, which take on the values $1/\epsilon$ and $1/(2\epsilon)$ are new and original critical exponents that emerge because of the effect of the rapid periodic drive. They can not be related to any of the \IRD exponents, and can only be observed in the presence of a rapid periodic drive. This is in direct opposition to the exponents that emerge in the adiabatic regime, and are related to the equilibrium critical exponents by the \KZ mechanism.

We conclude this section by pointing out that the drive can also affect the phase transition in a non-universal way. Indeed, the sign of $c_1$ controls the macroscopic phase of the system (see \Fig{fig_phase_diagram}{b}). In the \IRD case, we find that $\Delta\sim c_1$ (see \App{app_scaling_behaviour}). When the system is driven however,  $c_1$ becomes a linear combination of $\Delta$ and the drive couplings. Then, the position of the phase transition in the parameter space is shifted (see \Fig{fig_phase_diagram}{a}, tilted black line). The drive protocol can be used to either enhance or suppress the emergence of a condensate. See, \eg \cite{Bukov2015b,Knap2016,Babadi2017,Murakami2017}, where similar effects were studied. This is a static effect where the \IRD description remains valid, but is renormalized by the drive.

\section{Discussion}
\label{sec_discussion}

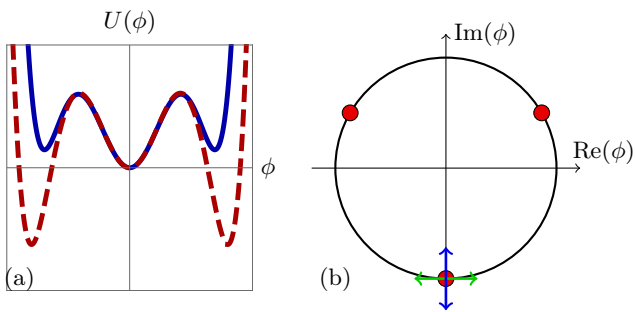
\begin{figure}[t]
\begin{center}
\begin{tikzpicture}[scale=1.05]
 \node[] at (-4,0) {\includegraphics[width=0.38\columnwidth]{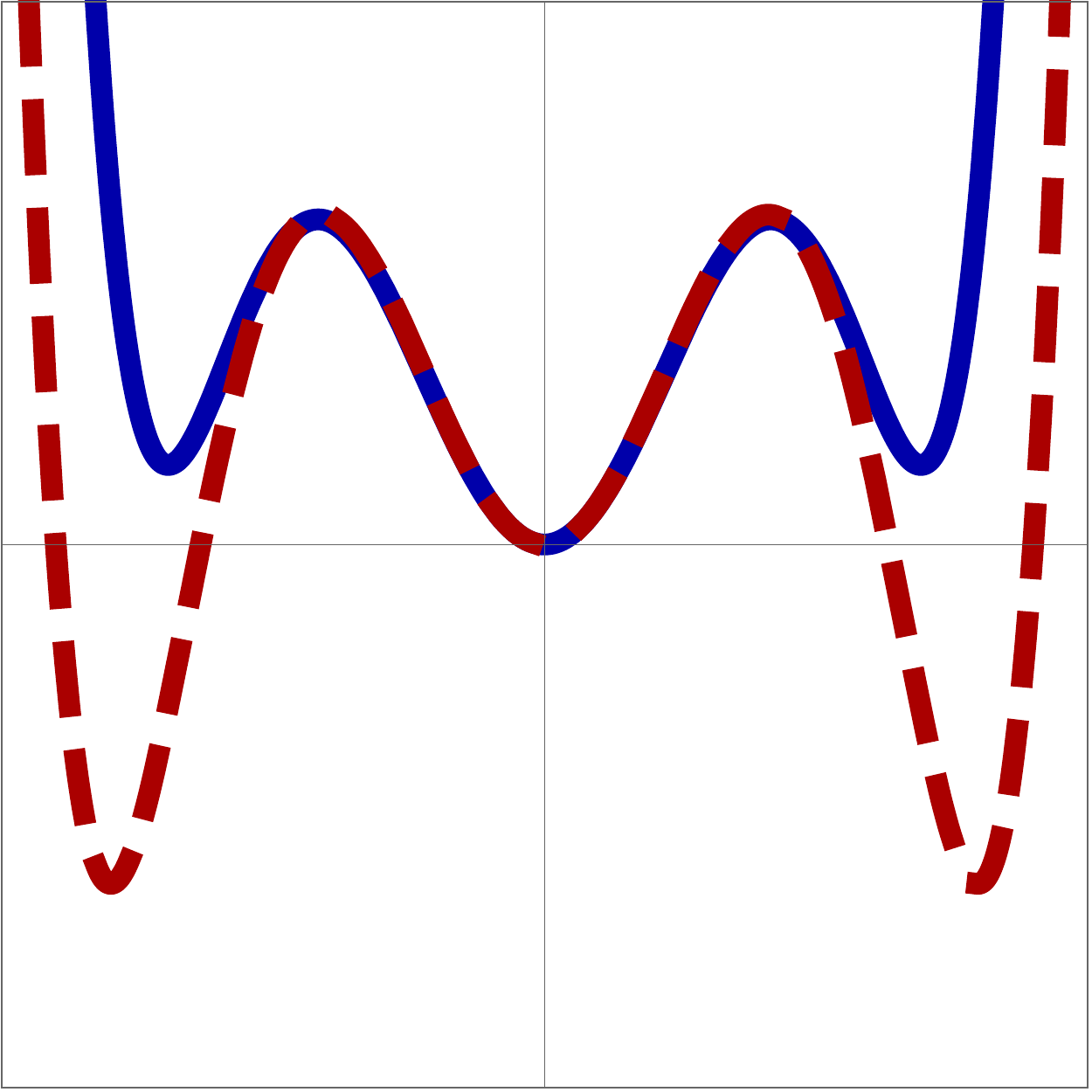}};
\node[] at (-2.25,0) {$\phi$};
\node[] at (-4,1.85) {$U(\phi)$};
\node[] at (-5.4,-1.4) {(a)};
\draw[thick] (0,0) circle (1.4);
\draw[->] (-1.7,0) -- (1.7,0);
\draw[->] (0,-1.7) -- (0,1.7);
\node at (2,0.2) {$\text{Re}(\phi)$};
\node at (0.5,1.7) {$\text{Im}(\phi)$};
\draw[fill=black!10!red] (1.4*0.86603,1.4*0.5) circle (0.1cm);
\draw[fill=black!10!red] (-1.4*0.86603,1.4*0.5) circle (0.1cm);
\draw[fill=black!10!red] (0,-1.4) circle (0.1cm);
\draw[<->,black!20!green,line width = 1] (-0.4,-1.4) -- (0.4,-1.4);
\draw[<->,black!10!blue,line width = 1] (0,-1) -- (0,-1.8);
\node[] at (-1.4,-1.4) {(b)};
\end{tikzpicture}
\end{center}
\caption{Analogous fluctuation induced first order transitions. (a), Coleman-Weinberg or Halperin-Lubensky-Ma mechanism: The blue continuous and red dashed curves represent the potential as a function of the order parameter in the symmetric and ordered phase respectively. The usual ``Sombrero'' potential for the order parameter is modified when additional gapless modes are integrated out and behaves as a $\phi^6$ potential, where the order parameter jumps from being zero to being finite as $\mu_0^I$ is lowered below a threshold. (b), $3$-state Potts model: The black ring represents the minimum of the potential of the fully $U(1)$ symmetric system. In the Potts model this symmetry is explicitly broken down to a discrete $3$-fold symmetry $Z_3$, so that the potential has $3$ minima (shown as red dots) and two different masses (blue and green arrows). As the system goes through the phase transition only one of these masses (vertical blue arrow) vanishes while the other stays finite and imposes a finite correlation length on the system.}
\label{fig_first_order_examples}
\end{figure}

In the previous section, we find that non-vanishing drive coefficients inhibit criticality by imposing a finite correlation length onto the system as it changes its phase. We interpret this as signaling the presence of a fluctuation-induced weakly first order transition as we argue in the following. While first order transitions commonly rely on an explicit symmetry breaking to impose a finite scale as the system changes phase, here the transition is turned first order as a result of the strong fluctuations of the periodic degrees of freedom. The system goes through a first order transition where the $U(1)$ symmetry is spontaneously broken (see \Fig{fig_first_order_examples}{a}).

Far away from the phase boundary (blue areas in \Fig{fig_phase_diagram}{a}), we expect the system to depend smoothly on the drive protocol. Close to the \IRD limit, the effect of the drive is small, and there are two different phases also when $\Omega$ is finite. Then our calculation shows that, when the drive is switched on, it is possible for the system to break the $U(1)$ symmetry spontaneously (with the order parameter changing non-analytically), without going through a critical point. Indeed, this is the phenomenology of a first order phase transition.
 
A more detailed physical picture can be obtained by drawing a parallel with phase transitions that are driven from second to first order by virtue of strong fluctuations (see \Fig{fig_first_order_examples}{a}). The present mechanism is analogous to the Coleman-Weinberg or Halperin-Lubensky-Ma mechanism, where additional gapless modes -- such as gauge fields \cite{Coleman1973,Halperin1974} or Goldstone modes \cite{Fisher1974,Nelson1974} -- compete with the critical ones in the vicinity of a phase transition, and change it from second to first order. In the driven case, the fluctuating fields can be decomposed in discrete Floquet components $\Phi_n(\omega) = \Phi(\omega-n\Omega)$ (with $|\omega|<\Omega/2$) that describe the occupation of the different Fourier modes of the order parameter [see \Eq{eq_floquet_modes}]. The analogy lies in the fact that the fluctuations of $\Phi_0$ compete with the fluctuations of $\Phi_{n\neq 0}$ and are not able to become critical any more.
 
Undriven systems are invariant under continuous time translations ${\Phi (t) \to \Phi (t + \Delta t)}$, for arbitrary $\Delta t$. This continuous symmetry is however broken down to a discrete one ${\Phi (t) \to \Phi(t + 2\pi/\Omega)}$, in the presence of a periodic drive. The continuous symmetry is trivially restored when the drive is switched off ${\mu_{n \neq 0}= g_{n \neq 0}=0}$, but also in the \IRD limit $\Omega^{-1} =0$, where the rotating wave approximation is applicable. This is manifest in \Eq{eq_green_omega_t}, where the effect of the drive enters through the ratio $E/\Omega$, which vanishes when $\Omega^{-1}=0$. Even when they do not vanish, the Fourier modes $\mu_{n\neq0}$ and $g_{n\neq0}$ play no role in this limit. This is reflected in the \RG flow equations \Eq{eq_dimensionful_flow_equations_Omega2}, where the Fourier modes $\mu_{n\neq0}$ and $g_{n\neq0}$ decouple from the rest of the problem when $\Omega^{-1}=0$. Conversely, the explicit breaking of time-translation symmetry allows for the presence of additional dimensionful couplings $\mu_{n \neq 0}$ and $g_{n \neq 0}$. These are not compatible with the undriven dynamical $\phi^4$ theory, and lead to new relevant couplings at the \WF fixed point.

This allows us to draw a different parallel with the Potts model, this time. There, a continuous \emph{external} (order parameter) symmetry is explicitly broken down to a non-trivial discrete subgroup (\eg $U(1) \to Z_3$ in the Potts model \cite{Golner1973,Wu1982}, or similar phenomena in $O(N)$ models \cite{Carmona2000,Aharony2003}). See \Fig{fig_first_order_examples}{b}. This allows for new relevant operators to emerge. The analogy with our system is that, while we do not break the external phase rotation symmetry $U(1) \simeq O(2)$, we explicitly break time translation invariance down to its discrete version. The corresponding newly relevant operators, which are given by the drive coefficients, emerge as a result of the non-vanishing Fourier modes $\mu_{n\neq0}$ and $g_{\neq0}$.

This interpretation in terms of a first order transition is further supported by the fact that, in the presence of a periodic drive, there is nothing stopping $g_0^I$ from becoming negative along the \RG flow (see, \eg \cite{Hnatic2020}). Indeed, the flow of $g_0^I$ does not stop and even decreases (as $k$ is lowered) when $g_0^I=0$ [see \Eq{eq_dimensionful_flow_equations_Omega2}] since $X_0^2$ and/or $G_0$ are never negative and $Q_0 \cong 0$. This means that the effective large-scale description can become unstable unless higher order couplings are taken into account. This is exactly what happens in $\phi^6$ theory, where the two-particle coupling is allowed to be negative from the outset, and only the three particle coupling (coefficient of the sextic term) stabilizes the theory. There the $U(1)$ symmetry breaks spontaneously through a first order transition. We emphasize that this is most likely to happen close to the driven critical point, where $g_0^I \cong k^\epsilon g^*$ becomes very small while $X_0$ and $G_0$ are held fixed.

A fluctuation-induced first order transition presents a qualitative effect of the drive that remains when the drive is (finite, but) very rapid. This means that the rotating wave approximation, which predicts a second order phase transition, must break down in the critical region.

We close this section by demonstrating that the mechanism discussed above is universal. For a generic drive protocol the drive coefficients will not be zero and will provide a finite correlation length. If we take the example of a monochromatic drive and purely imaginary couplings, we find that the drive parameters can vanish when either $\mu_{\pm}^I=0$ or $g_{\pm}^I=0$ [see \Eq{eq_drive_coefficients_monochromatic}]. This would suggest that criticality in a driven system remains possible for certain drive protocols. However, this is only possible when the drive is fine-tuned to a specific form. Indeed, we have chosen a mesoscopic description of the model, which is written in terms of the dynamics of the order parameter alone [see the discussion leading up to \Eq{eq_action}]. The connection between the microscopic and mesoscopic descriptions is extremely complex. When the system is driven at the microscopic level, this drive propagates through the full problem and produces a complex mesoscopic theory, where all the couplings are synchronized with the drive. In practice it is not possible to drive the system microscopically in such a way that $\mu(t)$ or $g(t)$ is an exact constant in time, and there is no reason to assume that this happens accidentally. We can actually even see an example of this in the general form of the \RG flow equations \Eq{eq_dimensionful_flow_equations_Omega2}. Indeed, when either $\mu$ or $g$ is undriven ($\mu_{n\neq 0}=0$ or $g_{n \neq 0 }=0$) then the coupling set to zero will be generated by the other as the running cut-off scale is lowered.\footnote{This is not explicitly  visible in \Sect{sec_scaling_behaviour}, where setting $g_{\pm 1}^I=0$ removes all the drive coefficients, and these are not generated in \Eq{eq_flow_equations_Omega2_zero}. This apparent discrepancy with \Eq{eq_dimensionful_flow_equations_Omega2} lies in the fact the $g_{\pm 1}^I$ is generated at $\mathcal{O}(\Omega^{-2})$ only and these contributions disappear when this is inserted in \Eq{eq_dimensionful_flow_drive_parameters_final}. There would be no discrepancy if higher orders in $\Omega^{-1}$ were taken into account.} This means that when the \RG flow reaches the vicinity of the fixed point (and our analysis applies) the relevant drive coefficients can only vanish if there is a physical mechanism forcing them to be zero. Although we identify such a mechanism in \App{app_equilibrium_symmetry} as a symmetry of our action, this reflects the fine tuning anticipated above since the symmetry of \App{app_equilibrium_symmetry} will not be realized generically.

\section{Conclusion}

We have combined the Floquet and Keldysh formalisms together into a dynamic \RG approach that consist of jointly accounting for the renormalization of the static (Fourier mode $n=0$) and periodic sectors together with their interplay. This has enabled us to account for strong critical fluctuations and include the effects of a periodic drive in the dynamics of a driven open gas of bosons at critical point. In the presence of a rapid drive, the system goes through a phase transition, where the $U(1)$ symmetry spontaneously breaks. This is a second order transition in the \IRD limit, where the system is invariant under time translations. Our main physics result is that the periodic drive enables new couplings to enter, which in turn regularize the divergence of the correlation length that takes place at this critical point.

This is a universal mechanism that mainly relies on the existence of an interacting critical point and could be realized in varied physical systems. There are nevertheless important conditions for its possible realization. First, it takes place in systems that have reached a Floquet steady state, where the drive and dissipation compensate each other and all the Floquet modes are populated. The system must be allowed to fully synchronize with the drive. Dissipation is therefore an essential ingredient. Second, there must be a non-trivial critical point. \Ie the underlying \IRD \RG fixed point must not be Gaussian ($\hat{g}^* \neq 0$). Indeed, in the absence of interaction, the drive decouples from the \IRD physics (the stability matrix becomes fully block diagonal) and can not affect the critical physics any more. This is reflected in our calculation by the fact that all the newly relevant couplings become irrelevant if $\epsilon<0$ (for $d>4$), when the \WF fixed point becomes trivial. In particular, this excludes integrable systems and most classical $1d$ systems.

At this stage, several directions of research await their exploration. A first one concerns a more precise computation of the new critical exponents emerging in the rapid drive limit, $\Omega^{-1} \to 0$. This could be achieved via our dynamic \RG approach within a $2$-loop calculation, systematic to order $\epsilon^2$ \cite{tauber2014critical}, or via a non-perturbative \RG approach \cite{Wetterich:1992yh}.

Furthermore, it will be interesting to explore the existence of a novel Floquet \RG fixed point at intermediate values of $\Omega$, now of comparable size to other scales in the problem. We expect such a fixed point to be out of reach of our $\Omega^{-1}$-expansion. Therefore, a different kind of expansion, for example involving a re-summation of the present asymptotic expansion, is necessary. In this respect, the calculation of \Sect{sec_poles} could be a good starting point.

Finally, recent findings for symmetry broken Floquet steady states with qualitatively new and original properties \cite{Walldorf2018}, motivate the development of a more analytical understanding. This calls for a more direct approach to the effect of a rapid drive on phases with spontaneously broken continuous symmetries, or more generally, with gapless modes. The presence of such gapless modes will most likely lead to a breakdown of the rotating wave approximation in the many-body system in a similar way as seen here near criticality. This could be accommodated within a dynamic effective action approach, that is capable of handling the steady state order parameter in the symmetry broken phase. In particular, one lesson learned in this work suggests that the order parameter modes $\Phi_0$ and $\Phi_{n\neq 0}$ need to be treated on an equal footing in the presence of gapless modes. In particular, applied to our model, such an approach would enable us to better understand the interplay between the periodic drive and the Goldstone modes that emerge when the $U(1)$ symmetry is spontaneously broken.

\textit{Acknowledgments - } We thank A. Altland, C.-E. Bardyn, M. Buchhold, C. Duclut, A. Gambassi, M. Heyl, A. Lazarides, G. Loza, J. Marino, R. Moessner, F. Piazza, A. Polkovnikov, G. Refael, A. Rosch, D. Roscher, M. Scherer, K. Seetharam, N. Sharma, U. T\"auber, M. Toulouse and J. Wilson for useful and inspiring discussions.  We acknowledge support by the funding from the European Research Council (ERC) under the Horizon 2020 research and innovation program, Grant Agreement No. 647434 (DOQS), and by the DFG Collaborative Research Center (CRC) 1238 Project No. 277146847 - project C04. This research was supported in part by the National Science Foundation under Grant No. NSF PHY-1748958.

\appendix

\section{Slow drive}
\label{app_slow_drive}

Here, we briefly analyze the case of a slow drive. See also \cite{RodriguezVega2018}, where a fully quantum system is analyzed. In particular we show how the Green functions can be expanded in powers of $\Omega$ leading to an adiabatic approximation of the problem, \Eq{eq_green_smallOmega}.

In the case of a slow drive, an expansion in powers of $\Omega$ can be carried out straightforwardly. We can directly expand the Wigner and/or Floquet Green functions. We start with the Floquet retarded Green function that is expanded as
\begin{align}
 G_{R} & = \frac{1}{\omega + M_0 + \hat{n}\Omega + E} \label{eq_smallOmega_Floquet} \\
 &  \cong \frac{1}{\omega + M_0 + E} - \frac{1}{\omega + M_0 + E} \hat{n}\Omega \frac{1}{\omega + M_0 + E}  \, , \nonumber
\end{align}
with $[\hat{n}]_{nm} = \delta_{nm} n$ a matrix in Floquet space. $M_0$ and $E$ are defined in \Eq{eq_Mt_fourier}. Converting to the Wigner representation provides
\begin{align}
& G_{R;n}(\omega) \cong \left[\frac{1}{\omega + M_0 + E}\right. + m \Omega \left(\frac{1}{\omega + M_0 + E}\right)^2 \nonumber \\
& - \left.\frac{1}{\omega + M_0 + E} \hat{n}\Omega \frac{1}{\omega + M_0 + E}\right]_{m+n/2,m-n/2} \, .
\label{eq_smallOmega_Wigner}
\end{align}
The second term in the above equation is obtained from the first term of \Eq{eq_smallOmega_Floquet} through the $\Omega$-dependent shift in frequency that is necessary to go from the Floquet to the Wigner representation [\Eq{eq_Gnm_to_Gn}]. The above expression is only valid for even values of $n$. When $n$ is odd, $m$ has to be replaced by $m\pm1/2$. In any case, $m$ must drop out at the end of the calculation since the Wigner Green functions only have one index. For this reason, we only work out the case of $n$ even. Odd values of $n$ are analogous.

The above expressions are still formal since they are written in terms of the inverse of a non-diagonal matrix. They can however be computed exactly, term by term. To this end, we use \Eq{eq_series} without truncating the sum, convert each term to its real-time representation and re-sum the obtained expressions. We use the following relations
\begin{align}
 & \left[ E^N \right]_{m+n/2,m-n/2} = \fint_t \text{e}^{\I n \Omega t} E(t)^N \nonumber \\
  & \left[A \hat{n}\Omega E B\right]_{st} = \left[A E \hat{n}\Omega B\right]_{st} + \I \left[A E' B\right]_{st} \nonumber \\
 & \left[A \hat{n}\Omega\right]_{st} = \left[A\right]_{st} t \Omega \, ,
\end{align}
where $N$, $s$ and $t$ are integers and $A$ and $B$ are Floquet matrices with the same structure as $E$, $A_{nm} = A_{n-m}$ and $B_{nm} = B_{n-m}$. $[E']_{nm} = (E')_{n-m} = \fint_t \text{e}^{\I(n-m) \Omega t} E'(t)$ is the Floquet matrix obtained from the time-derivative of $E$. To order $\mathcal{O}(\Omega)$, the obtained Green functions are
\begin{align}
 & G_{K;n}(\omega) = \fint_t \text{e}^{\I n \Omega t} \left[\frac{- \I \gamma}{\left|\omega+M(t)\right|^2} + \frac{\gamma}{2} \frac{(E(t)^*)'-E(t)'}{\left|\omega+M(t)\right|^4} \right] \nonumber \\ 
  & G_{R;n}(\omega) = \fint_t \text{e}^{\I n \Omega t} \frac{1}{\omega+M(t)} \, .
 \label{eq_green_smallOmega}
\end{align}
The correction to $G_{R;n}(\omega)$ start at $\mathcal{O}(\Omega^2)$. $M(t) = M_0 + E(t)$ is defined above \Eq{eq_Mt_fourier}. It is clear from \Eq{eq_green_smallOmega} that the problem becomes adiabatic when $\Omega \rightarrow 0$. Indeed, the Green functions are obtained (to leading order) by computing them for a stationary system and restoring the time dependence at the end. $\Omega$ is even not explicit here. Instead, we have written \Eq{eq_green_smallOmega} in terms of the time derivative of the couplings, $E(t)' = dE/dt$. Even if this derivation is based on the Floquet formalism, it shows how a driven system with a very long period effectively looses its periodicity.

This expansion can be inserted in the real-time representation of the \RG flow equations \Eqs{eq_real_time_flow_equations_mu}, to produce an adiabatic expansion of the full problem. See \cite{mathey2019a}, where such an expansion is applied to the \KZ problem.

\section{Detailed balance}
\label{app_equilibrium_symmetry}

Detailed balance in stationary (undriven) systems can be framed in terms of a microscopic symmetry of the dynamic action \cite{Sieberer2013b,tauber2014critical,Sieberer2015,Aron2018}. In this section, we show that this statement can actually be generalized to periodically driven systems. In this case however, the system does not exhibit detailed balance, but instead, a generalized version of the \FDRs emerges [see \Eq{eq_fluctuation_dissipation}]. Here we will give an interpretation of this result, which we discuss at the end of \Sect{sec_discussion}.

Equilibrium driven open systems are found to display \FDRs. The converse is also true: A system that displays \FDRs can be said to be at thermal equilibrium. Displaying \FDRs involves a lot of information since these relations extend to all the correlation functions. This information can however be distilled into a microscopic symmetry [see \Eq{eq_equilibrium_symmetry}] that the system must obey in order to be at thermal equilibrium \cite{Sieberer2013b,tauber2014critical,Sieberer2015,Aron2018}. Then the \FDRs emerge naturally as Ward identities. While this symmetry was discovered for stationary systems, we can still ask under what conditions it is a symmetry of the periodically driven system. Although the simplest answer is that the drive must be turned off, we find that there are also specific choices of nontrivial drive protocols for which the system is invariant under the equilibrium symmetry. In these cases, the corresponding Ward identities are however not real \FDR relations any more [\eg see \Eq{eq_fluctuation_dissipation}] because they involve the Wigner Green functions with all values of $n$ independently. Nevertheless, although they remains out of equilibrium, these systems exhibit some of the properties of thermal equilibrium.

Although the microscopic symmetry is described \eg in \cite{tauber2014critical} for classical systems, we follow \cite{Sieberer2013b}, which is closest to our set-up. Before we give the transformation, we go to a generalized version of our model \Eq{eq_action},
\begin{align}
 S = \int_{t,\boldsymbol{x}} \tilde{\phi}^* & \left[ Z^* \I \partial_t \phi - \mathcal{K}(\left|\phi\right|,t) \phi  \right]  + \text{c.c.} + \I \gamma |\tilde{\phi}|^2 \, ,
 \label{eq_general_action}
\end{align}
where the interaction as well as the kinetic term are bundled in the operator ${\mathcal{K}(\left|\phi\right|,t) = K \nabla^2 - \mu - g\left|\phi\right|^2}$. All the parameters of the above equation are potentially periodic in time, and, except for $\gamma$, which is real and positive, they are all complex numbers. This model reduces to \Eq{eq_action} when $Z = 1$ and $\gamma$ is constant. Our symmetry is easiest to interpret if $\tilde{\phi}$ is rescaled as
\begin{align}
 \tilde{\phi}' = \frac{Z_1-r Z_2}{r-\I} \, \tilde{\phi} \, .
 \label{eq_rescaling_phiq}
\end{align}
$Z_{1,2}$ are the real and imaginary parts of $Z$ respectively, and $r$ is a real parameter yet to be determined. In general, both can be time-dependent. After this rescaling the parameters of \Eq{eq_general_action} become
\begin{align}
& \mathcal{K}' = \frac{r+\I}{Z_1-rZ_2} \mathcal{K} = \frac{r \mathcal{K}_1-\mathcal{K}_2+\I \left(r \mathcal{K}_2+\mathcal{K}_1\right)}{Z_1-r Z_2} \, , \nonumber \\
& Z' = \frac{Z_1 r + Z_2}{Z_1-rZ_2} - \I \, , \quad \gamma' = \gamma \frac{1+r^2}{\left(Z_1-rZ_2\right)^2} \, , 
\label{eq_rescaled_couplings}
\end{align}
with $\mathcal{K} = \mathcal{K}_1 + \I \mathcal{K}_2$. This is the most general rescaling that sets the imaginary part of $Z$ to minus one. We now define the following field transformation in terms of the rescaled variables\footnote{Note that the symmetry transform is constructed as the time-reversal of the linear combination $\tilde{\phi}+\I/(2T) \partial_t \phi$. This implies that the complex conjugate of $\tilde{\phi}$ is transformed as $\hat{\tilde{\phi}}^*(t) = \tilde{\phi}(-t) {+} \I/(2T) \partial_t \phi(-t)$ because time-reversal is a linear operator. See \cite{Sieberer2015} for further details.}
\begin{align}
\hat{\phi}(t) = \phi^*(-t) \, , && \hat{\tilde{\phi}}(t) = \tilde{\phi}^*(-t) + \frac{\I}{2T} \partial_t \phi^*(-t) \, .
\label{eq_equilibrium_symmetry}
\end{align}
$T$ is another real parameter. It was shown for time independent couplings \cite{Sieberer2013b,tauber2014critical,Sieberer2015,Aron2018}, that the Ward identities associated to this symmetry are thermal \FDRs. In that case, the system is at thermal equilibrium if and only if there exists two real and positive constants $r$ and $T$ such that \Eq{eq_equilibrium_symmetry} is a symmetry of the action. Then $T$ plays the role of the temperature. It is also possible in the case of a periodically driven system to find a periodic function of time $r(t)$ and a positive real constant $T$ such that \Eq{eq_equilibrium_symmetry} is a symmetry of the system. Then, although the system remains out of equilibrium, it has some of the properties of thermal equilibrium.

In particular, the fluctuation-dissipation relation for the two-point correlation functions emerge if our field transform has no effect on the correlation function with two $\tilde{\phi}(t)$'s
\begin{align}
  \langle \hat{\tilde{\phi}}^*(t) \hat{\tilde{\phi}}(t') \rangle =  \langle {\tilde{\phi}}^*(t) {\tilde{\phi}}(t') \rangle \, .
\end{align}
Indeed, inserting \Eq{eq_equilibrium_symmetry} and remembering that the right-hand-side of the above equation must vanish provides
\begin{align}
  \partial_t G_R(t,t') + \partial_t' G_A(t,t') = -\frac{\I}{2 T} \partial_t \partial_t' G_K(t,t') \, ,
  \label{eq_fluctuation_dissipation_real_time}
\end{align}
which is
\begin{align}
 \frac{G_{R;n}(\omega)}{\omega - \frac{n \Omega}{2}} - \frac{G_{A;n}(\omega)}{\omega + \frac{n \Omega}{2}} = \frac{1}{2T} G_{K;n}(\omega) \, ,
 \label{eq_fluctuation_dissipation}
\end{align}
in terms of the Wigner Green functions and the rescaled variables \Eqs{eq_rescaling_phiq} and \eq{eq_rescaled_couplings}. In the absence of drive, only the $n=0$ sector is physical. The Green functions are then time-translation invariant $G_n(\omega) = \delta_{n0} G_0(\omega)$, and the usual \FDR emerges. When the system is driven the equilibrium \FDRs remain in the $n=0$ sector. The Green functions obey \FDRs when they are averaged over one period. There are however additional relations for the other Green functions that are not \FDRs.

We will see (in \Sect{sec_discussion}) that it is actually possible to fine-tune the drive protocol (without turning it off) in order to allow the system to become critical. This can be understood in terms of the present discussion. Indeed, it is now clear that certain driven systems are closer to thermal equilibrium than others. Bringing the system as close to equilibrium as possible, then produces some of the properties of equilibrium such as \FDRs for the $n=0$ sector. Even though we can not directly connect the fine tuning of \Sect{sec_discussion} to the equilibrium symmetry within our $1$-loop approximation, we can bring our system closer to equilibrium by choosing $\mu(t)$ and $g(t)$ to oscillate in phase with each other [thus setting $x=0$ in \Eq{eq_flow_equations_Omega2_zero}]. Then the effect of the suppression of criticality goes from being $\mathcal{O}(\Omega^{-1})$ to $\mathcal{O}(\Omega^{-2})$. It becomes weaker. We believe that, extending our approximation scheme (to include the renormalization of $K$, $Z$ and $\gamma$) will equate the fine tuning of \Sect{sec_discussion} to being symmetric under \Eq{eq_equilibrium_symmetry}.

We conclude this section by listing the criteria that the parameters of \Eq{eq_general_action} must satisfy for \Eq{eq_equilibrium_symmetry} to be a symmetry. This will be useful to interpret the different couplings that emerge in \Sect{sec_scaling_behaviour}. We find that our driven system is symmetric under \Eq{eq_equilibrium_symmetry} when:
\begin{enumerate}[(i)]
 \item all the time-dependent couplings are even in $t$ (up to a global time shift).
 \item There is a single (possibly time-dependent) real number $r(t)$ such that the imaginary part of $\mathcal{K}'$ vanishes,
  \begin{align}
    \mathcal{K}_1 = - r \mathcal{K}_2 \, ,
    \label{eq_equilibrium_condition_1}
  \end{align}
  [see \Eq{eq_rescaled_couplings}]. This relation defines $r$. It is a generalization of the requirement (found in \cite{Sieberer2013b}) that all the couplings lie on the same ray of the complex plane. This means that if there is a real number $r$ satisfying \Eq{eq_equilibrium_condition_1}, then it is possible to rescale $\tilde{\phi}$ so that $\text{Im}(Z') = -1$ and $\text{Im}[\mathcal{K}'] = 0$.
  \item The time dependence of $Z$, $\gamma$ and $r$ are such that the temperature, which is defined as
\begin{align}
 T = \frac{\gamma (1+r^2)}{4(Z_1-r Z_2)^2} = \frac{\gamma'}{4} \, ,
\end{align}
does not depend on time.
\item The real part of $\mathcal{K}'$ does not depend on time. Inserting \Eq{eq_equilibrium_condition_1} then provides
\begin{align}
 \frac{\partial}{\partial t} \left[ \frac{1+r^2}{Z_1 - r Z_2} \text{Im}[\mathcal{K}(\left|\phi\right|,t)]\right] = 0 \, ,
\end{align}
where the time derivative does not act on the field. This equation must be valid for all values of $\phi$. It implies that there exists a time-independent real-valued operator $U(\left|\phi\right|)$ such that
\begin{align}
 \mathcal{K}(\left|\phi\right|,t) = \frac{Z_1-r Z_2}{r+\I} \, U(\left|\phi\right|) \, .
\end{align}
\Ie the time-dependent couplings all oscillate in phase with each other. We will see in \Sect{sec_scaling_behaviour} that the qualitative effect of the drive on the \IRD criticality becomes $\mathcal{O}(\Omega^{-2})$ when $\mu(t)$ and $g(t)$ oscillate in phase with each other. See \Eq{eq_drive_coefficients} [and \Eq{eq_drive_coefficients_monochromatic}], where $X_0$ vanishes in that case. The above equation provides the following interpretation: When the couplings are synchronized, the system is closer to thermal equilibrium (where the system can become critical) and the effect becomes weaker.
\end{enumerate}

All these conditions can be inserted into the action \Eq{eq_general_action}. Then, rescaling the response field as in \Eq{eq_rescaling_phiq}, and expressing it as an equivalent Langevin equation provides
\begin{align}
 & \I \left(Z_1' - \I \right) \partial_t \phi = \, K_1'(\left|\phi\right|) + \xi' \, , \nonumber \\
 & \langle \xi'(t,\vec{x}) \xi'^{*}(t',\vec{x'}) \rangle = \delta(t-t') \delta(\vec{x}-\vec{x}') 2 \gamma' \, ,
\end{align}
with all the time-periodicity residing in $Z_1' = \frac{Z_1r+Z_2}{Z_1-r Z_2}$. $\gamma'$ and $r$ are defined in \Eqs{eq_rescaled_couplings} and \eq{eq_equilibrium_condition_1} respectively. When our periodically driven system is symmetric under \Eq{eq_equilibrium_symmetry}, it is a dissipative system with periodically time-dependent reversible couplings.

\section{Expansion of the Green functions to \texorpdfstring{$\mathcal{O}(E^2)$}{O(E*E)}}
\label{app_second_order_Greens}

To second order in $E$ the retarded Wigner and Keldysh Green functions are,
\begin{align*}
 & G_{R;n}(\omega) = \frac{\delta_{n0}}{\omega+M_0} - \frac{E_n}{\left(\omega+M_0\right)^2-\left(\frac{n\Omega}{2}\right)^2} \nonumber \\
 & + \frac{\omega+M_0}{\left(\omega+M_0\right)^2-\left(\frac{n\Omega}{2}\right)^2} \sum_m \frac{E_{n-m} E_{m}}{\left(\omega+M_0\right)^2-\left[\left(m-\frac{n}{2}\right)\Omega\right]^2} \, ,
\end{align*}
and
\begin{align*}
 & G_{K;n}(\omega) = \I \gamma \left[\frac{ E_n}{\left(\omega+M_0+\frac{n}{2}\Omega\right) \left|\omega+M_0-\frac{n}{2}\Omega\right|^2} \right. \nonumber \\
 & +\frac{ (E_{-n})^*}{\left(\omega+M_0^*-\frac{n}{2}\Omega\right) \left|\omega+M_0+\frac{n}{2}\Omega\right|^2} - \frac{\delta_{n0}}{\left|\omega+M_0\right|^2}  \nonumber \\
 & - \frac{1}{\left(\omega+M_0 + \frac{n}{2}\Omega\right) \left(\omega+M_0^* - \frac{n}{2}\Omega\right)} \nonumber \\
 & \times \sum_m \left( \frac{\left(\omega+M_0\right)}{\omega+M_0 - \frac{n}{2}\Omega} \frac{E_{m}E_{n-m}}{\left(\omega+M_0\right)^2 - \left[\left(\frac{n}{2}-m\right)\Omega\right]^2} \right. \nonumber \\
 & \quad + \frac{\left(\omega+M_0^*\right)}{\omega+M_0^* + \frac{n}{2}\Omega} \frac{(E_{-m})^*(E_{m-n})^*}{\left(\omega+M_0^*\right)^2 - \left[\left(\frac{n}{2}-m\right)\Omega\right]^2} \nonumber \\
 & \quad \left. \left. +  \frac{E_{m}(E_{m-n})^*}{\left|\omega+M_0 + \left(\frac{n}{2}-m\right)\Omega\right|^2} \right) \right] \, .
\end{align*}

\section{Derivation of the \texorpdfstring{\RG}{RG} flow equations}

In this section, we give additional details on the derivation of the different versions of the \RG flow equations. We start with \App{app_flow_equations}, where we describe the technicalities leading to the real-time representation of the flow equations \Eqs{eq_real_time_flow_equations_mu} and \eq{eq_real_time_flow_equations_g}. These equations are derived with the help of the $1$-loop expansion, but contain no approximation with respect to the periodic drive. In \App{app_expansion_to_Omega2}, we explain how the Wigner representation of the flow equations \Eq{eq_dimensionful_flow_coupling}, is expanded to order $\mathcal{O}(\Omega^{-2})$ in the case of purely imaginary couplings, leading to \Eq{eq_dimensionful_flow_equations_Omega2}. Finally, we approximate the flow of the different drive coefficients in \App{app_flow_drive_coefficients} to obtain \Eq{eq_dimensionful_flow_drive_parameters_final}.

\subsection{General flow equations}
\label{app_flow_equations}

In this section, we give further details on the derivations of \Eqs{eq_real_time_flow_equations_mu} and \eq{eq_real_time_flow_equations_g}. We start by detailing the three approximations that enter their derivation:

\begin{inparaenum}[(i)]
\item We use perturbation theory, \Eq{eq_oneloop}. This approximation, which is in principle valid as long as the $k$-dependent coupling $g_n$ is small, is justified for $0<\epsilon = 4-d\ll1$ and at criticality, where the rescaled coupling [see \Eq{eq_rescaling}] are either asymptotically small (for $n\neq 0$) or proportional to $\epsilon = 4-d$ [for $n=0$, see \Eq{eq_wf_fixed_point}]. We extend our results to $d=3$, even if $\epsilon = 1$ in that case. Indeed loop perturbation theory relies on the fact that the dimensionality provides a parameter that can be continuously varied from $\epsilon\ll1$ to $\epsilon = 1$ with the results being smoothly deformed \cite{tauber2014critical}. Furthermore we ignore the \RG flow of higher order vertexes (\ie $\Gamma_k$ is expanded to fourth order in the fields) by setting them to zero on the right hand side of \Eq{eq_oneloop}. We assume that if $\Gamma^{(4)}$ is small, then $\Gamma^{(6)}$ must be even smaller since it is not present at $k=\Lambda$ [see \Eq{eq_action}].
\item We do not keep track of the full frequency and momentum dependence of the four-point vertex. Instead we restrict the real-time representation of $\Gamma^{(4)}$ to take a local form.\footnote{Although we do not write it out explicitly, the space dependence is assumed to be local in the same way.} The flowing two-body couplings, which is defined in \Eq{eq_proj_gamma4} is given by
\begin{align}
 \Gamma^{(4)}(t_1,t_2,t_3,t_4)  =  2 g(t_1) \delta(t_1-t_2) \delta(t_1-t_3) \delta(t_1-t_4) \, .
 \label{eq_gamma4_Lambda}
\end{align}
This is realized microscopically (\ie when $k=\Lambda$), as can by seen by differentiating the action $S$ as in \Eq{eq_proj_gamma4}. Although the \RG flow produces a rich space-time dependence [see \Eq{eq_full_flow_g}] the above expression stems from the leading term in an expansion of the $4$-point vertex in powers of the frequencies. Indeed, the above equation is obtained with
\begin{align}
 \Gamma_n(f_1,\vec{p}_1,f_2,\vec{p}_2,f_3,\vec{p}_3) \rightarrow \Gamma_n(0,\vec{0},0,\vec{0},0,\vec{0}) \, ,
\end{align}
inserted in
\begin{align}
 & \Gamma^{(4)}(t_1,t_2,t_3,t_4)  = \nonumber \\
 & \sum_n \int_{f_2,f_3,f_4} \text{e}^{-\I \left[n \Omega t_a + f_2 \tau_2+f_3 \tau_3 + f_4 \tau_4\right]} \, \Gamma_n(f_2,f_3,f_4) \, ,
 \label{eq_gamma4_general}
\end{align}
with the variables $t_a = (t_1+t_2+t_3+t_4)/4$ and $\tau_i = t_i -t_a$. This approximation is justified at criticality because the neglected couplings contain higher powers of $p$ and $\omega$ and are thus irrelevant.
\item We do not allow $\Gamma^{(4)}$ to develop an original tensor structure as $k$ is lowered. Since $\Gamma^{(4)}$ is the fourth field derivative of the effective action, it is a $4\times4\times4\times4$ symmetric tensor.\footnote{There are two fields each with real and imaginary parts.} We do not take this into account. Instead we impose that $\Gamma^{(4)}$ takes the same form as the fourth derivative of the original action $S$. See \Eq{eq_gamma4_tensor}, which shows the field-dependent part of the second field derivative of $\Gamma_k$. Instead of using a general form on the right-hand-side of the \RG flow equations, we replace $\Gamma^{(4)}$ by the second derivative of $\Gamma^I$. Then $\Gamma^{(4)}$ depends on a single complex parameter, $g(t)$. This approximation is partially justified by noting that terms of order larger than two in the quantum field are found to be irrelevant at criticality (see, \eg \cite{Sieberer2013b}). This is a semi-classical approximation \cite{tauber2014critical}, which is reflected in the structure of \Eq{eq_action}. Assuming that this remains the case here, we obtain that $\Gamma^{(4)}$ must vanish if more than two derivatives with respect to $\tilde{\phi}$ are taken and the resulting tensor structure is simplified. Moreover, the added tensor structure is absent at $k=\Lambda$. It is generated by the \RG flow and must therefore be small if $g(t)$ is small. Within perturbation theory, it can be neglected.
\end{inparaenum}

We emphasize that all these approximations are usually made within $1$-loop perturbation theory \cite{tauber2014critical}. We spell them out here for completeness.

$1$-loop perturbation theory is implemented in \Eq{eq_wetterich} by pulling out the derivative with respect to the running cut-off on the right-hand side and neglecting its effect on $\Gamma^{(2)}$ \cite{Papenbrock1995,Litim2002,Codello2014}
\begin{align}
 \text{Tr}\left[\frac{k\partial_k R}{\Gamma^{(2)}+R}\right] \rightarrow k\partial_k \text{Tr}\left[\text{ln}\left(\Gamma^{(2)}+R\right)\right] \, .
\end{align}
Then inserting the sharp cut-off operator provides \Eq{eq_oneloop},
\begin{align}
 k \partial_k \Gamma_k = - \frac{\I}{2} \text{Tr}\left\{\text{ln}\left(\Gamma_k^{(2)}[\tilde{\phi},\phi]\right)\right\}_k \, .
\label{eq_oneloop_app}
\end{align}
The trace on the right-hand side, which acts on operators (\ie objects with two sets of indexes), is defined as
\begin{align}
 \text{Tr}\left\{A\right\}_k = \int_{\vec{p},t} \sum_{i=1}^4 A_{ii}(\vec{p},t;-\vec{p},t) \delta(p-k) \, ,
 \label{eq_trace}
\end{align}
with the discrete index denoting the fields $\phi$ and $\tilde{\phi}$ and their complex conjugates. The delta function selects the momenta with modulus set to the running cut-off $p=k$. The logarithm in \Eq{eq_oneloop_app} is defined in terms of operator products,
\begin{align*}
 [AB]_{ij}(\vec{p},t;\vec{p}',t') = \int_{\vec{q},\tau} \hspace{-4pt} \sum_k A_{ik}(\vec{p},t;-\vec{q},\tau) B_{kj}(\vec{q},\tau;\vec{p}',t) ,
\end{align*}
through the Taylor expansion of the usual logarithm function. This product also serves to define the functional inverses [see, \eg \Eq{eq_rg_flow_gammaR}] through
\begin{align}
 [A A^{-1}]_{ij}(\vec{p},t;\vec{p}',t') = (2\pi)^d \delta_{ij} \delta(t-t') \delta(\vec{p}+\vec{p}') \, .
\end{align}

The \RG flow equations of $\mu$ and $g$ can now be extracted from \Eq{eq_oneloop_app}. We see that the flow equations are expressed in terms of the trace of the inverse of $\Gamma_k^{(2)}$. The second field derivative of $\Gamma_k$ is a $4\times4$ matrix defined as
\begin{align}
 \Gamma_k^{(2)}[\tilde{\phi},\phi] = \left. \frac{\delta^2 \Gamma_k}{\delta \phi_i(t) \delta \phi_j(t')} \right|_{\tilde{\phi},\phi} \, ,
\end{align}
with $\vec{\phi} = (\phi^*,\tilde{\phi}^*,\phi,\tilde{\phi})$. Under the above assumptions, $\Gamma^{(2)}[\tilde{\phi},\phi]$ takes the form
\begin{align}
\Gamma_k^{(2)}[\tilde{\phi},\phi] = \Gamma^0 +  \Gamma^I \delta(t-t')  \, .
\label{eq_gamma2_decomposition}
\end{align}
The first term is the kinetic term. It contains the inverse propagators and no field dependence,
\begin{align}
 \Gamma^{0} = \left(\begin{array}{cccc} 0& 0& 0 & G^{-1}_A(t,t') \\ 0 & 0 & G^{-1}_R(t,t') & P_K(t,t') \\ 0 & G^{-1}_R(t',t) & 0 & 0\\ G^{-1}_A(t',t) & P_K(t',t) & 0 & 0 \end{array} \right) \, .
\end{align}
The inverse propagators depend on time and momentum. They are defined in \Eq{eq_def_gamma}. The Green functions are given by the inverse of $\Gamma^{0}$. See \Eq{eq_greens}, where the lower left block of $[\Gamma^{0}]^{-1}$ is displayed. The second term of \Eq{eq_gamma2_decomposition} contains the contribution from the interaction,
\begin{align}
 & \Gamma^I = \nonumber \\
 & 2 \left( \begin{array}{cccc} g^* \tilde{\phi} \phi & \frac{g \phi \phi}{2} & g^* \tilde{\phi} \phi^* + g \tilde{\phi}^* \phi & g^* \phi^* \phi \\
                       \frac{g \phi \phi}{2} & 0 & g \phi \phi^* & 0 \\
                       g^* \tilde{\phi} \phi^* + g \tilde{\phi}^* \phi & g \phi \phi^* & g \tilde{\phi}^* \phi^* & \frac{g^* \phi^* \phi^*}{2} \\
                       g^* \phi \phi^* & 0 & \frac{g^* \phi^* \phi^*}{2} & 0
                      \end{array} \right) \, .
\label{eq_gamma4_tensor}
\end{align}
It depends on the fields and the complex time-dependent coupling $g$.

Now that we have set up all the necessary ingredients, we are ready to write down the \RG flow equations in real time. The flow equation of $\mu(t)$ can be extracted from the flow equation of $\Gamma_{R}$ by noting that (at $1$-loop in perturbation theory) the pre-factors of the space and time derivatives are not affected by the coarse-graining. \Ie the terms proportional to $K \nabla^2$ and $\I \partial_t$ do not depend on $k$. Then we have
\begin{align}
k \partial_k \Gamma_{R}(t,t') = k \partial_k \mu(t) \delta(t-t') \delta(\vec{p}-\vec{p}') \, .
\label{eq_gammar_to_mu}
\end{align}
The above equation will provide the left-hand side of \Eq{eq_real_time_flow_equations_mu}. The right-hand side is obtained by taking the second field derivative of the right-hand-side of \Eq{eq_oneloop_app},
\begin{align}
k\partial_k \Gamma_{R}(t,t') = - \frac{\I}{2} \text{Tr}\left\{ \frac{1}{\Gamma^{0}} \frac{\delta^2 \Gamma^{I}}{\delta \tilde{\phi}^*(t) \delta \phi(t') }\right\}_k \, .
\label{eq_rg_flow_gammaR}
\end{align}
The whole equation is evaluated at ${\phi=\tilde{\phi}=0}$. Multiplying the matrices, taking the trace and inserting \Eq{eq_gamma4_tensor} provides \Eq{eq_real_time_flow_equations_mu}
\begin{align}
 k \partial_k \mu(t) = -2 \I S_d k^d g(t) \, G_{K}(t,t) \, .
\end{align}
$g(t)$ is the time-dependent coupling and $G_K(t,t)$ is the Keldysh Green function. The momentum integration is performed according to \Eq{eq_trace} and reduced to the surface of a sphere of radius $k$ (leading to the pre-factor ${S_d = 2 \pi^{d/2}/[(d/2-1)! (2\pi)^{d}]}$) because the Green function only depends on the modulus of $\vec{p}$. Note that an overall factor of ${\delta(t-t') \delta(\vec{p}-\vec{p}')}$ was divided out from \Eq{eq_rg_flow_gammaR}.

The \RG flow equation of $g(t)$ [\Eq{eq_real_time_flow_equations_g}] is obtained in a similar way. The derivative of \Eq{eq_gamma4_Lambda} with respect to the running cut-off provides
\begin{align}
  k \partial_k & \Gamma^{(4)}(t_1,t_2,t_3,t_4) \nonumber  \\
  & =  2 k \partial_k g(t_1) \delta(t_1-t_2) \delta(t_1-t_3) \delta(t_1-t_4) \, ,
  \label{eq_gamma4_to_g}
\end{align}
while, taking four field derivatives [according to \Eq{eq_proj_gamma4}] of the right-hand side of \Eq{eq_oneloop_app} and then performing the matrix multiplications, momentum integrals and traces gives
\begin{align}
  & k \partial_k \Gamma^{(4)}(t_1,t_2,t_3,t_4) =  \label{eq_full_flow_g} \\
  = & \frac{\I}{2}\text{Tr}\left\{ \frac{1}{\Gamma^0} \frac{\delta^2 \Gamma^I}{\delta \tilde{\phi}^*(t_1) \delta \phi^*(t_2)} \frac{1}{\Gamma^0} \frac{\delta^2 \Gamma^I}{\delta \phi(t_3) \delta \phi(t_4)}\right\}_k \nonumber \\
 & + \frac{\I}{2}\text{Tr}\left\{\frac{1}{\Gamma^0} \frac{\delta^2 \Gamma^I}{\delta \tilde{\phi}^*(t_1) \delta \phi(t_3)} \frac{1}{\Gamma^0} \frac{\delta^2 \Gamma^I}{\delta \phi^*(t_2) \delta \phi(t_4)}\right\}_k \nonumber \\
&  + \frac{\I}{2}\text{Tr}\left\{\frac{1}{\Gamma^0} \frac{\delta^2 \Gamma^I}{\delta \tilde{\phi}^*(t_1) \delta \phi(t_4)} \frac{1}{\Gamma^0} \frac{\delta^2 \Gamma^I}{\delta \phi(t_3) \delta \phi^*(t_2)} \right\}_k \nonumber \\
 = &  S_d k^d 4 \I \nonumber \\
 & \times \big[ g(t_1) g(t_3) \delta(t_1-t_2) \delta(t_3-t_4) G_R(t_1,t_3) G_K(t_1,t_3) \nonumber \\ 
 & + g(t_1) g(t_2) \delta(t_1-t_3) \delta(t_4-t_2) G_R(t_1,t_4) G_K(t_4,t_1)  \nonumber \\
 & + g(t_1) g(t_2)^* \delta(t_1-t_3) \delta(t_2-t_4) G_A(t_4,t_1) G_K(t_1,t_4)  \nonumber \\
 & + g(t_1) g(t_3) \delta(t_1-t_4) \delta(t_3-t_2) G_R(t_1,t_2) G_K(t_2,t_1)  \nonumber \\
 & + g(t_1) g(t_3)^* \delta(t_1-t_4) \delta(t_2-t_3) G_A(t_2,t_1) G_K(t_1,t_2) \big] \, . \nonumber 
\end{align}
Finally we recover \Eq{eq_real_time_flow_equations_g} by integrating the above equation over the relative times $\tau_i = t_i -t_a$.

\subsection{Expansion of the flow to \texorpdfstring{$\mathcal{O}(\Omega^{-2})$}{O(1/(Omega*Omega))}}
\label{app_expansion_to_Omega2}

In this section, we give some details on the steps going from \Eq{eq_dimensionful_flow_coupling} to \Eq{eq_dimensionful_flow_equations_Omega2}. As we discuss in \Sect{sec_asymptotic_Omega_expansion}, the asymptotic expansion in powers of $\Omega^{-1}$ truncated to a given order is obtained from \Eq{eq_dimensionful_flow_coupling} by using the truncated expansion of the Green functions in powers of $E$ [\Eqs{eq_GRn} and \eq{eq_GKn}], performing the frequency integrals and expanding the result in powers of $\Omega^{-1}$. If the $E$-expansion is truncated at the same order than the desired order in the $\Omega^{-1}$ expansion, then the result is systematic in $\Omega^{-1}$.

This procedure is straightforward but can get quite long. We illustrate it for the flow of $\mu_n$ and to $\mathcal{O}(\Omega^{-1})$, and comment on the general case at the end of this section. To $\mathcal{O}(E^{1})$, \Eqs{eq_GRn} and \eq{eq_GKn} are inserted in \Eq{eq_dimensionful_flow_coupling}. Using the Residue theorem to perform the frequency integration provides
\begin{align}
 k\partial_k \mu_n = \frac{S_d k^d \gamma}{|M_0^I|} \hspace{-3pt} \left[\hspace{-2pt} -g_n \hspace{-3pt} + \hspace{-3pt}\sum_{m} \hspace{-2pt} \frac{g_m (E_{n-m}-E_{m-n}^*)}{2\I M_0^I+(n-m) \Omega} \right] \hspace{-3pt} .
 \label{eq_flow_mun_Eexpansion}
\end{align}
Expanding this equation to $\mathcal{O}(\Omega^{-1})$ provides the first of \Eqs{eq_dimensionful_flow_equations} of the main text.\footnote{It is necessary that $M_0^I>0$ for the theory to be stable. We assume that it is and remove the absolute values.} The definition of $E_n$ as being equal to $\mu_n$ if $n\neq 0 $ and $E_0=0$ plays an important role here. Indeed, without $E_0=0$, there would be an additional leading term arising when $n=m$, which would not depend on $\Omega$. This is straightforward (and not very interesting) in the above equation, but must be handled carefully in the flow of $g_n$, where such cancellations occur. For example, the flow of $g_n$ expanded to $\mathcal{O}(E^0)$ is given by
\begin{align}
 k\partial_k g_n = & \frac{S_d k^d \gamma}{|M_0^I|} \left[ \sum_m  \frac{g_m g_{n-m}}{2M_0 - (2m-n)\Omega/2} \right. \nonumber \\
 & \left. + \sum_m \frac{2(g_m -g_{-m}^*) g_{n-m}}{2\I M_0^I + (2m-n) \Omega/2} \right] \, .
 \label{eq_flow_g_E0}
\end{align}
We see that the leading term of the sum on the right-hand-side occurs when $2m=n$. This term only appears when $n$ is even and leads to the difference between the even and odd values of $n$ in \Eq{eq_dimensionful_flow_coupling}.

Both of the above equations contain terms of arbitrarily high order in $\Omega^{-1}$. These must be taken into account in the $\Omega^{-1}$-expansion, but if these equations are expanded to an order of $\Omega^{-1}$ that is larger than the corresponding order in the $E$-expansion, then there will be missing contributions to the $\Omega^{-1}$-expansion. For example, the expansion of \Eq{eq_flow_g_E0} to $\mathcal{O}(\Omega^{-2})$ is required to recover \Eq{eq_dimensionful_flow_equations_Omega2} and will produce all the terms that remain when $E$ is set to zero (\ie the terms with $[g_{r}^I]^2$ [$\mathcal{O}(\Omega^{0})$] and $G_{2r}$ [$\mathcal{O}(\Omega^{-2})$]). The additional terms are however still missing and come from the first and second order terms in the $E$-expansion.

We conclude this section with the comment that the asymptotic expansion to $\mathcal{O}(\Omega^{-1})$ [leading to \Eq{eq_dimensionful_flow_equations}] was performed manually, but that a computer was used to obtain the next order [\Eq{eq_dimensionful_flow_coupling}] because of the large amount of terms that emerge. The recipe (expand in powers of $E$, apply the residue theorem and expand in powers of $\Omega^{-1}$) remains the same, but the different steps together with the simplification of the obtained expressions were automated.

\subsection{Simplification of the flow equations}
\label{app_flow_drive_coefficients}

In this section, we use \Eq{eq_dimensionful_flow_equations_Omega2} to compute the \RG flow of the drive coefficients [\Eq{eq_drive_coefficients}] and show how these flow equations can be simplified in the presence of a monochromatic drive. This ultimately leads to \Eq{eq_dimensionful_flow_drive_parameters_final} that is used in the main text.

In this work, we are interested in the effect of the drive on the \IRD critical physics. For this reason we do not have to expand the flow equations of the drive parameters to $\mathcal{O}(\Omega^{-2})$. We keep only the terms that will lead to corrections of order $\mathcal{O}(\Omega^{-2})$ or lower in the flow of $\mu_0^I$ and $g_0^I$, \Eq{eq_dimensionful_flow_equations_Omega2}. Specifically, the flow of $X_0$ is truncated to $\mathcal{O}(\Omega^{-1})$ while only the leading term is included in the flow of the other parameters. Inserting \Eq{eq_dimensionful_flow_equations_Omega2} into the derivatives of the drive parameters \Eq{eq_drive_coefficients}, with respect to the running cut-off then provides
\begin{align}
 & k \partial_k M_2 \cong -\frac{2 S_d k^d \gamma}{M_0^I} R_0 \, , \nonumber \\
 & k \partial_k X_0 \cong \frac{2\I S_d k^d \gamma}{M_0^I \Omega} \sum_{m\neq0} \frac{g_{-m}^I  X_m}{m} \nonumber \\
 & \qquad +  \frac{5 \I S_d k^d \gamma}{2 [M_0^I]^2}\sum_{m\neq0} \frac{\mu_{-2m}^I g_m^I}{m} \left( g_m^I + \frac{X_m}{\Omega} \right) \, ,\nonumber \\
 & k \partial_k S_0 \cong \frac{2S_d k^d \gamma}{M_0^I} \sum_{\substack{m\neq0,m_1\neq 0\\m\neq-m_1}} \frac{g_{m_1}^I g_m^I \mu_{-m-m_1}^I}{m(m+m_1)} \nonumber \\
 & \qquad - \frac{5 S_d k^d \gamma}{2[M_0^I]^2} \sum_{\substack{m\neq0,m_1\neq0\\m\neq -2m_1}} \frac{\mu_m^I \mu_{-m-2m_1}^I [g_{m_1}^I]^2}{m(m+2m_1)} \, , \nonumber \\
 & k \partial_k R_0 \cong -\frac{4 S_d k^d \gamma}{M_0^I} G_0 + \frac{5 S_d k^d \gamma}{8 [M_0^I]^2} \sum_{m\neq0} \frac{\mu_{-2m}^I [g_m^I]^2}{m^2} \, ,\nonumber \\
 & k \partial_k G_0 \cong \frac{5 S_d k^d \gamma}{16 [M_0^I]^2} \sum_{m\neq0} \frac{g_{-2m}^I[g_m^I]^2}{m^2} \, ,\nonumber \\
 & k \partial_k Q_0 \cong \frac{5 S_d k^d \gamma}{8 [M_0^I]^2} \sum_{\substack{m\neq0,m_1\neq0\\2m\neq -m_1}} \frac{\mu_{-2m-m_1}^I g_{m_1}^I [g_{m}^I]^2}{m m_1}  \, .
 \label{eq_dimensionful_flow_drive_parameters_1}
\end{align}
In the derivation of the flow equation of $Q_0$ we used the fact that
\begin{align}
 \sum_{\substack{m\neq0,m_1\neq0\\m\neq-m_1}} \frac{g_m^I g_{m_1}^I g_{-m-m_1}^I}{m m_1} = 0 \, .
\end{align}
This identity emerges because of the symmetry of the internal indexes. It can be seen by tripling the sum and making two different changes of variables ($m_1 = -m-m_1'$ and $m = -m_1 -m'$) in the two additional expressions. Then the sum of the three terms vanishes.

The $k$-dependence of $g_{\pm1}^I$ can be neglected in \Eq{eq_dimensionful_flow_drive_parameters_1} because its flow starts at $\mathcal{O}(\Omega^{-2})$ only [see \Eq{eq_dimensionful_flow_equations_Omega2}]. This is particularly useful in the case of a monochromatic drive because it means that the Fourier components of $g_n^I$ are all generated from $g_{\pm1}^I$ in a very simple way: If $n$ is not a power of $2$ then $g_n^I= 0$. If $|n|=2^r$, then $g_{\pm2^r}^I = I_{\pm,r}(k) \, [g_{\pm1}^I]^{2^r}$. All the $k$-dependence is relegated to the pre-factor $I_r(k)$, which is a complicated multi-dimensional integral over the pre-factor of \Eq{eq_dimensionful_flow_equations_Omega2}. The simplest example of this (beyond $I_{\pm,0}(k) = 1$) is
\begin{align}
 g_{\pm 2} = [g_{\pm 1}^I]^2 \frac{5 S_d \gamma}{2} \int_{\Lambda}^k \frac{\tilde{k}^{d-1}}{[M_0^I(\tilde{k})]^2} \text{d}\tilde{k} + \mathcal{O}(\Omega^{-2}) \, ,
 \label{eq_gnot1}
\end{align}
as can be seen from \Eq{eq_dimensionful_flow_equations_Omega2}. The initial condition $g_{\pm2}(\Lambda) = 0$ is a consequence of the monochromaticity of the drive. $g_{\pm 4}$ is then obtained by inserting the above equation into \Eq{eq_dimensionful_flow_equations_Omega2} and leads to a similar equation, although with a different $k$-dependent pre-factor. This procedure can in principle be iterated up to any value of $r$, and it is clear that it does not depend on the sign of $n$, $I_{+,r}(k) = I_{-,r}(k)=I_{r}(k)$. The flow of $\mu_n^I$ behaves in the same way, $\mu_{\pm 2^r}^I = J_r(k) [g_{\pm1}^I]^{2^r}$, although up to $\mathcal{O}(\Omega^{0})$ and not for $r=0$. This leads to
\begin{align}
 & \sum_{m\neq0} \frac{\mu_{-2m}^I [g_m^I]^2}{m} \sim \Omega^{-1} \, . 
\end{align}
The flow of all the drive parameters actually starts at $\mathcal{O}(\Omega^{-1})$. In particular $X_0$ can be taken as a constant if the problem is truncated to $\mathcal{O}(\Omega^{-1})$, as in \cite{mathey2018a}.

Finally, we use the fact that critical physics is obtained by linearising the flow close to the \IRD \RG fixed point, where all the drive coefficients vanish. Then it is only necessary to account for the terms that are linear in the drive coefficients. The rest will have no effect on the critical properties. We implement this simplification by using \Eq{eq_gnot1} and its generalization, $g_{\pm2^r}^I = I_{r}(k) \, [g_{\pm1}^I]^{2^r}$ and neglecting all the terms of order $\mathcal{O}(g_{\pm1}^{3})$ and higher. Then we find that $Q_0$ together with its flow vanish. Moreover, the flow equations can be closed if we introduce the additional variables
\begin{align}
 U = \sum_{\substack{m\neq0,m_1\neq0\\m\neq -2m_1}} \frac{\mu_m^I \mu_{-m-2m_1}^I [g_{m_1}^I]^2}{m(m+2m_1)} \, ,
 \label{eq_defU}
\end{align}
that does not flow, $k\partial_k U = 0$. The end result is given by \Eq{eq_dimensionful_flow_drive_parameters_final}.

\section{Wilson--Fisher fixed point}
\label{app_wilson_fisher}

In this section, we show how thermal equilibrium and \WF physics is recovered in our set-up. In particular we recover the first order $\epsilon$-expansion (see, \eg \cite{tauber2014critical}) of the equilibrium $O(2)$ model.

Because the action \eq{eq_action} is at most quadratic in $\tilde{\phi}$, the corresponding dynamics can be formulated as stochastic driven open Gross--Pitaevskii dynamics,\footnote{We extract a factor $2$ on the right-hand-side because we are comparing our calculation to the $O(2)$ model by identifying the real and imaginary parts of $\phi$ with the two components. There the model A dynamics is defined by $\partial_t \phi_i = - \delta H/\delta \phi_i + \xi_i$. The factor $2$ cancels out the $2$ in the denominators of $\phi_1 = (\phi+\phi^*)/2$ and $\phi_2 = (\phi-\phi^*)/(2\I)$.}
\begin{align}
 \I \partial_t \phi = - 2\left( \frac{\delta H_c}{\delta \phi^*} + \I \frac{\delta H_d}{\delta \phi^*}\right) + \xi \, ,
 \label{eq_ddgpe_hohenberg_halperin}
\end{align}
with a zero-average stochastic noise term ($\xi = \xi_1 + \I \xi_2$)
\begin{align}
  \langle \xi_i(\tx) \xi_j(\txp) \rangle = 2 \gamma \delta_{ij} \delta(t-t') \delta(\vec{x}-\vec{x}') \, ,
\end{align}
and a complex Hamiltonian functional
\begin{align}
 H = \frac{1}{2} \int_{\tx} - K \phi^* \nabla^2 \phi + \mu \left|\phi\right|^2 + \frac{g}{2} \left|\phi\right|^4 \, .
\end{align}
$H_{c/d}$ are the real and imaginary parts of ${H = H_c+\I H_d}$. They model coherent and dissipative dynamics respectively. See \cite{Tauber2013a} and references therein.

The stochastic dissipative dynamics of a non-conserved complex field $\phi$, which is the equation of motion of the $2$-component model A of \cite{Hohenberg1977a} is recovered from \Eq{eq_ddgpe_hohenberg_halperin} by setting $H_c=0$ and inserting time-independent coupling (see, \eg \cite{Sieberer2013a,Sieberer2013b,Tauber2013a}). This is equivalent to choosing $\text{Re}[\mu] = \text{Re}[g] = \text{Re}[K] = 0$ (and keeping the couplings constant) in \Eq{eq_action}. Then we recover relaxational dynamics close to thermal equilibrium with the corresponding Hamiltonian is given by $H_d$.

We now show that the above correspondence holds down to the level of the \RG flow equations \Eq{eq_dimensionful_flow_coupling}. To this end we choose ${\mu(t) = \I \mu_0^I}$, ${g(t) = \I g_0^I}$ and ${K = \I/(2m)}$. Then the Wigner Green functions are computed straightforwardly,
\begin{align*}
 & G_{R;n} = \frac{\delta_{n0}}{\omega + K p^2 +\mu}\, , && G_{K;n} = \frac{-\I \delta_{n0} \gamma}{\left|\omega + K p^2 +\mu\right|^2} \, .
\end{align*}
The integrals of \Eq{eq_dimensionful_flow_coupling} can be performed analytically and provide,
\begin{align}
 && k & \partial_k \mu_0^I = -\gamma S_d k^d \frac{g_0^I}{\left|\frac{k^2}{2m}+\mu_0^I\right|}\, , \nonumber \\
 && k & \partial_k g_0^I = \frac{5\gamma S_d k^d}{2} \frac{(g_0^I)^2}{\left(\frac{k^2}{2m}+\mu_0^I\right)\left|\frac{k^2}{2m}+\mu_0^I\right|} \, .
 \label{eq_nondriven_dimensionfull}
\end{align}
Finally, we rescale the couplings according to
\begin{align}
 \hat{\mu} = \frac{2m}{k^2} \mu_0^I \, , && \hat{g} = \gamma m^2 k^{d-4} g_0^I \, ,
\end{align}
and write
\begin{align}
& k \partial_k \hat{\mu} = -2 \hat{\mu} - \frac{4 S_d \hat{g}}{\left|1+\hat{\mu}\right|}\, , \nonumber \\
& k \partial_k \hat{g} = (d-4) \hat{g} + \frac{10 S_d \hat{g}^2}{\left(1+\hat{\mu}\right)\left|1+\hat{\mu}\right|} \, .
\end{align}
To linear order in $\epsilon = 4-d$, we find the \WF fixed point with coordinates and critical exponent
\begin{align}
 \hat{\mu}^* \cong - \frac{\epsilon}{5} \, , && \hat{g}^* \cong \frac{4 \pi^2 \epsilon}{5} \, , && \nu = \frac{1}{2} + \frac{\epsilon}{10} \, .
\end{align}
See, \eg \cite{tauber2014critical}, where these results are obtained in the equilibrium case.

\section{Scaling behavior}
\label{app_scaling_behaviour}

In this section, we give additional details on the scaling analysis of the rapidly driven system. We start by giving an explicit expression for the rescaled drive parameters:
\begin{align}
&  x = \frac{\gamma k^{-\epsilon}}{4 [K^I]^2} \frac{X_0}{\Omega} \, , && m =  \frac{M_2}{\Omega^2} \, , \nonumber \\
& s = \frac{\gamma k^{-\epsilon}}{4 [K^I]^2} \frac{S_0}{\Omega^2} \, , &&  r = \frac{\gamma k^{d-2}}{4 K^I} \frac{R_0}{\Omega^2} \, , \nonumber \\
& g = \frac{\gamma^2 k^{2d-4}}{16 [K^I]^2} \frac{G_0}{\Omega^2} \, , && u = \frac{\gamma^2 k^{-2\epsilon}}{16 [K^I]^4} \frac{U}{\Omega^2} \, ,
\label{eq_rescaled_drive_parameters}
\end{align}
which are used in \Eq{eq_flow_equations_Omega2_zero}.

Next, we explicitly compute the stability matrix of the periodically driven system. We will show how the coefficients $c_i$ of \Eq{eq_flow_general} are related to the microscopic couplings. In particular we find that the relevant couplings $c_{6,7,8}$ can only vanish when the corresponding drive coefficients $x$, $u$ and $s$ vanish themselves. Moreover, we show how to derive the phase diagram of \Fig{fig_phase_diagram}{a}.

The stability matrix is the Jacobian of the flow equations evaluated at the fixed point \Eq{eq_wf_fixed_point}. It is obtained by taking partial derivatives of every element of \Eq{eq_flow_equations_Omega2_zero} with respect to all the couplings and evaluating them at $\hat{\mu}_0 = \hat{\mu}^*$, $\hat{g}_0 = \hat{g}^*$ and $m=r=g=s=u=x=0$. The gradients of the elements of \Eq{eq_flow_equations_Omega2_zero} are vectors, which provide the lines of $A$. To $\mathcal{O}(\epsilon)$ we obtain
\begin{widetext}
 \begin{align}
 & A =  \left( \begin{array}{cccccccc}
             -2 + \frac{2\epsilon}{5} & -\frac{10+\epsilon \Gamma }{20\pi^2} & -\frac{8 \epsilon}{5} & \frac{10+\epsilon\left(2+\Gamma\right)}{5\pi^2} & 0 & \frac{10+\epsilon \left(\Gamma+4\right)}{10\pi^2} & 0 & -\frac{10+\epsilon \Gamma}{20 \pi^2} \\
             0 & \epsilon & 0 & -8 \epsilon & \frac{10 \left[10+\epsilon\left(\Gamma+2\right)\right]}{5\pi^2} & -\frac{3\epsilon}{10} & 0 & \epsilon \\
             0 & 0 & 0 & \frac{10-\epsilon \Gamma}{10\pi^2} & 0 & 0 & 0 & 0 \\
             0 & 0 & 0 & 2-\epsilon & -\frac{10+\epsilon \Gamma}{5 \pi^2} & 0 & 0 & 0  \\
             0 & 0 & 0 & 0 & 4-2\epsilon & 0 & 0 & 0 \\
             0 & 0 & 0 & 0 & 0 & -\epsilon & -\frac{10+\epsilon\left(\Gamma-2\right)}{4\pi^2} & 0 \\
             0 & 0 & 0 & 0 & 0 & 0 & -2\epsilon & 0 \\
             0 & 0 & 0 & 0 & 0 & 0 & 0 & -\epsilon 
            \end{array}\right) \, ,
            \label{eq_stability_matrix}
\end{align}
\end{widetext}
with $\Gamma = 5[\ln(4\pi)-c]+3 \cong 12.77$ (and $c \cong 0.58$ Euler's constant).

The above upper-triangular form of the stability matrix makes it particularly easy to diagonalize. The eigenvalues [\Eq{eq_eigenvalues}] are simply the diagonal elements of $A$. The eigenvectors of $A$ can also be computed, and have a simple structure as a result of the upper-triangular structure: The first eigenvector (with eigenvalue $\lambda_1 = -2+2\epsilon/5$) is $\overrightarrow{v_1} = (1,0,0,0,0,0,0,0)$, the second eigenvector is $\overrightarrow{v_2} = (1,x_2,0,0,0,0,0,0)$ (with $x_2$ some constant), the third is $\overrightarrow{v_3} = (1,x_3,y_3,0,0,0,0,0)$, and so on. The constants $x_i$ can be complicated functions of $\epsilon$, but the important part is that $\overrightarrow{v_i}^j=0$ if $j>i$. This makes it easy to relate the coefficients of the linear combination
\begin{align}
 \overrightarrow{G} = \sum_{i=1,8} c_i(k) \overrightarrow{v_i} \, , && c_i(k) = c_i \, \left(\frac{k}{\Lambda}\right)^{\lambda_i} \, ,
\end{align}
[\Eq{eq_flow_general}] to the microscopic couplings. In particular, since the only eigenvector with a nonzero entry at the end is $\overrightarrow{v_8}$, we must have $c_8 \sim x$. Then $c_7$ must be a linear combination of $x$ and $u$ only because only $\overrightarrow{v_7}$ and $\overrightarrow{v_8}$ contribute to its determination. This goes on all the way up to $c_1$ which is in principle a linear combination of all the couplings. In particular we see that if we completely switch off the drive, then $c_{3,\dots ,8}=0$ and only the \IRD critical exponents are activated. Moreover, $c_{6,7,8}$ depend on $x$, $u$ and $s$ exclusively. Neither the driven irrelevant ($m$, $r$ and $g$) nor the undriven couplings ($\delta g$ and $\delta \mu$) enter in their determination. This means that when $c_{6,7,8}=0$, then $x=u=s=0$ and reciprocally.

The phase of the system  is determined by the sign of $c_1$ since it determines weather $\mu_0$ is positive or negative at large scales, see \Fig{fig_phase_diagram}{a}. With the eigenvectors normalized to one, we find that
\begin{align}
 c_1 & = \delta \mu + \frac{x + \delta g-2 r - 2 s}{4\pi^2} - \frac{8 g + 5 u}{8 \pi^4} + \mathcal{O}(\epsilon) \nonumber \\
 & = \frac{1}{4\pi^2} \left(\frac{\Delta}{a} + x -2 r - 2 s - \frac{8 g + 5 u}{2 \pi^4} \right) + \mathcal{O}(\epsilon) \, .
 \label{eq_c1}
\end{align}
We have inserted $\Delta = a(\delta g + 4\pi^2 \delta \mu)$ (with $a>0$ a non-universal constant) in the second line. In the absence of drive $\Delta$ would be identified with the reduced temperature. Setting $c_1=0$ produces a linear relation between $\Delta$ and the drive coefficients with a non-universal yet finite slope. This is represented in \Fig{fig_phase_diagram}{a} as a tilted black line.

%

\end{document}